\documentclass[%
 reprint,
 amsmath,amssymb,
 aps,
 prb,
]{revtex4-2}

\usepackage{graphicx}
\usepackage{dcolumn}
\usepackage{bm}
\usepackage{hyperref}
\usepackage{color}
\usepackage{braket}
\usepackage{physics}
\usepackage{booktabs}
\usepackage{tikz}
\usetikzlibrary{arrows.meta,calc,decorations.pathreplacing,shadings}

\definecolor{subA}{HTML}{C8423B}      
\definecolor{subB}{HTML}{2F6DB5}      
\definecolor{spinF}{HTML}{7A3E9D}     
\definecolor{fieldB}{HTML}{2E8B3E}    
\definecolor{ivcN}{HTML}{D9831F}      
\definecolor{potHi}{HTML}{C8423B}     
\definecolor{potLo}{HTML}{2F6DB5}     
\definecolor{bondgray}{gray}{0.55}
\definecolor{kekule}{gray}{0.18}

\newcommand{\tkview}[2]{%
  \pgfmathsetmacro{\tkvxX}{-sin(#1)}%
  \pgfmathsetmacro{\tkvxY}{-sin(#2)*cos(#1)}%
  \pgfmathsetmacro{\tkvyX}{cos(#1)}%
  \pgfmathsetmacro{\tkvyY}{-sin(#2)*sin(#1)}%
  \pgfmathsetmacro{\tkvzY}{cos(#2)}%
  \pgfmathsetmacro{\tkvgamma}{#1}%
}
\tikzset{
  tkview/.style={x={(\tkvxX cm,\tkvxY cm)},
                 y={(\tkvyX cm,\tkvyY cm)},
                 z={(0cm,\tkvzY cm)}},
  spin/.style={-{Stealth[length=4.2pt,width=3.6pt]}, line width=1.1pt},
  vec/.style={-{Stealth[length=5pt,width=4pt]}, line width=1.2pt},
  axis/.style={-{Stealth[length=3.5pt,width=2.8pt]}, line width=0.4pt},
  bond/.style={line width=0.6pt, bondgray, line cap=round},
  siteA/.style={circle, fill=subA, draw=white, line width=0.3pt,
                inner sep=0pt, minimum size=4.2pt},
  siteB/.style={circle, fill=subB, draw=white, line width=0.3pt,
                inner sep=0pt, minimum size=4.2pt},
  panel/.style={font=\small\bfseries, anchor=north west, inner sep=0pt},
  blochzaxis/.style={},   
}

\newcommand{\blochsphere}[4]{%
  \shade[inner color=white, outer color=gray!18] (0,0) circle (#1);
  \draw[line width=0.5pt] (0,0) circle (#1);
  \begin{scope}[tkview]
    \draw[gray!70, line width=0.35pt, dashed]
      plot[domain=\tkvgamma+90:\tkvgamma+270, samples=60, smooth]
        ({#1*cos(\x)},{#1*sin(\x)},0);
    \draw[gray!70!black, line width=0.4pt]
      plot[domain=\tkvgamma-90:\tkvgamma+90, samples=60, smooth]
        ({#1*cos(\x)},{#1*sin(\x)},0);
    \draw[axis] (0,0,0) -- (1.32*#1,0,0) node[anchor=north east, inner sep=1pt] {#2};
    \draw[axis] (0,0,0) -- (0,1.30*#1,0) node[anchor=west, inner sep=1.5pt] {#3};
    \draw[axis, blochzaxis] (0,0,0) -- (0,0,1.28*#1) node[anchor=south, inner sep=1.5pt] {#4};
  \end{scope}%
}

\newcommand{\hcset}[3]{\def\hcR{#1}\def\hcK{#2}\def\hcb{#3}}
\newcommand{\hcbonds}{%
  \foreach \hcr in {0,...,\hcR}{%
    \draw[bond] plot[samples at={0,...,\hcK}]
      ({-(1.5*\hcr + 0.5*mod(\x+\hcr,2))*\hcb}, {0.8660254*\x*\hcb}, 0);
  }%
  \pgfmathtruncatemacro{\hcRm}{\hcR-1}%
  \foreach \hcr in {0,...,\hcRm}{%
    \foreach \hck in {0,...,\hcK}{%
      \pgfmathtruncatemacro{\hcpar}{mod(\hck+\hcr,2)}%
      \ifnum\hcpar=1
        \draw[bond] ({-(1.5*\hcr+0.5)*\hcb},{0.8660254*\hck*\hcb},0)
                 -- ({-(1.5*\hcr+1.5)*\hcb},{0.8660254*\hck*\hcb},0);
      \fi
    }%
  }%
}
\newcommand{\hcsites}[2]{%
  \foreach \hcr in {\hcR,...,0}{%
    \foreach \hchi in {1,0}{
      \foreach \hck in {0,...,\hcK}{%
        \pgfmathtruncatemacro{\hcpar}{mod(\hck+\hcr,2)}%
        \ifnum\hcpar=\hchi
          \pgfmathsetmacro{\hcx}{-(1.5*\hcr+0.5*\hcpar)*\hcb}%
          \pgfmathsetmacro{\hcy}{0.8660254*\hck*\hcb}%
          \ifnum\hcpar=0 #1\else #2\fi
        \fi
      }%
    }%
  }%
}
\newcommand{\hcedge}[5]{%
  \pgfmathtruncatemacro{\hera}{#2}\pgfmathtruncatemacro{\heka}{#3}%
  \pgfmathtruncatemacro{\herb}{#4}\pgfmathtruncatemacro{\hekb}{#5}%
  \pgfmathtruncatemacro{\hein}{(\hera>=0)*(\hera<=\hcR)*(\herb>=0)*(\herb<=\hcR)*%
    (\heka>=0)*(\heka<=\hcK)*(\hekb>=0)*(\hekb<=\hcK)}%
  \ifnum\hein=1
    \draw[#1] ({-(1.5*\hera+0.5*mod(\heka+\hera,2))*\hcb},{0.8660254*\heka*\hcb},0)
           -- ({-(1.5*\herb+0.5*mod(\hekb+\herb,2))*\hcb},{0.8660254*\hekb*\hcb},0);
  \fi
}
\newcommand{\hckekule}[1]{%
  \pgfmathtruncatemacro{\hcKp}{\hcK+1}%
  \foreach \hcr in {-1,...,\hcR}{%
    \foreach \hck in {-1,...,\hcKp}{%
      \pgfmathtruncatemacro{\hcpar}{mod(\hck+\hcr+4,2)}%
      \pgfmathtruncatemacro{\kek}{mod(mod((\hck-3*\hcr)/2,3)+3,3)}%
      \ifnum\hcpar=0 \ifnum\kek=0
        \hcedge{#1}{\hcr}{\hck}{\hcr}{\hck+1}%
        \hcedge{#1}{\hcr}{\hck+1}{\hcr+1}{\hck+1}%
        \hcedge{#1}{\hcr+1}{\hck+1}{\hcr+1}{\hck}%
        \hcedge{#1}{\hcr+1}{\hck}{\hcr+1}{\hck-1}%
        \hcedge{#1}{\hcr+1}{\hck-1}{\hcr}{\hck-1}%
        \hcedge{#1}{\hcr}{\hck-1}{\hcr}{\hck}%
      \fi\fi
    }%
  }%
}

\providecommand{\tkpanelA}{a}
\providecommand{\tkpanelB}{b}

\hypersetup{
    colorlinks=true,
    linkcolor=blue,
    filecolor=magenta,      
    urlcolor=blue,
    citecolor=blue,
    pdfpagemode=None
}

\newcommand{\evalat}{\bigg|}

\providecommand{\norm}[1]{\left\Vert#1\right\Vert}

\providecommand{\hal}{\frac{1}{2}}

\providecommand{\ip}[2]{\langle#1|#2\rangle}

\DeclareMathOperator{\sgn}{sgn}

\def\Z{\mathbb{Z}}

\def\ep{\varepsilon}

\newcommand{\ee}{\mathrm{ee}}
\newcommand{\elph}{\mathrm{ep}}
\newcommand{\ph}{\mathrm{ph}}
\newcommand{\pp}{\mathrm{pp}}
\newcommand{\bg}{\mathbf g}

\begin{document}

\title{Superlattice Stabilization of the Topological Quantum Hall Ferromagnet in Graphene}

\author{Tiago Cecchi}
 \email{tcecchi@alumni.cmu.edu}
\author{Ahmed Khalifa}
\thanks{Corresponding authors: ahmedk@andrew.cmu.edu and shubhayuchatterjee@cmu.edu}
\author{Shubhayu Chatterjee}
\thanks{Corresponding authors: ahmedk@andrew.cmu.edu and shubhayuchatterjee@cmu.edu}
 \affiliation{Department of Physics, Carnegie Mellon University}

\date{\today}

\begin{abstract}
Competition between spin and valley orders makes graphene a versatile platform for engineering correlated topological phases in the quantum Hall regime. 
We show that a tunable superlattice potential reorganizes this competition, and stabilizes the ferromagnetic quantum Hall topological insulator at charge neutrality without requiring a large in-plane magnetic field or strong dielectric suppression of Coulomb interactions. 
The interaction anisotropies responsible for choosing between different competing states are first computed by a generalized renormalization group procedure that accounts for the superlattice potential.
We find that the potential suppresses anisotropies that compete with spin-polarization. 
At the same time, the potential introduces dispersion which prefers a partly spin-polarized compensated semi-metal. 
Therefore, we carry out a self-consistent momentum-space Hartree-Fock computation with renormalized aniostropies to establish the regime of stability of the insulating spin-polarized ferromagnet. 
Finally, we verify that this ferromagnet is adiabatically connected to the quantum Hall ferromagnet, thereby confirming its topological character.  Our results establish superlattice potentials as a means of controlling the phase diagram of interacting graphene and accessing its helical quantum Hall regime.
\end{abstract}

\maketitle

\section{Introduction}
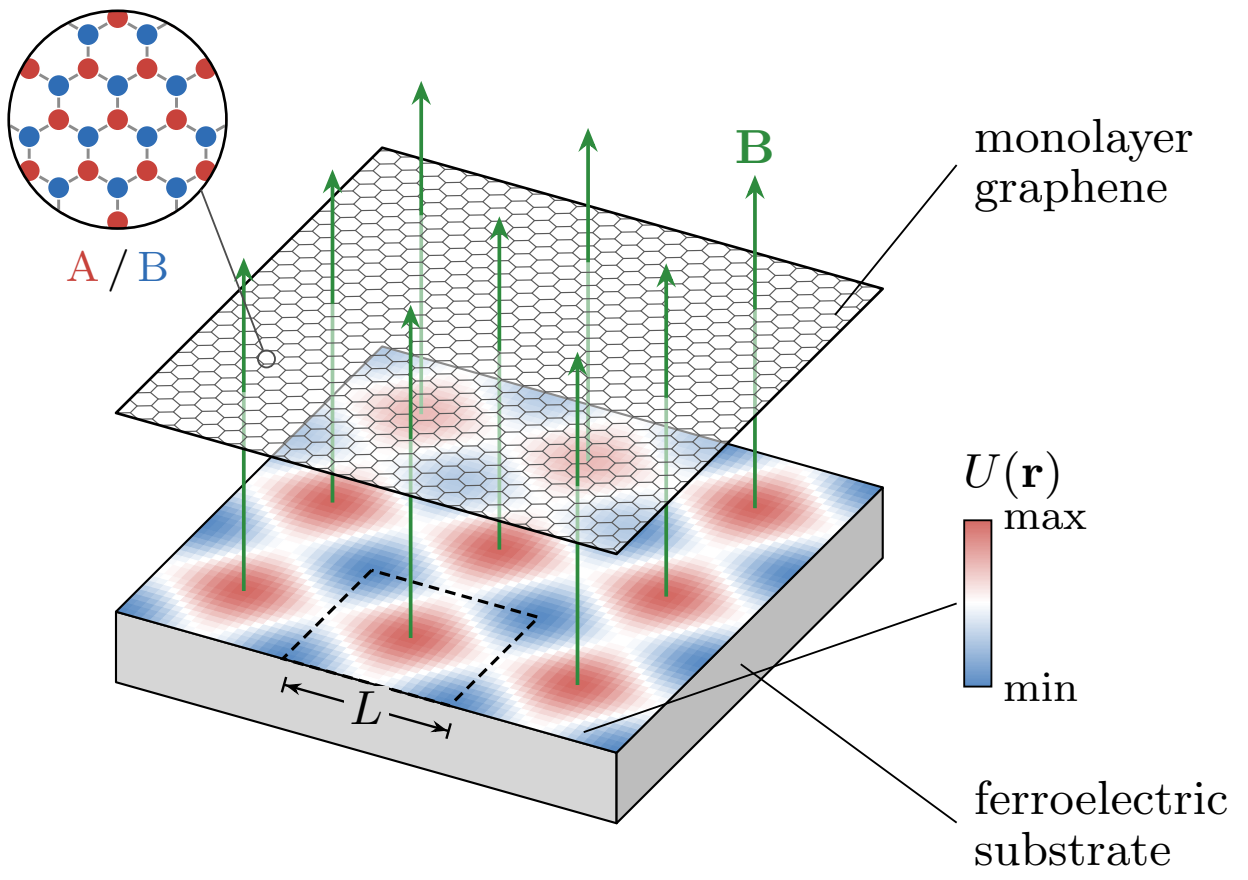
\begin{figure}[t]
    \centering
\begin{tikzpicture}[font=\small]
  \tkview{28}{32}
  \def\L{1.25}          
  \def\nL{3}            
  \def\Ncell{20}        
  \def\T{0.55}          
  \def\h{1.55}          
  \pgfmathsetmacro{\X}{\nL*\L}
  \pgfmathsetmacro{\Y}{\nL*\L}
  \pgfmathsetmacro{\d}{\L/\Ncell}
  \pgfmathtruncatemacro{\Ntot}{\nL*\Ncell-1}

  \begin{scope}[tkview]
    \filldraw[fill=gray!32, draw=black, line width=0.4pt, line join=round]
      (\X,0,0) -- (\X,\Y,0) -- (\X,\Y,-\T) -- (\X,0,-\T) -- cycle;
    \filldraw[fill=gray!52, draw=black, line width=0.4pt, line join=round]
      (0,\Y,0) -- (\X,\Y,0) -- (\X,\Y,-\T) -- (0,\Y,-\T) -- cycle;

    \foreach \i in {0,...,\Ntot}{
      \foreach \j in {0,...,\Ntot}{
        \pgfmathsetmacro{\U}{-0.5*(cos(360/\Ncell*(\i+0.5))+cos(360/\Ncell*(\j+0.5)))}
        \pgfmathtruncatemacro{\p}{round(80*abs(\U))}
        \ifdim\U pt>0pt \def\c{potHi!\p!white}\else\def\c{potLo!\p!white}\fi
        \filldraw[fill=\c, draw=\c, line width=0.15pt]
          (\i*\d,\j*\d,0) -- (\i*\d+\d,\j*\d,0) -- (\i*\d+\d,\j*\d+\d,0)
          -- (\i*\d,\j*\d+\d,0) -- cycle;
      }
    }
    \draw[line width=0.4pt, line join=round]
      (0,0,0) -- (\X,0,0) -- (\X,\Y,0) -- (0,\Y,0) -- cycle;

    \draw[line width=0.6pt, dash pattern=on 2pt off 1.2pt]
      (2*\L,\L,0) -- (3*\L,\L,0) -- (3*\L,2*\L,0) -- (2*\L,2*\L,0) -- cycle;
    \draw[{Bar[width=3pt] Stealth[length=3pt]}-{Stealth[length=3pt] Bar[width=3pt]},
          line width=0.45pt, black]
      (\X,\L,-0.2) -- node[pos=0.5, fill=gray!32, inner sep=1pt] {$L$}
      (\X,2*\L,-0.2);

    \foreach \i in {0.5,1.5,2.5}{
      \foreach \j in {0.5,1.5,2.5}{
        \draw[fieldB, line width=0.7pt] (\i*\L,\j*\L,0) -- (\i*\L,\j*\L,\h);
      }
    }

    \fill[white, opacity=0.55]
      (0,0,\h) -- (\X,0,\h) -- (\X,\Y,\h) -- (0,\Y,\h) -- cycle;
    \begin{scope}
      \clip (0,0,\h) -- (\X,0,\h) -- (\X,\Y,\h) -- (0,\Y,\h) -- cycle;
      \begin{scope}[shift={(\X,0,\h)},
                    bond/.style={line width=0.22pt, black!65, line cap=round}]
        \pgfmathtruncatemacro{\gR}{\X/(1.5*0.095)+1}
        \pgfmathtruncatemacro{\gK}{\Y/(0.8660254*0.095)+1}
        \hcset{\gR}{\gK}{0.095}
        \hcbonds
      \end{scope}
    \end{scope}
    \draw[line width=0.5pt, line join=round]
      (0,0,\h) -- (\X,0,\h) -- (\X,\Y,\h) -- (0,\Y,\h) -- cycle;

    \foreach \i in {0.5,1.5,2.5}{
      \foreach \j in {0.5,1.5,2.5}{
        \draw[fieldB, line width=0.7pt, -{Stealth[length=4.5pt,width=3.6pt]}]
          (\i*\L,\j*\L,\h) -- (\i*\L,\j*\L,\h+1.05);
      }
    }
    \node[fieldB, anchor=south, inner sep=2pt] at (0.5*\L,2.5*\L,\h+1.05) {$\mathbf B$};

    \coordinate (zoomspot) at (2.15*\L,0.45*\L,\h);
    \coordinate (graphL)   at (0.35*\L,2.9*\L,\h);
    \coordinate (subsL)    at (1.6*\L,\Y,-0.5*\T);
    \coordinate (potL)     at (2.9*\L,2.75*\L,0);
  \end{scope}

  \begin{scope}[shift={(-1.75,1.5)}]
    \fill[white] (0,0) circle (0.72);
    \begin{scope}
      \clip (0,0) circle (0.72);
      \foreach \m in {-5,...,5}{
        \foreach \n in {-3,...,3}{
          \pgfmathsetmacro{\cx}{0.19486*(2*\m+\n)}  
          \pgfmathsetmacro{\cy}{0.3375*\n}          
          \draw[bondgray, line width=0.55pt]
            (\cx,\cy) -- ++(0,0.225)
            (\cx,\cy) -- ++(0.1949,-0.1125)
            (\cx,\cy) -- ++(-0.1949,-0.1125);
        }
      }
      \foreach \m in {-5,...,5}{
        \foreach \n in {-3,...,3}{
          \pgfmathsetmacro{\cx}{0.19486*(2*\m+\n)}
          \pgfmathsetmacro{\cy}{0.3375*\n}
          \node[siteA] at (\cx,\cy) {};
          \node[siteB] at (\cx,\cy+0.225) {};
        }
      }
    \end{scope}
    \draw[line width=0.5pt] (0,0) circle (0.72);
    \node[font=\footnotesize, anchor=north] at (0,-0.74)
      {\textcolor{subA}{A}\,/\,\textcolor{subB}{B}};
    \coordinate (zoomc) at (0,0);
  \end{scope}
  \draw[line width=0.35pt, gray!60!black] (zoomspot) circle (1.6pt);
  \draw[line width=0.35pt, gray!60!black, shorten >=1.8pt]
    ($(zoomc)+(-40:0.72)$) -- (zoomspot);

  \def\lx{3.8}
  \node[anchor=west, align=left] (gl) at (\lx,1.2) {monolayer\\[-1pt]graphene};
  \draw[line width=0.35pt] (gl.west) -- (graphL);
  \node[anchor=west, align=left] (sl) at (\lx,-3.15) {ferroelectric\\[-1pt]substrate};
  \draw[line width=0.35pt] (sl.west) -- (subsL);

  \begin{scope}[shift={(\lx+0.05,-2.25)}]
    \shade[top color=potHi!80!white, bottom color=potLo!80!white, middle color=white]
      (0,0) rectangle (0.18,1.1);
    \draw[line width=0.35pt] (0,0) rectangle (0.18,1.1);
    \node[anchor=west, font=\footnotesize, inner sep=2pt] at (0.18,1.1) {max};
    \node[anchor=west, font=\footnotesize, inner sep=2pt] at (0.18,0) {min};
    \node[anchor=south west, inner sep=1pt] (ul) at (-0.05,1.2) {$U(\mathbf r)$};
  \end{scope}
  \draw[line width=0.35pt] (\lx,-1.7) -- (potL);
\end{tikzpicture}
    \caption{Schematic of the system (exploded view, not to scale). Monolayer graphene sits on a ferroelectric substrate patterned with a square superlattice potential $U(\mathbf r)$ of period $L$ (colour map; the dashed square marks one superlattice unit cell). The magnetic field $\mathbf B$ is perpendicular to the graphene sheet. Inset: the honeycomb lattice with sublattices $A$ and $B$.}
    \label{fig:system}
\end{figure}

The charge-neutral quantum Hall regime of monolayer graphene provides a setting in which weak short-range interactions stabilize qualitatively distinct symmetry-broken electronic phases. At filling factor $\nu=0$, the fourfold spin and valley degenerate zeroth Landau level is half filled. Coulomb exchange favors ordering in this internal space, while lattice-scale interaction anisotropies and the Zeeman coupling select among competing orders~\cite{nomura2006,alicea2006,kharitonov2012}. These include a canted antiferromagnet (CAF), a spin-singlet intervalley coherent (IVC) state associated with Kekul\'e bond order, and a fully spin-polarized ferromagnet (F). Understanding how experimentally accessible parameters reorganize this competition is central to controlling the electronic properties of graphene at charge neutrality.

The ferromagnetic state is particularly appealing because it realizes a quantum Hall topological insulator: an insulating bulk supports counter-propagating edge modes with opposite spin polarization~\cite{abanin2006,young2014,veyrat2020,BreyFertig2006}. Their protection relies on conservation of spin ($S^z$) along the magnetic-field direction ($\hat{z}$), rather than time-reversal symmetry, which is broken by the applied field~\cite{abanin2006,kharitonov2016qhti}. Experimentally, helical edge transport has been obtained by applying a large in-plane field to enhance the Zeeman energy relative to the interaction anisotropies~\cite{young2014}. A complementary route uses a high-dielectric-constant SrTiO$_3$ substrate to screen the Coulomb interaction and modify the anisotropy balance, allowing helical transport at perpendicular fields as low as $1$ T~\cite{veyrat2020}. These results motivate additional ways to favor the ferromagnet by controlling the competition between spin alignment and interaction-driven ordering.

Long-wavelength superlattice potentials offer such a control parameter. Moir\'e superlattices reconstruct graphene's Dirac spectrum and, in a magnetic field, generate magnetic minibands and Hofstadter gaps~\cite{dean2013,ponomarenko2013,hunt2013}. Artificially patterned dielectric environments extend this approach to designed lattice geometries and tunable potential strengths~\cite{forsythe2018}. Programmable ferroelectric substrates provide a further route to nanoscale electrostatic potentials in graphene heterostructures~\cite{yang2024programming}, as sketched in Fig.~\ref{fig:system}. Beyond engineering the one-electron spectrum, these platforms raise an important question about the interacting many-body phase: can a smooth, flavor-independent potential reshape the small interaction anisotropies that select the $\nu=0$ ground state, and thereby stabilize the quantum Hall ferromagnet?
Here we address this question for monolayer graphene in a square scalar superlattice potential and a magnetic field. We extend the renormalization-group treatment of the anisotropic interactions~\cite{xuhuang2025} to include the reconstructed zero-field Bloch states, and use the resulting couplings in self-consistent Hartree-Fock calculations retaining multiple Landau levels. This approach expresses the phase competition in terms of magnetic field, dielectric environment, and superlattice amplitude and period. Although the potential is independent of spin and valley, it changes the orbital wavefunctions and the virtual transitions that renormalize the anisotropies. In the parameter range studied, the relevant anisotropy couplings are suppressed relative to their homogeneous values, while the bare Zeeman coupling is unchanged. The superlattice therefore changes the balance between these energy scales without introducing an additional flavor-symmetry-breaking field.

We find that the superlattice reorganizes the competition among IVC, CAF, and F. The reduced anisotropies favor phases that gain Zeeman energy, expanding the ferromagnetic regime and favoring CAF over the spin-singlet IVC state. In the full Hartree-Fock phase diagram, a gapped ferromagnet occurs for superlattice amplitudes of the order of tens of meV and out of plane fields of the order of 1 Tesla, even without an in-plane field. Superlattice engineering thus provides a route to stabilizing the ferromagnetic state at modest magnetic fields and reducing the reliance on a large in-plane component.

The reconstructed magnetic minibands also produce a distinct outcome when occupied and unoccupied bands overlap at charge neutrality. We find a partially spin-polarized compensated semimetal, with electron and hole pockets of equal total area. The superlattice consequently controls both the selection of insulating order and the onset of a metallic bulk. In the gapped ferromagnetic regime, the same band reconstruction motivates exploring how a periodic potential modifies the dispersion of helical edge modes while preserving their symmetry protection. These two effects, the reorganization of the ordered phase diagram and the formation of dispersive magnetic minibands, provide the organizing principles for the results below.

\section{Model and Symmetries}
\begin{figure}[t]
    \centering
    \includegraphics[width=\columnwidth]{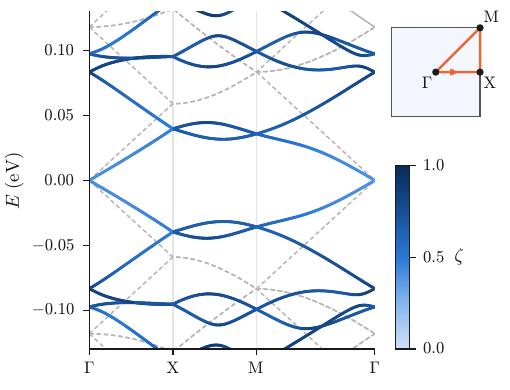}
    \caption{Zero-field bands of graphene (valley $K$) in the square superlattice potential, along the $\Gamma$--X--M--$\Gamma$ path of the moir\'e Brillouin zone shown in the inset. We use a superlattice period $L = 35$ nm and first-harmonic amplitude $V = 35$ meV (i.e.\ $U(\mathbf r) = -2V[\cos(Gx)+\cos(Gy)]$, $G = 2\pi/L$) throughout our numerics. The color of each band gives its momentum mixing, $\zeta = 1 - w_{\mathbf k+\mathbf G_0}$, where $w_{\mathbf k+\mathbf G_0}$ is the weight of the Bloch state on its dominant plane wave $\mathbf k+\mathbf G_0$. Dashed gray lines: the folded Dirac cone at $V = 0$, for which every state is a single plane wave.}
    \label{fig:zero_field_bands}
\end{figure}

We consider monolayer graphene, whose lattice is taken to be a continuum, in the presence of a square symmetric moir\'e potential and a magnetic field, as depicted in Fig.~\ref{fig:system}. Since typical superlattice periods $L$ are tens of nanometers, much larger than the graphene lattice constant $a = 2.46$ \AA, the continuum approximation is well justified for capturing the low-energy physics (the parameters used in our numerics are given with the zero-field band structure in Fig.~\ref{fig:zero_field_bands}). We thus work in a mini-Brillouin zone defined by the superlattice's reciprocal lattice vectors and containing linearly dispersing electrons. The magnetic field modulates the periodicity of the system at finite flux $p/q$ for $p,q$ integers, leading us to work in the mini-magnetic-Brillouin zone (mini-MBZ), which is compressed by a factor $q$ relative to the zero-field zone. In addition to the usual reorganization of the bands into relativistic Landau levels by the magnetic field, we also include the Zeeman effect in our Hamiltonian. \\\\In keeping with our aim of exploring experiment-facing mechanisms, we wish to formulate the phase diagram in terms of more concrete physical parameters, in particular dielectric constant and magnetic field. To that end, we utilize the renormalization-group (RG) program developed in \cite{xuhuang2025}, which obtains an RG flow for the anisotropy energy as a function of the magnetic field and dielectric constant. With this final piece in hand, we may write down the full Hamiltonian:
\begin{equation}
    H = H_0 + H_C + H_{an}, \qquad H_0 = H_D + H_U + H_Z .
    \label{eq:H_total}
\end{equation}
Here $H_D$ is the continuum Dirac Hamiltonian of graphene including the orbital coupling of the magnetic field, $H_U$ is the superlattice potential, $H_Z$ the Zeeman coupling, $H_C$ the long-range Coulomb interaction and $H_{an}$ the short-range anisotropic interactions. The non-interacting part $H_0$ is written out explicitly in Eqs.~\eqref{eq:H_D}--\eqref{eq:H_Z}, $H_C$ in Eq.~\eqref{eq:H_C} and $H_{an}$ in Eq.~\eqref{eq:H_an}. The symmetries of $H$ are listed in Table~\ref{tab:symmetries} by their action on the electron field. The terms $H_D$, $H_U$ and $H_C$ are invariant under arbitrary $SU(4)$ rotations of the spin-valley flavor; $H_Z$ and $H_{an}$ reduce this to $U(1)_s\times U(1)_v$.

\begin{table}[tb]
\caption{\label{tab:symmetries}Symmetries of the Hamiltonian and their associated actions on the  electron field $\tilde\psi$ of Eq.~\eqref{eq:psi_tilde} (in the valley-symmetric convention). $\tau_\alpha$, $\sigma_\alpha$ and $s_\alpha$ are Pauli matrices in valley, sublattice and spin space acting on $\tilde\psi$, with $s_z$ along the total field $\mathbf B_{\rm tot}$. $\mathbf a$ is a superlattice vector and $\chi_{\mathbf a}$ is defined by $\mathbf A(\mathbf r+\mathbf a) = \mathbf A(\mathbf r) + \nabla\chi_{\mathbf a}(\mathbf r)$. For primitive vectors, $T_{\mathbf a_1}T_{\mathbf a_2} = e^{2\pi i p/q}\,T_{\mathbf a_2}T_{\mathbf a_1}$, so translations of the $q$-fold enlarged magnetic unit cell commute. $R_4$ is a rotation by $\pi/2$ about an extremum of $U(\mathbf r)$. $M_x\mathcal T$ is written in the Landau gauge $\mathbf A = B_\perp x\hat{\mathbf y}$; neither the mirror $M_x$ nor the spinless time reversal $\mathcal T$ is a symmetry on its own, since each reverses $B_\perp$. The particle-hole operator $S$ acts on the single-particle Hamiltonian. With $\mathbf d = \frac L2(\hat{\mathbf x}+\hat{\mathbf y})$, it anticommutes with the single-particle Hamiltonian, $Sh_0S^{-1} = -h_0$, and gives the particle-hole symmetric spectrum of Fig.~\ref{fig:butterfly}. It requires $U(\mathbf r+\mathbf d) = -U(\mathbf r)$ and thus holds when only the first harmonic is kept.}
\begin{ruledtabular}
\begin{tabular}{lll}
Symmetry & Action on $\tilde\psi(\mathbf r)$ & Group\\
\colrule
Charge & $e^{i\theta}\tilde\psi(\mathbf r)$ & $U(1)_c$\\
Spin rotation & $e^{i\theta_s s_z/2}\tilde\psi(\mathbf r)$ & $U(1)_s$\\
Valley rotation & $e^{i\theta_v\tau_z/2}\tilde\psi(\mathbf r)$ & $U(1)_v$\\
Magnetic translation $T_{\mathbf a}$ & $e^{ie\chi_{\mathbf a}/\hbar}\,\tilde\psi(\mathbf r+\mathbf a)$ & MTG\\
Rotation $C_4$ & $e^{-i\pi\sigma_z/4}\,\tilde\psi(R_4^{-1}\mathbf r)$ & $\Z_4$\\
Inversion $I$ & $\tau_y\sigma_z\,\tilde\psi(-\mathbf r)$ & $\Z_2$\\
$M_x\mathcal T$ (antiunitary) & $\tau_z\tilde\psi(-x,y)$ & $\Z_2$\\
Particle-hole $S$ & $\tau_z\sigma_z s_x\,e^{ie\chi_{\mathbf d}/\hbar}\,\tilde\psi(\mathbf r+\mathbf d)$ & $\Z_2$
\end{tabular}
\end{ruledtabular}
\end{table}

\subsection{Non-Interacting Hamiltonian}

We describe the low-energy electrons of graphene by the continuum (envelope) field $\psi_{\xi s}(\mathbf r)$, a two-component spinor in sublattice space ($A$, $B$) with labels for valley $\xi = \pm$ (i.e.\ $K$, $K'$) and spin $s = \pm$. The magnetic field couples to the electrons in two ways: through its orbital coupling, which in the continuum description is implemented by minimal coupling, $\mathbf p \mapsto \boldsymbol\pi = \mathbf p + e\mathbf A(\mathbf r)$ with $\nabla\times\mathbf A = B_\perp\hat{\mathbf z}$, and through the Zeeman coupling to the spin. The three terms of the non-interacting Hamiltonian $H_0 = H_D + H_U + H_Z$ are
\begin{align}
    H_D &= \sum_{\xi,s}\int\dd^2r\;\psi^\dagger_{\xi s}(\mathbf r)\, v_F\left(\xi\pi_x\sigma_x + \pi_y\sigma_y\right)\psi_{\xi s}(\mathbf r), \label{eq:H_D}\\
    H_U &= \sum_{\xi,s}\int\dd^2r\;\psi^\dagger_{\xi s}(\mathbf r)\, U(\mathbf r)\,\psi_{\xi s}(\mathbf r), \label{eq:H_U}\\
    H_Z &= -\frac{g\mu_B B_{\rm tot}}{2}\sum_{\xi,s}s\int\dd^2r\;\psi^\dagger_{\xi s}(\mathbf r)\psi_{\xi s}(\mathbf r), \label{eq:H_Z}
\end{align}
where $\sigma_{x,y}$ are Pauli matrices in sublattice space, $\boldsymbol\pi = -i\hbar\nabla + e\mathbf A(\mathbf r)$, and where $B_{\rm tot} = (B_\perp^2 + B_\parallel^2)^{1/2}$. An in-plane field $B_\parallel$ enters only through $H_Z$. The potential $U(\mathbf r)$ is independent of spin and valley and acts as the identity in sublattice space.

For the symmetries and the anisotropic interactions it is convenient to use the valley-symmetric field \cite{kharitonov2012}
\begin{equation}
    \tilde\psi_{Ks} = \psi_{Ks}, \qquad \tilde\psi_{K's} = \sigma_y\,\psi_{K's},
    \label{eq:psi_tilde}
\end{equation}
in which the Dirac Hamiltonian takes the same form in both valleys, $H_D = v_F\sum_{\xi,s}\int\dd^2r\,\tilde\psi^\dagger_{\xi s}(\pi_x\sigma_x + \pi_y\sigma_y)\tilde\psi_{\xi s}$. We write $\tilde\psi = (\tilde\psi_{\xi s})$ for the eight-component field, and from here on $\tau_\alpha$, $\sigma_\alpha$ and $s_\alpha$ denote Pauli matrices in valley, sublattice and spin space acting on $\tilde\psi$. Acting on the original field $\psi$, the sublattice matrices of this basis are $(\sigma_0,\sigma_x,\sigma_y,\sigma_z) \to (1,\tau_z\sigma_x,\sigma_y,\tau_z\sigma_z)$ and the valley matrices are $(\tau_x,\tau_y,\tau_z) \to (\tau_x\sigma_y,\tau_y\sigma_y,\tau_z)$. In particular, the physical sublattice polarization is   $\langle\tau_z\sigma_z\rangle$. The zeroth Landau levels of the two valleys lie on opposite physical sublattices but on the same component of $\tilde\psi$, so within the zeroth Landau level $\sigma_z$ is fixed and valley and sublattice are locked. At $U = 0$, $H_D$ is diagonalized by the relativistic Landau levels, with energies $E_n = \sgn(n)\,\hbar v_F\sqrt{2|n|}/\ell_B$ with $\ell_B = \sqrt{\hbar/eB_\perp}$ \cite{goerbig2011} (see SI for a more detailed derivation). The $H_U$ term is the contribution from the long wavelength square symmetric moir\'e potential, which can be decomposed into its Fourier modes as 
\begin{equation}
    U(\mathbf r) = \sum_{m\in \textrm{r.l}}U_m\exp(i\mathbf G_m\cdot \mathbf r) \, ,
\end{equation}
\begin{figure}[t]
    \centering
    \includegraphics[width=\columnwidth]{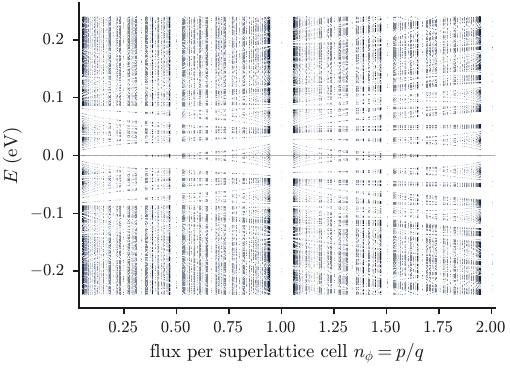}
    \caption{Non-interacting energy (Hofstatder) spectrum in electron volts at the $\Gamma$ point of the mini-MBZ as a function of magnetic flux ratio. First harmonic of potential strength fixed to be $35$ meV, all other harmonics set to zero. As explained in the table of symmetries, retaining only the first harmonic yields a particle-hole symmetric spectrum.}
    \label{fig:butterfly}
\end{figure}

\begin{figure*}[t]
    \centering
    \includegraphics[width=\textwidth]{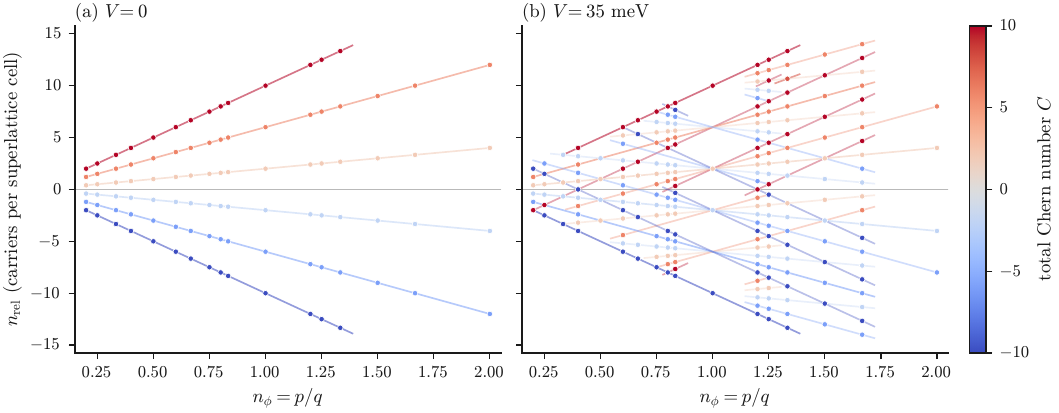}
    \caption{Wannier diagrams \cite{wannier1978,streda1982} of the non-interacting spectrum for (a) $V = 0$ and (b) $V = 35$ meV, at $L = 35$ nm. Each point is a gap of the spectrum at flux ratio $n_\phi = p/q$ and density $n_{\rm rel}$ (carriers per superlattice cell, measured from charge neutrality), colored by its total Chern number $C$, i.e.\ its Hall conductance $\sigma_{xy} = Ce^2/h$ summed over all four spin-valley flavors. Lines are $n_{\rm rel} = C n_\phi + s$ (St\v{r}eda formula). At $V = 0$ only the Landau-level gaps with $s = 0$ appear; the potential opens Hofstadter gaps with $s\neq 0$.}
    \label{fig:wannier}
\end{figure*}

where the sum is over reciprocal lattice vectors of the (square) moir\'e lattice, i.e those of the form $\frac{2\pi}{L}\langle n_x,n_y\rangle$ where $n_x,n_y$ are each elements of $\Z$. That is, the potential is such that the minima and maxima of the potential strength form a square lattice. The square symmetry in fact further constrains the potential to be of the form
\begin{equation}
    U(\mathbf r) = \sum_{j}2U_j\cos(\mathbf G_j\cdot \mathbf r)
\end{equation}
where the sum now ranges over equivalence classes of reciprocal lattice vectors modulo C$_4$ rotations.

We assume that the magnetic flux ratio is of the form $\frac{\Phi}{\Phi_0} = p/q$, where $p,q$ are co-prime integers, $\Phi = BL^2$ and $\Phi_0 = \frac{h}{e}$ is the magnetic flux quantum \cite{hofstadter1976}. In such a situation, the moir\'e unit cell becomes the magnetic unit cell, which is effectively enlarged by a factor of $q$ such that it admits an integer $p$ number of flux quanta \cite{hofstadter1976,zak1964}. Correspondingly, the moir\'e Brillouin zone becomes the magnetic Brillouin zone, which is shrunk by a factor of $q$ \cite{zak1964}. Then the Hamiltonian no longer commutes with the usual translation operator algebra, but rather with the magnetic translation algebra \cite{peierls1933,luttinger1951,zak1964,brown1964}. Thus, in analogy with Bloch's theorem, the eigenstates of the magnetic part of the Hamiltonian will be eigenstates of the generically non-abelian magnetic translation group \cite{brown1964}. Indeed,  to resolve this Hamiltonian, we use the basis of (relativistic) magnetic translation group (MTG) eigenstates and treat the moir\'e potential as a perturbation. Due to the enlargement of the unit cell by a factor of $q$, in the presence of a magnetic field with a flux ratio $p/q$, these MTG eigenstates acquire a $p$-fold degeneracy on top of the fourfold spin and valley degeneracy (discussed further in the Interacting Model section) \cite{wangvafek2022,wangvafek2023}. A pedagogical introduction to the MTG is given in the appendix. \\\\
Since the moir\'e potential is a linear combination of plane waves, one may build up the matrix element for the moir\'e potential from the Fourier matrix element by taking appropriate linear combinations. For the zero-field spectrum figure, we use the plane-wave basis, where the Fourier matrix element is a delta function. For the finite-field spectrum, we use the magnetic translation group eigenstates discussed above, whose Fourier matrix element is derived in the supplemental information. Using these matrix elements, we diagonalize this Hamiltonian exactly up to some Landau level index cutoff to obtain the non-interacting "Hofstader Butterfly" spectrum (see Fig.~\ref{fig:butterfly}). \\\\
The gaps of this spectrum are summarized by the Wannier diagram in Fig.~\ref{fig:wannier} \cite{wannier1978,streda1982}. Each point marks a spectral gap, at the density $n_{\rm rel}$ (carriers per superlattice cell, measured from charge neutrality) that fills all states below it, for a flux ratio $n_\phi = p/q$. By the St\v{r}eda formula, each gap lies on a line $n_{\rm rel} = C\,n_\phi + s$. The slope $C = \partial n_{\rm rel}/\partial n_\phi$ is the Hall conductance of the gap, $\sigma_{xy} = C e^2/h$. Equivalently, it is the total Chern number of the filled states: the sum of the Chern numbers of all occupied bands of all four spin-valley flavors~\cite{thouless1982}, or equivalently the many-body Chern number of the corresponding Slater determinant~\cite{niu1985}. The intercept $s$ is the number of carriers per superlattice cell, also an integer. At $V = 0$ only lines through the origin appear, with $C = \pm2,\pm6,\pm10$. These are the relativistic Landau-level gaps of graphene, $C = \pm4(N+\tfrac12)$ with $N = 0,1,2$~\cite{novoselov2005,zhang2005,gusynin2005}, where the factor $4$ arises from the spin-valley degeneracy. At $V = 35$ meV the potential opens additional gaps with $s\neq0$, which split the Landau levels into Hofstadter subbands. \\\\
For the purposes of this work, we kept only the first harmonic; that is, we took only $U_1$, corresponding to the the equivalence classes of reciprocal lattice vectors under C$_4$ orbits nearest the origin, to be non-zero. Previous work has shown that a second harmonic can be used to create flat band structures and induce an anomalous quantum Hall effect in TMDs \cite{suri2023}. However, this requires the second harmonic to be larger than the first; for a potential imprinted electrostatically from a patterned dielectric, the harmonics decay as $e^{-|G|d}$ with distance $d$ from the pattern~\cite{forsythe2018}. 
We therefore expect any configuration where the second harmonic is larger than the first to be fine-tuned. 
When keeping just the first harmonic, the band structure is found to be particle hole symmetric, as seen in the Hofstadter butterfly.
\\\\
The moir\'e potential will generically break the massive Landau level degeneracy. Generically one would expect the potential to couple different momentum states, in particular the valleys, and hence the potential should break SU(4) symmetry. However, the nature of the length scales of the problem means that the momenta in the moir\'e Brillouin zone are many orders of magnitude smaller than the valleys, and hence inter-valley scattering would be an extremely high-order Umklapp effect. Therefore, any contribution of the potential to valley symmetry breaking from this effect would be highly suppressed.\\\\
Finally, the Wannier diagrams of Fig.~\ref{fig:wannier} show no single-particle gap at charge neutrality, $n_{\rm rel} = 0$, either at $V = 0$ or at $V = 35$ meV; there the zeroth Landau level is half filled. The $\nu = 0$ gap must therefore be opened by interactions, which we now introduce.
\\\\

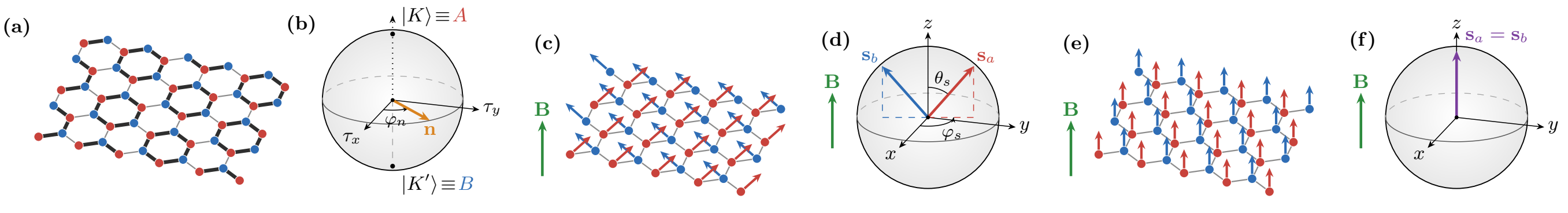
\begin{figure*}[t]
    \centering
    \begin{minipage}[b]{0.325\textwidth}\centering
        \resizebox{\linewidth}{!}{\def\tkpanelA{a}\def\tkpanelB{b}
\begin{tikzpicture}[font=\small]
  \def\phn{50}     
  \tkview{18}{42}
  \begin{scope}[tkview]
    \hcset{4}{10}{0.4}
    \hcbonds
    \hckekule{kekule, line width=1.7pt, line cap=round}
    \hcsites{\node[siteA] at (\hcx,\hcy,0) {};}{\node[siteB] at (\hcx,\hcy,0) {};}
  \end{scope}
  \node[panel] at (-0.6,1.95) {(\tkpanelA)};

  \begin{scope}[shift={(5.8,0.6)}]
    \def\R{1.15}
    \tkview{18}{20}
    \tikzset{blochzaxis/.style={dotted, line width=0.6pt}}
    \blochsphere{\R}{$\tau_x$}{$\tau_y$}{}
    \begin{scope}[tkview]
      \draw[line width=0.4pt, dashed, gray!70] (0,0,0) -- (0,0,-\R);
      \fill (0,0,\R) circle (1.1pt);
      \fill (0,0,-\R) circle (1.1pt);
      \draw[line width=0.4pt, -{Stealth[length=2.5pt]}]
        plot[domain=0:\phn, samples=20, smooth] ({0.42*\R*cos(\x)},{0.42*\R*sin(\x)},0);
      \node[font=\footnotesize, inner sep=0pt]
        at ({0.42*\R*cos(\phn/2)},{0.42*\R*sin(\phn/2)},-0.2) {$\varphi_n$};
      \draw[vec, ivcN] (0,0,0) -- ({\R*cos(\phn)},{\R*sin(\phn)},0)
        node[anchor=north, inner sep=2pt] {$\mathbf n$};
    \end{scope}
    \fill (0,0) circle (0.9pt);
    \node[anchor=south west, inner sep=1pt] at (0.1,\R+0.02) {$|K\rangle\!\equiv\!\textcolor{subA}{A}$};
    \node[anchor=north west, inner sep=1pt] at (0.1,-\R-0.02) {$|K'\rangle\!\equiv\!\textcolor{subB}{B}$};
    \node[panel] at (-1.75,1.4) {(\tkpanelB)};
  \end{scope}
\end{tikzpicture}}
    \end{minipage}\hfill
    \begin{minipage}[b]{0.325\textwidth}\centering
        \resizebox{\linewidth}{!}{\def\tkpanelA{c}\def\tkpanelB{d}
\begin{tikzpicture}[font=\small]
  \def\ths{40}     
  \def\phs{108}    
  \pgfmathsetmacro{\sx}{sin(\ths)*cos(\phs)}
  \pgfmathsetmacro{\sy}{sin(\ths)*sin(\phs)}
  \pgfmathsetmacro{\sz}{cos(\ths)}

  \tkview{18}{42}
  \begin{scope}[tkview]
    \hcset{3}{8}{0.42}
    \hcbonds
    \def\ls{0.52}
    \hcsites
      {\draw[spin, subA] (\hcx,\hcy,0) -- ++(\ls*\sx,\ls*\sy,\ls*\sz);
       \node[siteA] at (\hcx,\hcy,0) {};}
      {\draw[spin, subB] (\hcx,\hcy,0) -- ++(-\ls*\sx,-\ls*\sy,\ls*\sz);
       \node[siteB] at (\hcx,\hcy,0) {};}
  \end{scope}
  \draw[vec, fieldB] (-0.45,-0.35) -- (-0.45,0.55)
    node[anchor=south, inner sep=1.5pt] {$\mathbf B$};
  \node[panel] at (-0.6,1.95) {(\tkpanelA)};

  \begin{scope}[shift={(5.8,0.6)}]
    \def\R{1.15}
    \tkview{18}{20}
    \blochsphere{\R}{$x$}{$y$}{$z$}
    \begin{scope}[tkview]
      \draw[subA, line width=0.4pt, dashed]
        (0,0,0) -- (\R*\sx,\R*\sy,0) -- (\R*\sx,\R*\sy,\R*\sz);
      \draw[subB, line width=0.4pt, dashed]
        (0,0,0) -- (-\R*\sx,-\R*\sy,0) -- (-\R*\sx,-\R*\sy,\R*\sz);
      \draw[line width=0.4pt, -{Stealth[length=2.5pt]}]
        plot[domain=0:\phs, samples=30, smooth] ({0.36*\R*cos(\x)},{0.36*\R*sin(\x)},0);
      \node[font=\footnotesize, inner sep=0pt]
        at ({0.58*\R*cos(\phs/2)},{0.58*\R*sin(\phs/2)},-0.1) {$\varphi_s$};
      \draw[line width=0.4pt]
        plot[domain=0:\ths, samples=20, smooth]
          ({0.45*\R*sin(\x)*cos(\phs)},{0.45*\R*sin(\x)*sin(\phs)},{0.45*\R*cos(\x)});
      \node[font=\footnotesize, inner sep=0pt]
        at ({0.62*\R*sin(\ths/2)*cos(\phs)},{0.62*\R*sin(\ths/2)*sin(\phs)},{0.62*\R*cos(\ths/2)})
        {$\theta_s$};
      \draw[vec, subA] (0,0,0) -- (\R*\sx,\R*\sy,\R*\sz)
        node[anchor=south west, inner sep=1pt] {$\mathbf s_a$};
      \draw[vec, subB] (0,0,0) -- (-\R*\sx,-\R*\sy,\R*\sz)
        node[anchor=south east, inner sep=1pt] {$\mathbf s_b$};
    \end{scope}
    \fill (0,0) circle (0.9pt);
    \draw[vec, fieldB] (-1.55,-0.5) -- (-1.55,0.4)
      node[anchor=south, inner sep=1.5pt] {$\mathbf B$};
    \node[panel] at (-1.75,1.4) {(\tkpanelB)};
  \end{scope}
\end{tikzpicture}}
    \end{minipage}\hfill
    \begin{minipage}[b]{0.325\textwidth}\centering
        \resizebox{\linewidth}{!}{\def\tkpanelA{e}\def\tkpanelB{f}
\begin{tikzpicture}[font=\small]
  \tkview{18}{42}
  \begin{scope}[tkview]
    \hcset{3}{8}{0.42}
    \hcbonds
    \def\ls{0.52}
    \hcsites
      {\draw[spin, subA] (\hcx,\hcy,0) -- ++(0,0,\ls);
       \node[siteA] at (\hcx,\hcy,0) {};}
      {\draw[spin, subB] (\hcx,\hcy,0) -- ++(0,0,\ls);
       \node[siteB] at (\hcx,\hcy,0) {};}
  \end{scope}
  \draw[vec, fieldB] (-0.45,-0.35) -- (-0.45,0.55)
    node[anchor=south, inner sep=1.5pt] {$\mathbf B$};
  \node[panel] at (-0.6,1.95) {(\tkpanelA)};

  \begin{scope}[shift={(5.8,0.6)}]
    \def\R{1.15}
    \tkview{18}{20}
    \blochsphere{\R}{$x$}{$y$}{$z$}
    \begin{scope}[tkview]
      \draw[vec, spinF] (0,0,0) -- (0,0,\R);
    \end{scope}
    \fill (0,0) circle (0.9pt);
    \node[anchor=south west, inner sep=1pt, text=spinF] at (0.1,\R+0.02) {$\mathbf s_a=\mathbf s_b$};
    \draw[vec, fieldB] (-1.55,-0.5) -- (-1.55,0.4)
      node[anchor=south, inner sep=1.5pt] {$\mathbf B$};
    \node[panel] at (-1.75,1.4) {(\tkpanelB)};
  \end{scope}
\end{tikzpicture}}
    \end{minipage}
    \caption{Schematics of the insulating ground states. (a,b) Intervalley coherent state. (a) On the lattice it appears as a $\sqrt3\times\sqrt3$ Kekul\'e bond order (thick bonds) with equal $A$ and $B$ densities; one representative of the $U(1)$ family is shown. (b) The valley pseudospin $\mathbf n$ lies on the equator of the valley Bloch sphere, whose poles $\ket{K}$, $\ket{K'}$ coincide with sublattices $A$, $B$ in the zeroth Landau level; the valley $U(1)$ symmetry is parameterized by the angle $\varphi_n$. Both spin species occupy the same valley state (spin singlet). (c,d) Canted antiferromagnet. (c) Spins on the honeycomb lattice: the in-plane components on sublattices $A$ and $B$ are antiparallel, while the out of plane components cant along $\mathbf B$. (d) The sublattice spins $\mathbf s_a$, $\mathbf s_b$ on the spin Bloch sphere ($z\parallel\mathbf B$), canted by $\theta_s$ from $\mathbf B$ (exaggerated for clarity); the spin $U(1)$ symmetry is parameterized by the azimuth $\varphi_s$. (e,f) Ferromagnet. (e) Spins on both sublattices align with $\mathbf B$. (f) On the spin Bloch sphere $\mathbf s_a=\mathbf s_b$ point along $\mathbf B$; the occupied states are $\ket{\uparrow}\otimes\{\ket{K},\ket{K'}\}$.}
    \label{fig:states}
\end{figure*}

\subsection{Long-range Coulomb Interaction}
\label{sec:coulomb}
Including the long-range Coulomb interaction in our model is essential to capture the competition between the phases \cite{nomura2006,alicea2006}. Despite the fact that its SU(4) symmetry \cite{alicea2006,yang2006} prevents it from directly selecting the phase, we shall see that it plays a large role in the renormalization of the anisotropies and in modulating the effects of the filled Dirac sea, which do select the winning phase.\\\\
We treat the Coulomb interaction in the mini-MBZ via the self-consistent Hartree-Fock method. However, due to our continuum approximation, the inter-valley mixing effects induced by both the superlattice potential and the long-range Coulomb interaction are highly suppressed Umklapp processes at these scales, making both of these parts of the Hamiltonian totally blind to the spin and valley degrees of freedom; that is, they are flavor-SU(4) symmetric. Although the Zeeman field breaks the degeneracy between the spin up and spin down states, to correctly capture the flavorful phase diagram in monolayer graphene systems one must include some form of the phenomenological Kharitonov anisotropy energies \cite{kharitonov2012}. The usual approach to the monolayer graphene phase diagram with anisotropies involves approximating the dynamics as completely projected into the zeroth Landau level, as in \cite{kharitonov2012,nomura2006,alicea2006,herbut2007,jung2009}. However, the effects we are studying are driven in large part by mixing of remote Landau levels, so we seek a framework that retains the anisotropies in the full Landau level Hilbert space.\\\\
The long-range Coulomb interaction is given as
\begin{equation}
    H_C = \frac12\int\dd^2r\,\dd^2r'\,
    \frac{e^2\,:\!\rho(\mathbf r)\rho(\mathbf r')\!:}{4\pi\epsilon_0\epsilon_r|\mathbf r-\mathbf r'|} ,
    \label{eq:H_C}
\end{equation}
where $\rho(\mathbf r) = \sum_{\xi,s}\psi^\dagger_{\xi s}(\mathbf r)\psi_{\xi s}(\mathbf r)$ is the total electron density. In momentum space,
\begin{equation}
    H_C = \frac{1}{2A}\sum_{\mathbf q}V(\mathbf q)\,:\!\rho(\mathbf q)\rho(-\mathbf q)\!:, \qquad V(\mathbf q) = \frac{e^2}{2\epsilon_0\epsilon_r|\mathbf q|},
    \label{eq:H_C_q}
\end{equation}
where $A$ is the system area and $\rho(\mathbf q) = \int\dd^2r\,e^{-i\mathbf q\cdot\mathbf r}\rho(\mathbf r)$. We expand the field in the MTG eigenstates $\ket{\Psi_a(\mathbf k)}$ of $H_D$ (Appendix~\ref{app:MTG}), with $\mathbf k$ in the mini-MBZ and $a$ a combined index for the Landau level, magnetic subband, band sign, valley and spin. With $c_{\mathbf k a}$ the corresponding annihilation operators and $c_{\mathbf k} = (c_{\mathbf k a})_a$, the density is
\begin{gather}
    \rho(\mathbf q+\mathbf G) = \sum_{\mathbf k} c^\dagger_{\mathbf k}\,\Lambda_{\mathbf q+\mathbf G}(\mathbf k)\,c_{\mathbf k+\mathbf q},
    \label{eq:rho_q}\\
    \big[\Lambda_{\mathbf q+\mathbf G}(\mathbf k)\big]_{ab} = \bra{\Psi_a(\mathbf k)}e^{-i(\mathbf q+\mathbf G)\cdot\mathbf r}\ket{\Psi_b(\mathbf k+\mathbf q)},
    \label{eq:form_factor}
\end{gather}
where we have split a general momentum into $\mathbf q$ in the mini-MBZ and a reciprocal lattice vector $\mathbf G$ of the mini-MBZ. Momenta $\mathbf k+\mathbf q$ outside the mini-MBZ are folded back into it, and the corresponding reciprocal lattice vector is absorbed into $\Lambda$. The form factors (structure factors) $\Lambda$ are overlaps of the magnetic Bloch wavefunctions, and their explicit form in the MTG basis is derived in Appendix~\ref{app:structure_factor}. They are diagonal in, and independent of, the spin-valley flavor. This makes the $SU(4)$ symmetry of $H_C$ manifest: $\rho(\mathbf q)$ is a sum over all four flavors with equal weight. The momentum scattering due to the Coulomb repulsion in principle couples the valleys, but for the long-range interaction these are highly suppressed Umklapp processes on the scale of the moir\'e Brillouin zone. We do not explicitly screen the Coulomb interaction, in keeping with the rest of the monolayer graphene literature.
\\\\ In the Hartree-Fock approximation, we make a single Slater determinant ansatz for the $N$-particle state \cite{szabo1996,ringschuck1980}, $\ket{\Omega} = \prod_{\alpha=1}^{N}c_\alpha^\dagger\ket{0}$, where $\ket{0}$ is the empty (vacuum) state and $c^\dagger_\alpha$ creates the $\alpha$-th occupied Hartree-Fock orbital. We restrict to uniform states, i.e those that preserve the magnetic translations, so that $\ket{\Omega}$ is fully characterized by the one-body density matrix
\begin{equation}
    \big[\Delta(\mathbf k)\big]_{ab} = \bra{\Omega}c^\dagger_{\mathbf k b}c_{\mathbf k a}\ket{\Omega}
    \label{eq:Delta_k}
\end{equation}
\begin{table}[tb]
\begin{ruledtabular}
\begin{tabular}{lll}
Phase & Order Parameters & Spectrum\\
\colrule
IVC & $\langle\tilde\psi^\dagger\tau_\pm\tilde\psi\rangle$ & gapped\\
CAF & $\langle\tilde\psi^\dagger\tau_z\sigma_z s_{x,y}\tilde\psi\rangle$, $\langle\tilde\psi^\dagger s_z\tilde\psi\rangle$ & gapped\\
F & $\langle\tilde\psi^\dagger s_z\tilde\psi\rangle$, $m = 1$ & gapped\\
SM-F & $\langle\tilde\psi^\dagger s_z\tilde\psi\rangle$, $m < 1$ & compensated pockets
\end{tabular}
\end{ruledtabular}
\caption{\label{tab:order_params}Ground-state phases at $\nu = 0$ and the fermion bilinear order parameters whose uniform expectation values are non-zero in each. $\tau_\alpha$, $\sigma_\alpha$ and $s_\alpha$ are Pauli matrices in valley, sublattice and spin space acting on the valley-symmetric field $\tilde\psi$ of Eq.~\eqref{eq:psi_tilde}, $\tau_\pm = (\tau_x\pm i\tau_y)/2$, and $s_z$ points along the total field. The CAF bilinear involves the physical sublattice polarization $\tau_z\sigma_z$, which reduces to $\tau_z$ within the zeroth Landau level. The insulating phases (IVC, CAF, F) have a gapped HF spectrum and a momentum-independent flavor occupation. The compensated semimetal SM-F has the same order parameter as F but no gap: spin-up hole pockets and spin-down electron pockets of equal area coexist (Sec.~\ref{sec:HF_results}), so its spin polarization $m$ (normalized to its value in the insulating F phase) is smaller than one. }
\end{table}
(translation-symmetry breaking states are discussed in Appendix~\ref{app:translation_breaking}). Applying Wick's theorem to Eq.~\eqref{eq:H_C_q}  gives the Hartree-Fock Hamiltonian
\begin{gather}
    H_{HF} = \sum_{\mathbf k} c^\dagger_{\mathbf k}\big[h_0(\mathbf k) + h_H(\mathbf k) + h_F(\mathbf k)\big]c_{\mathbf k},
    \label{eq:HF_hamiltonian}\\
    \begin{split}
    h_H(\mathbf k) = {}&\frac1A\sum_{\mathbf G\neq 0}V(\mathbf G)\,\Lambda_{\mathbf G}(\mathbf k)^\dagger\\
    &\times\sum_{\mathbf k'}\tr\big(\Lambda_{\mathbf G}(\mathbf k')\,[\Delta(\mathbf k')-\Delta_0(\mathbf k')]\big),
    \end{split}
    \label{eq:h_H}\\
    h_F(\mathbf k) = -\frac1A\sum_{\mathbf q,\mathbf G}V(\mathbf q+\mathbf G)\,\Lambda_{\mathbf q+\mathbf G}(\mathbf k)\,\Delta(\mathbf k+\mathbf q)\,\Lambda_{\mathbf q+\mathbf G}(\mathbf k)^\dagger,
    \label{eq:h_F}
\end{gather}
where $h_0(\mathbf k)$ is the non-interacting Hamiltonian $H_0$ in the MTG basis. The Hartree term $h_H$ couples to the density modulation $\langle\rho(\mathbf G)\rangle$ at the superlattice reciprocal lattice vectors. Its $\mathbf G = 0$ component is cancelled by the neutralizing background. To account for the background charge of the ions, the density matrix in the Hartree term is measured from charge neutrality, $\Delta \to \Delta - \Delta_0$, where $\Delta_0$ is the non-interacting density matrix at half filling (Appendix~\ref{app:background}). The Fock term $h_F$ is the exchange self-energy, which favors flavor polarization. The anisotropy contribution of Eq.~\eqref{eq:H_an_MF_matrix} below is added to the bracket in Eq.~\eqref{eq:HF_hamiltonian}. Requiring the energy to be stationary among all Slater determinants yields the self-consistency condition \cite{ringschuck1980}
\begin{equation}
    \big[\Delta(\mathbf k), h_{HF}(\mathbf k)\big] = 0,
    \label{eq:selfconsistency}
\end{equation}
where $h_{HF}(\mathbf k)$ is the bracket in Eq.~\eqref{eq:HF_hamiltonian}, together with the requirement that $\Delta$ project onto the $N$ lowest eigenstates of $h_{HF}$ across the mini-MBZ. The equivalent expressions in terms of general two-body matrix elements $V_{ijk\ell}$ are given in Appendix~\ref{app:HF} \cite{cote1990,blanchette2018}. This condition is solved iteratively to obtain the ground state density matrix. The iterative algorithm for finding the ground state is described in great detail in the appendix, including methods for accelerating convergence. Upon convergence, we extract the ground state and evaluate the order parameters of Table~\ref{tab:order_params} on it.\\\\Each point of the phase diagram is the lowest-energy converged state among six families of starting ansatzes for the iterative procedure. These are the gapped F, CAF and IVC phases (Fig.~\ref{fig:states}) as well as their semimetallic variants. The runs are also supplemented by warm-starting with converged density matrices from adjacent points of the phase diagram. The dielectric constant $\epsilon_r$, and hence $\kappa_0$, is varied in the phase diagram. At the flux $p/q = 1/2$ of our phase diagram, we keep all Landau levels with $|n|\le 19$ ($39$ levels per spin and valley) and $N_k = 32$ momentum points in the mini-MBZ, i.e  total of $64$ points in the unfolded zone. The grid contains the high-symmetry points of the mini-MBZ: the $\Gamma$ point and the midpoints and corners of the zone. Various convergence acceleration algorithms were used, such as direct inversion in the iterative subspace (DIIS) method \cite{pulay1980,pulay1982} and the optimal damping algorithm (ODA) \cite{cances2000} \footnote{For the full breadth of implementation details, the reader is invited to the GitHub page associated with this paper.}.

 We generically expect to find symmetry-broken ground states induced by the long-range Coulomb interaction \cite{nomura2006}. For example, in the absence of any anisotropy, the Coulomb exchange energy favors the emergence of a generalized ferromagnetic ground state \cite{yang2006}. However, the interacting monolayer graphene system at half filling under just a magnetic field has been extensively studied and found to harbor a richer class of symmetry broken ground states than the generalized ferromagnet \cite{kharitonov2012,de2023,stefanidis2023}. To capture these correctly, we must include the anisotropy energies.

\subsection{Anisotropy Energies}

The long-range Coulomb interaction preserves the SU(4) flavor symmetry in monolayer graphene \cite{kharitonov2012,alicea2006}, and thus to select a flavor phase there must be a term in the Hamiltonian that breaks this symmetry and selects among the ordered states sketched in Fig.~\ref{fig:states}. The canonical approach to this involves phenomenological terms understood to come from the much weaker short-range interactions, with contributions from both electron-electron and electron-phonon interactions. In monolayer graphene the general anisotropy Hamiltonian is a sum of local four-fermion operators \cite{kharitonov2012,de2023},
\begin{equation}
    H_{an} = \frac12\sum_{\alpha,\beta} g_{\alpha\beta}\int\dd^2r\; :\!\left[\tilde\psi^\dagger(\mathbf r)\,\tau_\alpha\sigma_\beta\,\tilde\psi(\mathbf r)\right]^2\!: ,
    \label{eq:H_an}
\end{equation}
where $\tilde\psi$ is the eight-component valley-symmetric field of Eq.~\eqref{eq:psi_tilde}, $\tau_\alpha$ and $\sigma_\beta$ ($\alpha,\beta = 0,x,y,z$, with $\tau_0 = \sigma_0 = 1$) are Pauli matrices in valley and sublattice space respectively, and $:\ldots:$ denotes normal ordering. The $(\alpha,\beta) = (0,0)$ term is a flavor-symmetric contact density-density interaction and is omitted ($g_{00} \equiv 0$). The in-plane couplings are equal, $g_{x\beta} = g_{y\beta} \equiv g_{\perp\beta}$ and $g_{\alpha x} = g_{\alpha y} \equiv g_{\alpha\perp}$, as required by the valley $U(1)_v$ and the rotation symmetry of Table~\ref{tab:symmetries}, respectively. Of the $8$ remaining couplings, only $4$ are allowed by zero-field time-reversal symmetry \cite{de2023}, namely $g_{\perp\perp},g_{\perp z},g_{z\perp},g_{zz}$.

\begin{figure*}[t]
    \centering
    \includegraphics[width=0.9\textwidth]{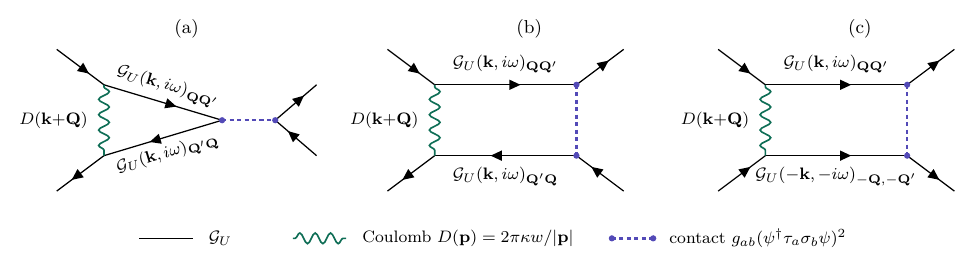}
    \caption{Diagrams for the renormalization-group flow of the anisotropies, leaving aside the renormalization of the Coulomb interaction itself. We keep only the diagrams that are first order in both the Coulomb and the contact interaction: (a) vertex correction, (b) double exchange and (c) BCS diagram. Solid lines, external and internal, are the superlattice propagator $\mathcal G_U$ of Eq.~\eqref{eq:spectral}, written in the plane-wave basis with reciprocal lattice vectors $\mathbf Q$, $\mathbf Q'$ of the superlattice. Wavy lines are the Coulomb interaction $D(\mathbf p) = 2\pi\kappa w/|\mathbf p|$, with $w = \hbar v$ (Appendix~\ref{app:RG_setup}); at each of its two ends it couples to the electron density, i.e.\ it acts as the identity in sublattice, valley and spin. The dashed line is the contact interaction $g_{ab}$ of Eq.~\eqref{eq:H_an}. Each of its two ends (dots) is a bilinear vertex $\tilde\psi^\dagger\tau_a\sigma_b\tilde\psi$, which carries the valley and sublattice matrix $\tau_a\sigma_b$.}
    \label{fig:placeholder}
    \vspace{2ex}
    \includegraphics[width=\textwidth]{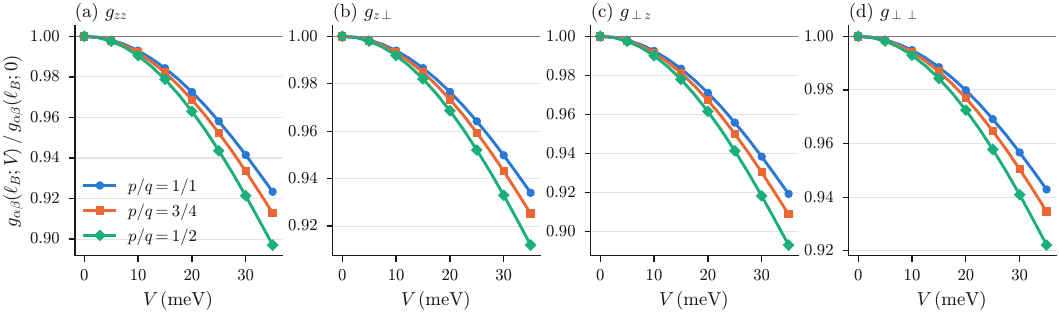}
    \caption{Suppression of the anisotropy couplings by the superlattice potential, obtained from our numerical finite-potential RG (see text and SI). Each panel shows one coupling, (a) $g_{zz}$, (b) $g_{z\perp}$, (c) $g_{\perp z}$ and (d) $g_{\perp\perp}$, evaluated at the end of the flow, $\ell = \ell_B$, and divided by its value for $V = 0$ (where the flow reduces to that of Ref.~\cite{xuhuang2025}), as a function of the first-harmonic amplitude $V$. The couplings include both the electron-electron and electron-phonon contributions. Parameters: $L = 35$ nm and bare interaction strength $\kappa_0 = 0.5$, for flux ratios $p/q = 1/1$, $3/4$ and $1/2$ per superlattice cell, i.e.\ $B_\perp = 3.38$, $2.53$ and $1.69$ T ($\ell_B = 14.0$, $16.1$ and $19.7$ nm). At $V = 35$ meV the couplings are reduced by $6$--$11\%$, more strongly at lower field, where the flow runs longer.}
    \label{fig:flowlines}
\end{figure*}

Within HF, the expectation value of Eq.~\eqref{eq:H_an} depends on the state only through the local one-body density matrix $R_{ab}(\mathbf r) = \langle\tilde\psi^\dagger_b(\mathbf r)\tilde\psi_a(\mathbf r)\rangle$, where $a,b$ are combined valley, spin and sublattice indices. We resolve its sublattice structure as $R(\mathbf r) = \frac12\sum_\gamma \rho_\gamma(\mathbf r)\otimes\sigma_\gamma$, where $[\rho_\gamma(\mathbf r)]_{ff'} = \langle\tilde\psi^\dagger_{f'}(\mathbf r)\,\sigma_\gamma\,\tilde\psi_f(\mathbf r)\rangle$ is a $4\times4$ matrix in spin-valley (flavor) space, with Fourier components $\rho_\gamma(\mathbf G) = A^{-1}\int\dd^2r\, e^{-i\mathbf G\cdot\mathbf r}\rho_\gamma(\mathbf r)$ at the reciprocal lattice vectors $\mathbf G$ of the superlattice ($A$ is the system area). Wick's theorem gives $\langle:\!(\tilde\psi^\dagger M\tilde\psi)^2\!:\rangle = (\tr MR)^2 - \tr(MRMR)$, and with $\sigma_\beta\sigma_\gamma\sigma_\beta = s_{\beta\gamma}\sigma_\gamma$ the anisotropy energy becomes
\begin{equation}
\begin{split}
    E_{an} = \frac{A}{2}\sum_{\mathbf G}\sum_{\alpha,\beta} g_{\alpha\beta}
    \Big[&\big|\tr\big(\tau_\alpha\rho_\beta(\mathbf G)\big)\big|^2 \\
    &- \frac12\sum_{\gamma} s_{\beta\gamma}\tr\big(\tau_\alpha\rho_\gamma(\mathbf G)\,\tau_\alpha\rho^\dagger_\gamma(\mathbf G)\big)\Big],
\end{split}
    \label{eq:E_an}
\end{equation}
where $s_{\beta\gamma} = +1$ if $\beta = 0$, $\gamma = 0$ or $\beta = \gamma$, and $s_{\beta\gamma} = -1$ otherwise. The direct (Hartree) term only involves the sublattice channel $\beta$ of each coupling, whereas the exchange (Fock) term involves all four sublattice components of the density. Keeping all four sublattice components in the exchange term matters outside the zeroth Landau level, where the sublattice is no longer locked to the valley.

Varying Eq.~\eqref{eq:E_an} with respect to $\rho_\gamma(\mathbf G)$ gives the anisotropy part of the mean-field Hamiltonian. In terms of the electron operators,
\begin{gather}
    H^{MF}_{an} = \sum_{\mathbf G}\sum_{\gamma}\int\dd^2r\; e^{-i\mathbf G\cdot\mathbf r}\,
    \tilde\psi^\dagger(\mathbf r)\left[\mathcal K_\gamma(\mathbf G)^\dagger\otimes\sigma_\gamma\right]\tilde\psi(\mathbf r),
    \label{eq:H_an_MF}\\
    \mathcal K_\gamma(\mathbf G) = \sum_{\alpha}\Big[g_{\alpha\gamma}\tr\big(\tau_\alpha\rho_\gamma(\mathbf G)\big)\,\tau_\alpha
    - \frac12\sum_{\beta} g_{\alpha\beta}\,s_{\beta\gamma}\,\tau_\alpha\rho_\gamma(\mathbf G)\,\tau_\alpha\Big].
    \label{eq:K_gamma}
\end{gather}
We can equivalently reformulate this in the basis of single-particle MTG eigenstates $\ket{n}$ used in the HF calculation (fermion operators $c_n$, flavor $f_n$, and density matrix $\Delta_{nm} = \langle c^\dagger_m c_n\rangle$, which is block diagonal in momentum with blocks $\Delta(\mathbf k)$ of Eq.~\eqref{eq:Delta_k}), with the form factors $\Lambda^{\gamma}_{mn}(\mathbf G) = \bra{m}e^{-i\mathbf G\cdot\mathbf r}\sigma_\gamma\ket{n}$. Here the sublattice spinors of $\ket{m}$ and $\ket{n}$ are written in the valley-symmetric basis, i.e.\ the $K'$ spinors of Appendix~\ref{app:MTG} are multiplied by $\sigma_y$. Relative to the original spinors, this flips the sign of the $\sigma_x$ and $\sigma_z$ form factors within $K'$, and inserts a factor $\sigma_y$ in the inter-valley form factors that enter the $\tau_{x,y}$ channels. We then have
\begin{align}
    \big[\rho_\gamma(\mathbf G)\big]_{ff'} &= \frac1A\sum_{m\in f',\,n\in f}\Lambda^{\gamma}_{mn}(\mathbf G)\,\Delta_{nm},
    \label{eq:rho_matrix}\\
    \big(H^{MF}_{an}\big)_{mn} &= \sum_{\mathbf G,\gamma}\big[\mathcal K_\gamma(\mathbf G)^\dagger\big]_{f_m f_n}\Lambda^{\gamma}_{mn}(\mathbf G),
    \label{eq:H_an_MF_matrix}
\end{align}
where $m\in f'$ denotes the states $m$ of flavor $f_m = f'$, so that $H^{MF}_{an} = \sum_{mn}(H^{MF}_{an})_{mn}c^\dagger_m c_n$ is added to the HF Hamiltonian of Eq.~\eqref{eq:HF_hamiltonian}. The anisotropies select among the flavor-ordered insulating states sketched in Fig.~\ref{fig:states}, whose order parameters are listed in Table~\ref{tab:order_params}.

\subsection{Numerical renormalization group in finite potential}
\label{sec:finiteV_method}

The superlattice changes the anisotropy interactions before they are used to select the ordered state. Our methodological extension is to evaluate this renormalization with the finite potential miniband dispersion: we evaluate the diagrams of Fig.~\ref{fig:placeholder} with superlattice propagators, which yields the suppressed couplings of Fig.~\ref{fig:flowlines}. Although $U(\mathbf r)$ is independent of flavor, the virtual transitions that dress a pre-existing anisotropy depend on these orbital wavefunctions. Thus a flavor-symmetric potential can change the competition between ordered phases without itself providing a new flavor-symmetry-breaking field. Conceptually, this is similar to how the SU(4)-symmetric long-range Coulomb interaction also renormalizes the anisotropies despite being flavor blind.

Following Refs.~\cite{kharitonov2012,xuhuang2025}, we integrate out short-distance modes from the microscopic input scale $a_0\simeq0.25$ nm, of the order of the graphene lattice constant $a$, to the graphene magnetic length $\ell_B$. The running length is $\ell=a_0e^t$, with stopping time $t_B=\ln(\ell_B/a_0)$. The couplings at $a_0$ are specified in Eq.~\eqref{eq:inputs}. We use zero-field miniband propagators for this part of the calculation. That is, we use the same diagrams as the zero potential literature, but with finite-potential propagator lines as opposed to zero-potential propagator lines.

For each valley channel $a=z,\perp$, collect the electron-electron contacts into $\mathbf g_a^{\ee}=(g_{az}^{\ee},g_{a\perp}^{\ee})^T$. The equations used in the numerical flow are
\begin{subequations}\label{eq:RG}
\begin{align}
 \frac{\dd\mathbf g_a^{\ee}}{\dd t}
 &=4A[\lambda(t)]\widehat{\mathsf K}[u(t),s(t)]\mathbf g_a^{\ee},\label{eq:RG_ee}\\
 \frac{\dd\ln v}{\dd t}&=A[\lambda(t)]
 \label{eq:RG_v}
\end{align}
\end{subequations}
Here $v(0)=v_F$, $\kappa_0=e^2/(4\pi\epsilon_0\epsilon_r\hbar v_F)$ is the bare interaction strength, and
\begin{equation}
 u(t)=\frac{V}{\hbar v(t)G},
 s(t)=\frac{1}{G\ell}, G=\frac{2\pi}{L}, \lambda(t)=\frac{\pi\kappa_0}{2}\frac{v_F}{v(t)}.
 \label{eq:RG_scales}
\end{equation}
The variable $s$ measures the running energy cutoff $\hbar v(t)/\ell$ relative to the superlattice energy scale $\hbar v(t)G$. The matrix $\widehat{\mathsf K}$ mixes the $z$ and $\perp$ sublattice channels of the contact interaction.

The two equations correspond to the twofold one-loop corrections. The Coulomb exchange self-energy renormalizes $v$, whereas diagrams dressing a four-fermion vertex renormalize the contacts. We retain the homogeneous RPA velocity rate $A(\lambda)$ \cite{gonzalez1994,kotov2012} of Appendix~\ref{app:RG_velocity}, whose weak-coupling limit is $A\simeq\kappa_0v_F/(4v)$. Increasing $v$ reduces the running Coulomb strength and also reduces $u(t)$ for fixed $V$.

To obtain $\widehat{\mathsf K}$, we diagonalize the zero-field square-superlattice Hamiltonian in a plane-wave basis and construct its spectral propagator from the band energies and two-component Bloch spinors. That is, we use the following propagator:
\begin{equation}
 G_U(\mathbf k,i\omega)=\sum_n
 \frac{|u_{n\mathbf k}\rangle\langle u_{n\mathbf k}|}{i\omega-E_{n\mathbf k}}.\label{eq:spectral}
\end{equation}
where $|u_{n\mathbf k}\rangle$ denotes the Bloch states, and $E_{n\mathbf k}$ denotes the Bloch energies. The renormalization of the anisotropies is understood to come from the long-range Coulomb interaction, via diagrams with both contact and Coulomb lines. The contact-contact diagrams are neglected as they are second order in the small anisotropy energy scale. The relevant diagrams are shown in Fig.~\ref{fig:placeholder}.
The regulator prescription is the same one that is used in the literature when the superlattice potential is absent~\cite{xuhuang2025}, i.e the contribution from the $\Omega$ frequency shell is regulated by 
\begin{equation}
   R_{\Omega}(\omega) =  \frac{2\Omega^2\omega^4}{(\omega^2+\Omega^2)^3}
\end{equation}
The regulated frequency integrals are evaluated analytically, while the band sums and mini-Brillouin-zone integrals are numerical (Appendix~\ref{app:RG_finiteV}). We project onto the retained local contact channels at zero external momenta and frequencies. The totals $g_{\alpha\beta}=g_{\alpha\beta}^{\ee}+g_{\alpha\beta}^{\elph}$ are evaluated at $\ell_B$.

To compute the matrix $\widehat{\mathsf K}$ at finite $V$, we evaluate the diagrams of Fig.~\ref{fig:placeholder} with the unscreened Coulomb interaction. The overall strength of the flow in Eq.~\eqref{eq:RG_ee}, the prefactor $4A(\lambda)$, is taken from the homogeneous theory. There, the Coulomb interaction is screened by particle-hole excitations of the filled Dirac sea, which we treat in the random-phase approximation (RPA, Eq.~\eqref{eq:RPAline}). The same rate $A(\lambda)$ sets the Fermi-velocity flow of Eq.~\eqref{eq:RG_v} (Appendix~\ref{app:RG_velocity}). The potential therefore changes only the matrix $\widehat{\mathsf K}$, and at $V = 0$ the flow reduces exactly to that of Ref.~\cite{xuhuang2025}. We do not include the change in this screening caused by the superlattice itself, i.e.\ the full dynamical dielectric response of the modulated electron system. Momentum-dependent interactions and channels outside the retained basis are not accounted for.

At $V=0$, the numerical construction reduces to
\begin{equation}
 \mathsf K_0=\begin{pmatrix}1&-1\\-\tfrac12&\tfrac12\end{pmatrix}.
\end{equation}
Its eigenvalues are $0$ and $3/2$: $g_{az}^{\ee}+2g_{a\perp}^{\ee}$ is conserved, while $g_{az}^{\ee}-g_{a\perp}^{\ee}$ is multiplied by
\begin{equation}
 E(t)=\exp\left[6\int_0^t A[\lambda(t')]\,\dd t'\right]
     =\left[\frac{v(t)}{v_F}\right]^6.
 \label{eq:RG_E}
\end{equation}
In the vertex-phonon convention used for the figures, the homogeneous factors multiplying $g_{z\perp}^{\elph}(0)$ and $g_{\perp z}^{\elph}(0)$ are respectively $[v(t)/v_F]^2=E^{1/3}$ and $[v(t)/v_F]^4=E^{2/3}$. At finite $V$, their vertex rates are evaluated with the superlattice propagators. The resulting end-point couplings, relative to $V=0$, are shown in Fig.~\ref{fig:flowlines}.

Due to finite-mesh truncation errors, our finite plane-wave basis does not attain the exact continuum result, even when the potential-dependent correction is already converged with respect to the mesh fineness. To separate this truncation error from the superlattice effect, we evaluate finite- and zero-potential loops with identical harmonic cutoffs and quadratures, and restore the exact homogeneous result:
\begin{equation}
 I(u,s)=I^{\infty}(0,s)+I^{\rm cut}(u,s)-I^{\rm cut}(0,s).
 \label{eq:UVloops}
\end{equation}
where $I^\infty, I^{\rm cut}$ denote the continuum and discretized loop integrals respectively. Linearity of the contact projection then gives
\begin{equation}
 \begin{split}
 \widehat{\mathsf K}(u,s)&=\mathsf K_0+\Delta\mathsf K_{\rm cut}(u,s),\\
 \Delta\mathsf K_{\rm cut}(u,s)&=\widehat{\mathsf K}_{\rm cut}(u,s)
                         -\widehat{\mathsf K}_{\rm cut}(0,s).
 \end{split}
 \label{eq:UV}
\end{equation}
Here $\widehat{\mathsf K}_{\rm cut}(u,s)$ is the contact matrix constructed from the truncated loops $I^{\rm cut}(u,s)$, evaluated at the dimensionless potential $u$ and running scale $s$ of Eq.~\eqref{eq:RG_scales}. As specified before, the homogeneous reference is supplied analytically. The zero-potential limit is therefore exact by construction. The numerical convergence at finite $V$ is established empirically.

\begin{figure*}[t]
    \centering
    \includegraphics[width=\textwidth]{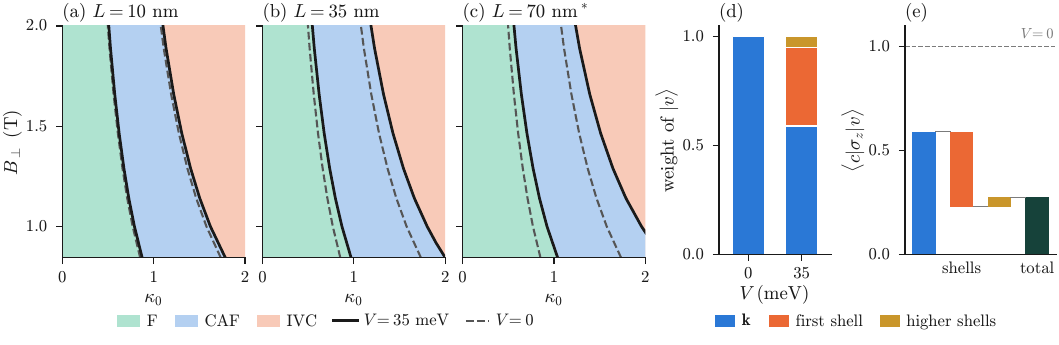}
    \caption{(a--c) Zeroth-Landau-level phase diagram obtained from the renormalized anisotropy couplings alone (without full Hartree-Fock), in the plane of bare interaction strength $\kappa_0$ and perpendicular field $B_\perp$, for superlattice periods (a) $L = 10$, (b) $35$ and (c) $70$ nm at fixed $V = 35$ meV and in-plane field $B_\parallel = 15$ T (which enters only through the Zeeman term). Solid lines: the F/CAF and CAF/IVC phase boundaries with the potential included in our finite-potential RG; dashed lines: the same boundaries at $V = 0$ \cite{xuhuang2025}. Shading indicates the phases with the potential. The potential moves both boundaries to larger $\kappa_0$, enlarging the F and CAF regions at the expense of the IVC phase, and the shift grows with $L$, i.e.\ with the dimensionless amplitude $u = V/\hbar v_F G$. The asterisk marks $u>0.4$, where the finite-potential approximation is less reliable. (d,e) Momentum mixing by the potential suppresses the sublattice transition amplitude. $\ket{v}$ and $\ket{c}$ are the highest negative-energy and lowest positive-energy zero-field Bloch states of one valley at the same momentum $\mathbf k$, obtained by exact diagonalization in the plane-wave basis $\mathbf k+\mathbf Q$ ($|\mathbf Q|\le 6G$, $G = 2\pi/L$) at $L = 35$ nm ($u = 0.30$ at $V=35$ meV). We take $|\mathbf k| = 0.02\,G$, at an angle of $0.3$ rad to the $x$ axis. (d) Weight of $\ket{v}$ on its original momentum $\mathbf k$, on the first shell $\mathbf k\pm\mathbf G_{x,y}$, and on higher shells, at $V=0$ and $V=35$ meV. (e) Contribution of each shell to $\bra{c}\sigma_z\ket{v}$ at $V=35$ meV, Eq.~\eqref{eq:RG_transition} (overall phase chosen so that the total is real and positive; dashed line: the $V=0$ value). By Eq.~\eqref{eq:RG_signed_weights}, each plane wave contributes $\pm$ its weight in (d). The first shell contributes with the opposite sign to the original momentum, so the amplitude drops from $1$ at $V = 0$ to $0.27$ at $V=35$ meV.}
    \label{fig:RGPHASEDIAGRAM}
\end{figure*}

\section{Results}
\label{sec:results}

Our central result is the self-consistent Hartree-Fock phase diagram of Sec.~\ref{sec:HF_results}. To disentangle the two ways in which the potential enters it, we first isolate the renormalization of the anisotropies in an RG plus zeroth-Landau-level calculation (Sec.~\ref{sec:RG_results}), and then return to the full Hartree-Fock treatment, which adds the potential to the Hamiltonian as a one-body term(Sec.~\ref{sec:HF_results}), and allows the density to relax in a momentum-dependent way.

\subsection{Renormalization Group $\nu = 0$ Phase Diagram}
\label{sec:RG_results}

The RG phase diagram of Fig.~\ref{fig:RGPHASEDIAGRAM} isolates how the potential changes the effective anisotropy interactions. We first evolve the contacts to $\ell_B$, and then select a spatially uniform ZLL state using those couplings and the Zeeman energy. The potential enters this calculation through the RG alone: the low-energy variational states retain the ideal ZLL orbital structure and have no miniband dispersion. This separation makes the interaction-renormalization mechanism visible before including the additional effects treated by Hartree-Fock. Due to the sublattice-valley locking of the ZLL, there are only two anisotropy energies in this projection:
\begin{equation}
 \begin{gathered}
 u_z=\frac{g_{zz}(\ell_B)}{2\pi\ell_B^2},\qquad
 u_\perp=\frac{g_{\perp z}(\ell_B)}{2\pi\ell_B^2},\\
 \epsilon_Z=\frac{g\mu_B}{2}\sqrt{B_\perp^2+B_\parallel^2}.
 \end{gathered}
 \label{eq:RG_ZLL_scales}
\end{equation}
 Up to a common flavor-symmetric contribution, the candidate energies are \cite{kharitonov2012}
\begin{align}
 \mathcal E_{\rm F}&=-u_z-2u_\perp-2\epsilon_Z,\nonumber\\
 \mathcal E_{\rm CAF}&=-u_z-\frac{\epsilon_Z^2}{2|u_\perp|},\nonumber\\
 \mathcal E_{\rm IVC}&=u_\perp,\qquad
 \mathcal E_{\rm CDW}=u_z.
 \label{eq:RG_phase_energies}
\end{align}
The CAF expression applies for $u_\perp<0$ and $\epsilon_Z<2|u_\perp|$; otherwise the canted solution saturates into F. In the CAF state the spin projection along the total field is $\cos\theta=\epsilon_Z/(2|u_\perp|)$. Thus the F/CAF boundary is $\epsilon_Z=2|u_\perp|$, while an unsaturated CAF meets IVC when $u_z+u_\perp+\epsilon_Z^2/(2|u_\perp|)=0$. \\\\

Beyond just the empirical RG flow lines, one can understand the zero potential phase diagram in a more intuitive way in terms of the influence of the filled Dirac sea. Increasing $\kappa$ increases the influence of the Dirac sea by lowering the non-interacting gaps and amplifying interaction induced Landau level mixing respectively. Inside the ZLL, the single-particle eigenstates are sublattice eigenstates owing to the sublattice valley locking. However, outside the ZLL, the $n$th Landau level acquires a two-component spinor structure of the form
\begin{equation}
    \begin{bmatrix}
        \ket{n-1}\\
        \lambda\ket{n}
    \end{bmatrix}
\end{equation}
where $\lambda = \pm 1$ labels whether the state is a valence or conduction level. This sign difference means forming sublattice eigenstates requires a superposition of Landau levels with indices $\pm n$; therefore, forming the CAF order parameter in remote levels comes at a kinetic energy cost. The greater the mixing of the ZLL with remote levels, the greater this energy cost will be, thus favoring IVC. Moreover, at very small interaction strength, the anisotropy interactions are strongly suppressed in favor of the non-interacting Zeeman term, thus favoring F.\\\\ At finite potential, the balance of the phases is shifted towards favoring the phases who gain energy from the Zeeman term, those being CAF and F, as opposed to the spin-singlet IVC phase. The zero-potential and finite-potential curves nearly coincide at short length and separate as the flow reaches scales sensitive to the superlattice.  At fixed $B_\perp$, increasing $\kappa_0$ strengthens the interaction corrections. Turning on the potential reduces the magnitudes of the relevant anisotropies relative to the homogeneous flow (Fig.~\ref{fig:flowlines}), while $\epsilon_Z$ is unchanged. The F/CAF condition then requires a larger bare interaction strength to overcome spin alignment. The CAF can also gain Zeeman energy by canting, whereas the spin-singlet IVC cannot. Both boundaries in Fig.~\ref{fig:RGPHASEDIAGRAM} move to larger $\kappa_0$, and their displacement increases with $L$ as the dimensionless potential grows. However, the F/CAF boundary shift is smaller, due to CAF also benefiting from Zeeman energy.\\\\ Now, we turn our attention to a physical picture of why the potential suppresses the anisotropies. Consider the zero-field Bloch states of one valley and spin, obtained by diagonalizing the Dirac Hamiltonian with $U(\mathbf r)$ in the plane-wave basis $\mathbf k+\mathbf Q$, where $\mathbf Q = G(n_x,n_y)$ runs over the superlattice reciprocal lattice. We define $\ket{v}$ ($\ket{c}$) as the highest negative-energy (lowest positive-energy) eigenstate at a momentum $\mathbf k$ near the Dirac point, and $u_{v\mathbf k}(\mathbf Q)$, $u_{c\mathbf k}(\mathbf Q)$ as their two-component sublattice spinors on the plane wave $\mathbf k+\mathbf Q$. At charge neutrality $\ket{v}$ is the highest occupied and $\ket{c}$ the lowest empty state at $\mathbf k$. At $V=0$ each is a single plane wave with $\mathbf Q = 0$, and the spinors are the Dirac spinors $(1,\pm e^{i\theta_{\mathbf k}})/\sqrt2$, with $\theta_{\mathbf k}$ the direction of $\mathbf k$. The first harmonic mixes in $\mathbf k\pm G\hat{\mathbf x}$ and $\mathbf k\pm G\hat{\mathbf y}$ (the first shell), with a smaller contribution from higher shells. Because the Dirac spinor is locked to the direction of momentum, a potential proportional to the sublattice identity nevertheless changes the sublattice composition of the Bloch states.

\begin{figure*}[t]
    \centering
    \includegraphics[width=\textwidth]{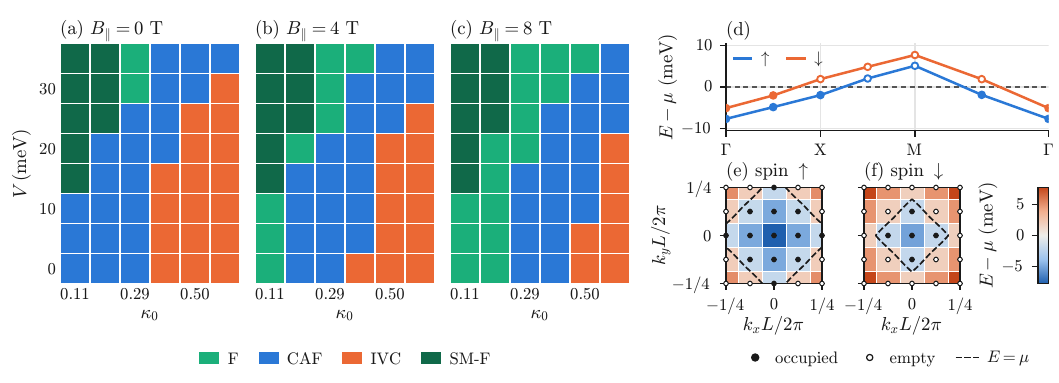}
    \caption{Hartree-Fock results at $\nu = 0$, $L = 35$ nm and flux ratio $p/q = 1/2$ per superlattice cell ($B_\perp = 1.69$ T). (a--c) Ground-state phase diagram in the plane of bare interaction strength $\kappa_0$ (columns: $\kappa_0 = 0.11$, $0.20$, $0.29$, $0.41$, $0.50$, $0.59$) and superlattice amplitude $V$ (rows: $0$ to $35$ meV in steps of $5$ meV), for in-plane fields (a) $B_\parallel = 0$, (b) $4$ and (c) $8$ T, which enter only through the Zeeman term. At each $V$ the anisotropy couplings are those of our finite-potential RG. Each cell shows the lowest-energy self-consistent state among the starting ansatzes described in Sec.~\ref{sec:coulomb}. F, CAF and IVC are insulating (Fig.~\ref{fig:states}); SM-F is the spin-polarized compensated semimetal; see Table~\ref{tab:order_params} for the order parameters. (d--f) Example of the SM-F state: the ground state at $\kappa_0 = 0.20$, $B_\parallel = 0$ and $V = 35$ meV. (d) Spin-$\uparrow$ and spin-$\downarrow$ HF bands nearest the Fermi level $\mu$ along $\Gamma$--X--M--$\Gamma$ (filled symbols: occupied, open: empty). (e,f) The same bands over one period of the mini-MBZ, $|k_{x,y}|\le\pi/2L$ (filled dots: occupied mesh points, open: empty; dashed: interpolated $E=\mu$ contour). A spin-$\uparrow$ hole pocket around M (split over the four corners) and a spin-$\downarrow$ electron pocket around $\Gamma$ coexist. Each covers $5$ of the $16$ mesh points of the period, i.e.\ the two pockets have equal area, as required by charge neutrality, and the spin polarization is reduced to $m = 0.375$.}
    \label{fig:phase_diagram}
\end{figure*}

The two anisotropies that select the $\nu=0$ state, $g_{zz}$ and $g_{\perp z}$, have contact vertices that carry the sublattice matrix $\sigma_z$. In the diagrams of Fig.~\ref{fig:placeholder}, these vertices are dressed by Coulomb-mediated virtual processes wherein the contact vertex scatters an electron from an occupied state into an empty one. How strongly a $\sigma_z$ vertex connects the occupied state $\ket v$ to the empty state $\ket c$ is measured by the sublattice transition amplitude
\begin{equation}
 M_z=\langle c|\sigma_z|v\rangle
 =\sum_{\mathbf Q}u_{c\mathbf k}^{\dagger}(\mathbf Q)
 \sigma_z u_{v\mathbf k}(\mathbf Q).
 \label{eq:RG_transition}
\end{equation}
Here $\sigma_z$ is the Pauli-z operator in the sublattice space. At $V=0$, $\sigma_z$ maps the valence Dirac spinor exactly onto the conduction one, so $|M_z| = 1$.

With the potential, the plane-wave components in Eq.~\eqref{eq:RG_transition} interfere destructively. For the first-harmonic potential this can be stated exactly. The orbital part of the particle-hole operator $S$ of Table~\ref{tab:symmetries}, $\sigma_z$ combined with the translation by $\mathbf d$, anticommutes with the Hamiltonian at fixed $\mathbf k$, and therefore maps $\ket v$ onto $\ket c$ up to a phase. Since the translation multiplies the plane wave $\mathbf k+\mathbf Q$ by $e^{i\mathbf k\cdot\mathbf d}(-1)^{n_x+n_y}$, this gives
\begin{equation}
 |M_z| = \Big|\sum_{\mathbf Q}(-1)^{n_x+n_y}\,w_{\mathbf Q}\Big|,
 \label{eq:RG_signed_weights}
\end{equation}
where $w_{\mathbf Q} = |u_{v\mathbf k}(\mathbf Q)|^2$ is the weight of $\ket v$ on the plane wave $\mathbf k+\mathbf Q$. The original momentum contributes with a positive sign and the four first-shell momenta ($n_x+n_y = \pm1$) with a negative one. Equivalently, weak mixing into a first-shell momentum rotates both spinors by $\boldsymbol\sigma\cdot\hat{\mathbf Q}$, and $(\boldsymbol\sigma\cdot\hat{\mathbf Q})\sigma_z(\boldsymbol\sigma\cdot\hat{\mathbf Q}) = -\sigma_z$. Figure~\ref{fig:RGPHASEDIAGRAM}(d,e) shows the result at $L = 35$ nm and $V = 35$ meV. The state $\ket v$ keeps $59\%$ of its weight on $\mathbf k$, but the $36\%$ on the first shell enters with the opposite sign, and $|M_z|$ drops from $1$ to $0.27$.

This cancellation weakens the RG enhancement of the anisotropies. A virtual pair $v\to c$ contributes to the dressing of a $\sigma_z$ vertex in proportion to $|M_z|^2/(E_c-E_v)$. Thus we are dealing with two competing effects. The potential lowers the particle-hole excitation energy to $0.64$ of its $V=0$ value at the momentum of Fig.~\ref{fig:RGPHASEDIAGRAM}(d,e), which alone would increase the contribution. However, $|M_z|^2$ falls to $0.075$, so the contribution drops to $0.12$ of its $V=0$ value. Since $\sum_n|\bra{n}\sigma_z\ket{v}|^2 = 1$, the missing weight moves to higher empty bands, which have larger energy denominators, and to occupied bands, which are Pauli blocked. Summed over all bands, the particle-hole part of the kernel $\widehat{\mathsf K}$ of Eq.~\eqref{eq:RG_ee} decreases, while the particle-particle part increases only slightly.  The contacts therefore grow more slowly than in the homogeneous flow.

\subsection{Full Hartree-Fock $\nu = 0$ Phase Diagram}
\label{sec:HF_results}

In the full Hartree-Fock treatment, the potential enters as a one-body term, and the bands are free to disperse across the mini-MBZ. Explicitly, the Hamiltonian that we solve self-consistently is
\begin{equation}
    H = \underbrace{H_D + H_U + H_Z}_{H_0} + H_C + H_{an}\big[g_{\alpha\beta}(\ell_B)\big].
    \label{eq:H_HF_model}
\end{equation}
Here $H_D$, $H_U$ and $H_Z$ are the Dirac, superlattice and Zeeman terms of Eqs.~\eqref{eq:H_D}--\eqref{eq:H_Z}, with the Zeeman energy set by the total field $B_{\rm tot}$, so that $B_\parallel$ enters only through $H_Z$. $H_C$ is the unscreened Coulomb interaction of Eq.~\eqref{eq:H_C}, whose strength is set by $\kappa_0$. $H_{an}$ is the contact interaction of Eq.~\eqref{eq:H_an}, with the couplings $g_{\alpha\beta}(\ell_B)$ obtained from our finite-potential RG flow (Sec.~\ref{sec:finiteV_method}) stopped at the magnetic length, evaluated separately at each potential amplitude $V$. The potential therefore enters both through $H_U$ and through the reduced couplings of Fig.~\ref{fig:flowlines}; at $V=0$ the couplings reduce to those of Ref.~\cite{xuhuang2025}. Repeating the calculation with the homogeneous ($V=0$) couplings at every $V$ changes the ground state in only $3$ of the $144$ cells of Fig.~\ref{fig:phase_diagram}(a--c). In each of these cells the two lowest phases are within $0.05$ meV of each other with homogeneous couplings, and the reduced couplings resolve the near-degeneracy in favor of F or CAF. The reorganization of the phase diagram is therefore driven mainly by the one-body potential $H_U$. In the MTG basis, we solve the self-consistency condition~\eqref{eq:selfconsistency} for the mean-field Hamiltonian
\begin{equation}
    h_{HF}(\mathbf k) = h_0(\mathbf k) + h_H(\mathbf k) + h_F(\mathbf k) + h^{MF}_{an}(\mathbf k).
    \label{eq:h_HF_full}
\end{equation}
Here $h_0$ is $H_0$ in the MTG basis, which contains the superlattice and Zeeman terms, $h_H$ and $h_F$ are the Hartree and Fock terms of Eqs.~\eqref{eq:h_H} and~\eqref{eq:h_F}, and $h^{MF}_{an}$ is the anisotropy mean field of Eq.~\eqref{eq:H_an_MF_matrix}. All four terms are evaluated with the same density matrix $\Delta(\mathbf k)$. Because $h_0$ now includes $H_U$, the Landau levels broaden into magnetic minibands of finite bandwidth. The resulting phase diagram is shown in Fig.~\ref{fig:phase_diagram}(a--c); the phases and their order parameters are listed in Table~\ref{tab:order_params} and sketched in Fig.~\ref{fig:states}. \\\\
More importantly, keeping the one-body term in the Hartree-Fock Hamiltonian allows the density to relax in a way that takes advantage of the momentum-dependent potential energy in the mini-MBZ.  When the potential strength is comparable to the inter-Landau-level spacing, the bands overlap and the insulating character of the zero-potential phases is destroyed. Pockets of the mini-MBZ fill all four flavors, inducing a partial polarization of the phase. The phases created in this transition are generically found to be compensated semimetals: since the filling is fixed at charge neutrality, the electron and hole pockets have equal area (Fig.~\ref{fig:phase_diagram}(d--f)). The flavor in these semimetallic phases prefers to be spin-polarized. The Zeeman term that stabilizes the ferromagnetic phase is a one-body term and hence linear in the magnetization, whereas the anisotropies are two-body terms and thus quadratic in the magnetization. The small anisotropy/Zeeman balance is thus shifted towards Zeeman for magnetizations smaller than unity. However, we make no claim that a partially polarized CAF or IVC phase can not exist. Indeed, the insulating to semimetallic transitions are continuous. Therefore, for sufficiently small potential-induced perturbations to a point where CAF is stable, the stable phase should be a semimetallic CAF, and ditto for IVC.

\subsection{Adiabatic Continuity Between Insulating Ferromagnetic State and Quantum Hall Ferromagnet}

In this section, we demonstrate that the insulating ferromagnetic state obtained in the Hartree-Fock numerics (F) is equivalent to the spin-polarized quantum Hall ferromagnet (QHFM) \cite{nomura2006,abanin2006,kharitonov2012}. To do this, we witness a path in parameter space connecting the two states along which the energy gap never closes. Two gapped states connected by such a path belong to the same phase \cite{chen2010}; the same strategy was used to show the topological equivalence between fractional Chern insulators and fractional quantum Hall states \cite{scaffidi2012,liu2013}. 

\begin{figure}[tb]
    \centering
    \includegraphics[width=\columnwidth]{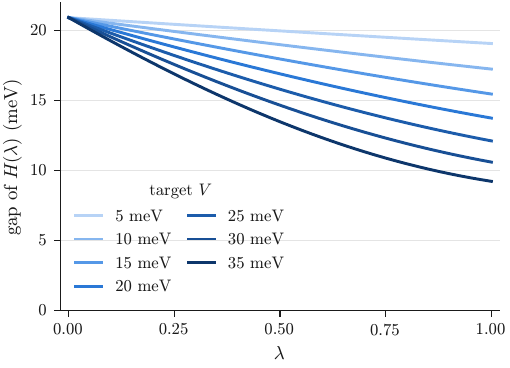}
    \caption{Gap of the straight line interpolation between the Hartree-Fock Hamiltonians for the zero potential and finite potential spin-polarized states at $L=35$ nm, half flux ($B_\perp = 1.69$ T), zero in-plane field and interaction strength $\kappa_0 = 0.5$.}
    \label{fig:adiabatic}
\end{figure}

We find that the straight-line linear interpolation path between the Hamiltonians suffices to witness adiabatic continuity. That is, for each potential strength $V$, interaction strength $\kappa_0$ and $\lambda \in [0,1]$, we form
\begin{equation}
    (1-\lambda)H_{HF}(0) + \lambda H_{HF}(V)
\end{equation}
where $H_{HF}(0)$ is the Hartree-Fock Hamiltonian corresponding to the zero-potential ferromagnet and $H_{HF}(V)$ is the Hartree-Fock Hamiltonian corresponding to the finite-potential ferromagnet. Each of the Hartree-Fock Hamiltonians used in the calculation are formed by a self-consistent converged spin-polarized ground state density matrix. Upon obtaining this interpolated Hamiltonian, we calculate its gap. At $V=0$, i.e $\lambda = 0$, we know that the spin-polarized ground state is a QHFM \cite{kharitonov2012,young2014}. 
Therefore, if the gap does not vanish along this straight line path, we are ensured of an adiabatic connection between our F state and the QHFM, since topological invariants such as the (spin) Chern number can only change when the gap closes \cite{thouless1982,sheng2006}. This is indeed what we find in our calculation, as shown in Fig.~\ref{fig:adiabatic}.

\section{Discussion and Conclusion}

We have shown that a smooth superlattice potential can reorganize the $\nu=0$ phase diagram of monolayer graphene through two connected mechanisms. First, the reconstructed Bloch wavefunctions modify the virtual processes that renormalize the short-range anisotropic interactions. The resulting suppression of the relevant couplings relative to the homogeneous system increases the importance of the Zeeman energy, favoring ferromagnetism and canted antiferromagnetism over spin-singlet intervalley coherence. Second, the potential gives the magnetic bands a finite dispersion, allowing their occupations to vary across the magnetic Brillouin zone. Combining these effects within self-consistent Hartree-Fock theory yields transitions among distinct insulating orders and a partially polarized compensated semimetal.

The stabilization of the gapped ferromagnet is a central consequence of this reorganization. Our phase diagram contains a ferromagnetic regime at zero in-plane field, for suitable potential amplitudes and interaction strengths. This result identifies superlattice engineering as a way to reduce the reliance on the large in-plane fields used to access the helical quantum Hall phase~\cite{young2014}. It complements dielectric screening~\cite{veyrat2020} by changing the orbital states that enter the interaction renormalization, and introduces the potential amplitude and period as additional controls over the ordered state. A direct experimental test would tune the potential at fixed perpendicular field and dielectric environment, using bulk compressibility together with nonlocal transport to distinguish an insulating bulk with conducting edges from a conducting bulk.

We establish the topological character of the tested ferromagnetic Hartree-Fock solutions by continuously interpolating between their Hamiltonians and a homogeneous ferromagnetic reference. For the tested parameter sets, the bulk gap remains positive along the path. Both endpoints conserve the spin component along the magnetic field, so the interpolation preserves the protecting spin $U(1)$ symmetry. This gapped, spin-conserving path establishes, within Hartree-Fock theory, that the finite-potential ferromagnetic solutions belong to the same topological phase as the conventional $\nu=0$ quantum Hall topological insulator~\cite{kharitonov2016qhti}. 

The detailed edge spectrum remains a natural direction for further study. A complementary characterization would compute the spin-resolved Chern numbers, followed by an edge-spectrum calculation to determine the reconstructed dispersions~\cite{khalifa2023absence,khalifa2023lattice}. A scalar superlattice potential can modify the edge states dispersions and localization while preserving the helical phase along a gapped, spin-conserving deformation. At the transition to the compensated semimetal, electron and hole pockets introduce bulk conduction, and the insulating-bulk description of the helical phase ceases to apply.

The semimetallic regime is itself a manifestation of the interplay between superlattice-induced dispersion and interaction-driven order. In the states found here, overlapping spin-resolved bands retain a net spin polarization while developing compensated Fermi pockets. Mapping this regime alongside the insulating phases is therefore necessary to identify where superlattice tuning can realize helical edge transport. 

More broadly, our work demonstrates that changing the superlattice geometry or its higher Fourier harmonics offers a route to controlling both the orbital matrix elements that dress the anisotropies and the magnetic miniband dispersion. Superlattice potentials thus provide a means to engineer the ordered states of a quantum Hall system through their underlying electronic wavefunctions.

\begin{acknowledgments}
T.C acknowledges funding from the Pennsylvania Space Grant Consortium (PSGC). We also acknowledge helpful conversations with Benjamin Hunt, Colin Hanes, Dihn Von Tuan, Oskar Vafek and Rokas Veitas. Computations in this work were done using SDSC through allocation PHY250361 from the Advanced Cyberinfrastructure Co-ordination Ecosystem: Services support (ACCESS) program \cite{access2023}, which is supported by U.S. National Science Foundation grants No. 2138259, No. 2138286, No. 2138307, No. 2137603, and No. 2138296.
\end{acknowledgments}

\nocite{pyatkovskiy2011}
\bibliography{references}

\clearpage
\onecolumngrid
\appendix

\section{Adding Interaction effects via the Hartree-Fock method}
\label{app:HF}

We seek to add electronic interaction effects into our model. We seek to achieve this by using a self-consistent Hartree-Fock approach; that is, we are looking for single Slater determinant (SSD) that minimizes the energy of the system. The derivation of the result can be found in many good resources \cite{szabo1996,ringschuck1980}; here we outline the derivation but focus on the algorithm itself.\\\\
Suppose one starts with a Hamiltonian of the form 
\begin{equation}
    H = \mathcal{H}_0 + \frac{1}{2}V_{ijk\ell}c_i^\dagger c_j^\dagger c_kc_\ell
\end{equation}
where $\mathcal{H}_0 = h_{ij}c_i^\dagger c_j$ is a one body Hamiltonian. Suppose further that 
\begin{equation}
    \ket{\Omega} = \prod_{\alpha=1}^{N}c_\alpha^\dagger \ket{0}
\end{equation}
minimizes energy among SSDs. Then we need a unitary matrix $U$ such that $c_\alpha = U_{\alpha,i}c_i$. Since $\ket{\Omega}$ is a SSD, we can write
\begin{equation}
    \bra{\Omega}c_i^\dagger c_j\ket{\Omega} = \Delta_{ji}
\end{equation}
The general approach is to take these $\Delta$ matrices as variational parameters in our search for the optimal SSD. In particular, we wish to minimize
\begin{equation}
    E = \bra{\Omega}H\ket{\Omega} = h_{ij}\Delta_{ji} + \frac{1}{2}V_{ijkl}(\Delta_{kj}\Delta_{\ell i} - \Delta_{ki}\Delta_{\ell j})
\end{equation}
Thus we must set
\begin{align}
    0 &= \delta E\\
    &= h_{ij}\delta\Delta_{ji} + \frac{1}{2}V_{ijk\ell}(\delta\Delta_{kj}\delta_{\ell i} + \Delta_{kj}\delta\Delta_{\ell i} - \delta\Delta_{ki}\Delta_{\ell j} - \Delta_{ki}\delta\Delta_{\ell i})\\
    &= [h_{ij} + V_{ik\ell j}\Delta_{\ell k} - V_{ijk\ell}\Delta_{\ell k}]\delta\Delta_{j\ell}
\end{align}
With this in mind, we define
\begin{equation}
    (H_{HF})_{ij} = h_{ij} + V_{ik\ell j}\Delta_{\ell k} - V_{ikj\ell}\Delta_{\ell k}
\end{equation}
We now want to find the Hartree-Fock basis; that is, the basis in which energy is minimized by filling $N$ one particle states. Put mathematically, the unitary matrix $U$ associated to such a basis will satisfy
\begin{equation}
    \Delta = UN_FU^\dagger
\end{equation}
where $N_F$ is a matrix that is $1$ on the first $N$ elements of the diagonal and zero everywhere else. Since $U$ is unitary, we may compute
\begin{equation}
    \delta(UU^\dagger) = \delta I = 0 \implies (\delta U)U^\dagger = -U(\delta U^\dagger)
\end{equation}
and so
\begin{align}
    \delta\Delta &= (\delta U)N_FU^\dagger + UN_F(\delta U^\dagger)\\
    &= (\delta U) N_FU^\dagger - UN_FU^\dagger (\delta U)U^\dagger\\
    &=  (\delta U)U^\dagger U N_FU^\dagger- UN_FU^\dagger (\delta U)U^\dagger\\
    &=  (\delta U)U^\dagger \Delta - \Delta (\delta U)U^\dagger\\
    &= [(\delta U)U^\dagger,\Delta]
\end{align}
Then
\begin{equation}
    0 = \delta E = \Tr(H_{HF}\delta \Delta) = \Tr(H_{HF}[(\delta U)U^\dagger,\Delta]) = \Tr((\delta U)U^\dagger [\Delta, H_{HF}])
\end{equation}
For the right hand side to vanish for all variations $\delta U$, it must be the case that $[\Delta, H_{HF}]$ vanishes. Thus the Hartree-Fock basis is the eigenbasis of the Hartree-Fock Hamiltonian, as the name would suggest. We now outline the algorithm for solving for the $\Delta$ matrices and obtaining the ground state phases.
\begin{enumerate}
    \item Guess an inital $\Delta$. For instance, one can do this by choosing the non-interacting ground state density matrix to be the first $\Delta$ matrix, i.e 
    \begin{equation}
        \Delta_{mn} = \bra{\Omega}\ket{n}\bra{m}\ket{\Omega}
    \end{equation}
    where $\ket{n}$ labels some basis, which would probably the basis in which the Hamiltonian was originally resolved, in our case the lattice basis, and $\Omega$ denotes the ground state wave function.
    \item Using $\Delta_{mn}$, create $H_{HF}$ via
    \begin{equation}
        (H_{HF})_{ij} = h_{ij} + V_{ik\ell j}\Delta_{\ell k} - V_{ikj\ell}\Delta_{\ell k}
    \end{equation}
    \item Fill the lowest $n_F$ energies of $H_{HF}$ and compute the new $U$ matrix. That is to say, diagonalize $H_{HF}$, and extract $U$ as the basis of eigenvectors of $H_{HF}$.
    \item Create a new $\Delta$ using $U$ and $N_F$; that is, write
    \begin{equation}
        \Delta = UN_FU^\dagger
    \end{equation}
    \item Fix some $\ep$ accuracy to taste; if $\norm{\Delta_{new}-\Delta_{old}} < \ep$, then you're done, where the norm denotes the Frobenius norm. Otherwise return to step $2$ and iterate until convergence.
    \item Upon obtaining the final density matrix, extract the filled orbitals (i.e the eigenvectors of the density matrix with eigenvalue $1$).
    \item Evaluate various order parameters on this approximate ground state.
\end{enumerate}

\section{Accelerating HF convergence with direct inversion in the iterative subspace }
We implemented the direct inversion in the iterative subspace (DIIS) algorithm \cite{pulay1980,pulay1982} to insure and expedite the convergence of the self-consistent field method. We derive the DIIS condition for minimizing the error and then elucidate the algorithm itself. The DIIS method seeks to find the best possible approximation to the true error vector out of error vectors accumulated through each previous iteration of the program. To that end, write the result of the ith iteration of the program as
\begin{equation}
    p + e_i
\end{equation}
where $p$ is the true solution and $e_i$ is then an error vector associated with the $i$th solution.\ Now suppose we have
\begin{equation}
    p = \sum_i c_i(p+e_i) = p\sum_ic_i + \sum_ic_ie_i
\end{equation}
Clearly, if we want to minimize our error in this ansatz, we wish for $\sum_ic_ie_i$ to be minimized and for $\sum_ic_i$ to be $1$. This is a constrained minimization problem that can be solved by a Lagrange multipliers approach. The associated Lagrangian is
\begin{equation}
   L =  \sum_{ij}c_jB_{ji}c_i - 2\lambda(\sum_ic_i - 1)
\end{equation}
where $\lambda$ is the Lagrange multiplier and $B_{ij} \equiv \ip{e_j}{e_i}$. Setting all the partials of $L$ to zero, we obtain the following system of equations in matrix form:
\begin{equation}
    \begin{bmatrix}
            B_{11} & B_{12} &\cdots & B_{1n} & -1\\
            B_{21} & B_{22} &\cdots & B_{2n} & -1\\
            \vdots & \vdots & \ddots & \vdots & \vdots \\
            B_{n1} & B_{22} &\cdots & B_{2n} & -1\\
            -1 & -1 & \cdots & -1 & 0
        \end{bmatrix}
        \begin{bmatrix}
            c_1\\
            c_2\\
            \vdots\\
            c_n\\
            \lambda
        \end{bmatrix}
            = \begin{bmatrix}
        0\\
        0\\
        \vdots\\
        0\\
        -1
    \end{bmatrix}
\end{equation}
The algorithm is then as follows:
\begin{enumerate}
    \item Initialize an initial guess for the $\Delta$ parameter, as well as an (empty) list of previous iterations $\Delta_i$ and another empty list of previous error vectors $e_i$.
    \item Run two SCF iterations (as described in a previous section) and append the difference between the old density matrix and the updated one to the error vector list, as well as appending the old guess to the list of previous iterations.
    \item Upon obtaining the next iterative update of the $\Delta$ matrix, solve the central matrix vector equation for the coefficients $c_i$ (where the $B_{ij},\lambda$ are defined as above) for a list of $n$ errors:
    \begin{equation*}
        \begin{bmatrix}
            B_{11} & B_{12} &\cdots & B_{1n} & -1\\
            B_{21} & B_{22} &\cdots & B_{2n} & -1\\
            \vdots & \vdots & \ddots & \vdots & \vdots \\
            B_{n1} & B_{22} &\cdots & B_{2n} & -1\\
            -1 & -1 & \cdots & -1 & 0
        \end{bmatrix}
        \begin{bmatrix}
            c_1\\
            c_2\\
            \vdots\\
            c_n\\
            \lambda
        \end{bmatrix}
            = \begin{bmatrix}
        0\\
        0\\
        \vdots\\
        0\\
        -1
    \end{bmatrix}
    \end{equation*}
    \item Use these coefficients to extrapolate the Fock matrix in the following way:
    \begin{equation}
        \Delta_{\textrm{new}} = \sum_{i}c_i\Delta_i 
    \end{equation}
    \item Diagonalize the new $\Delta$ matrix and once again iterate the SCF procedure. If the process hasn't converged, return to step 3. Otherwise, you're done.
\end{enumerate}
We also note that one typically keeps a preset maximum number of error vectors, usually about $8$, and keeps those with the smallest moduli.

\section{Magnetic translation group}
\label{app:MTG}
Our presentation follows Ref.~\cite{girvinyang2019}. Though Landau levels carry a band-like index, they are not themselves bands. For the purposes of diagonalizing a Bloch Hamiltonian for this problem, we need to build Bloch states out of Landau levels. \\\\
In an ordinary lattice situation, the Hamiltonian commutes with the lattice translation operators, and hence its eigenstates are also eigenstates of the lattice translation operators $T_a$ (Bloch wave functions) with eigenvalue labeled by the (crystal) momentum. Indeed, Bloch eigenstates translated by a lattice vector are also eigenstates with the same eigenvalue. Suppose we have an eigenstate of the magnetic Hamiltonian $H$, with $H\psi = E\psi$. If we attempt the same conjugation by translation operators on the magnetic Hamiltonian, we find
\begin{equation}
 E T_a \psi = T_a H\psi =  T_a HT_a^\dagger T_a \psi = \frac{1}{2 _e}[p+\frac{e}{c}A(r+a)]^2T_a\psi
\end{equation}
Hence $T_a\psi$ is an eigenstate of a different Hamiltonian, that being
\begin{equation}
    H' = T_a HT_a^\dagger = \frac{1}{2m_e}[p+\frac{e}{c}A(r+a)]^2
\end{equation}
Of course, in the case of a uniform magnetic field, the vector potential and its translation have the same curl, that being the uniform magnetic field $B$. Then there is $\chi_a$ such that
\begin{equation}
    A(r+a) = A(r) + \nabla \chi_a(r)
\end{equation}
with the corresponding gauge transformation
\begin{equation}
    \exp(i\frac{e\chi_a}{\hbar c})H'\exp(-\frac{e\chi_a(r)}{\hbar c}) = H
\end{equation}
This leads us to the definition of the magnetic translation group (MTG):
\begin{equation}
    T^B_a =  \exp(i\frac{e\chi_a}{\hbar c})T_a
\end{equation}
with 
\begin{equation}
    [T^B_a,H] = 0
\end{equation}
Then the eigenstates of our Hamiltonian will be eigenstates of the magnetic translation group operators. From here-on out, we work in the Landau gauge for convenience; i.e, we have that $A = Bx\hat{y}$. In this gauge, we may compute the commutator subgroup of the MTG. For a square lattice with primtive vectors $a\hat{x},b\hat{y}$, we have that 
\begin{align}
    T^B_{-a}T_{-b}^BT_a^BT_b^B = \exp(2\pi i\frac{eB}{\hbar }ab)I
\end{align}
To see this, note that in the exponent we are adding up (integrating) the magnetic vector potential around a closed loop, giving the enclosed magnetic flux. We prefer abelian groups, as we can decompose them into a direct sum of $1$D (unitary) irreducible representations. From this calculation, it is obvious what the condition for the operators to commute; we would need the unit cell to be enlarged such that it contained an integer number of flux quanta.  Moreover, because the flux condition determines only the area of the unit cell and not the shape, it must be the case that the band dispersion is periodic for the reciprocal lattices associated to each of the magnetic unit cell construction; because we can continuously vary this family of lattices, the band dispersion is forced to be constant! This is the origin of the degenerate Landau levels of constant energy.\\\\
We now consider a weak periodic perturbation to the problem, represented by a term $U(r)$, which introduces a unit cell to our system but not necessarily a commensurate one (i.e with integer flux). However, we may construct such a commensurate magnetic unit cell; if there are $p/q$ flux quanta through the lattice unit cell, we may use $q$ copies of the lattice unit cell as our magnetic unit cell. Thus, each copy of the original unit cell contains $p$ (an integer) many flux quanta, and accordingly, the Landau levels split into $p$ magnetic sub-bands. In the presence of a potential, these subbands are in general not degenerate. Furthermore, the eigenfunctions of this Hamiltonian will not be Bloch functions (eigenfunctions of the lattice translation algebra) but rather they will be eigenfunctions of the magnetic translation algebra; that is, they will satisfy the following generalized Bloch conditions (in the Landau gauge):
\begin{equation}
    u_k(x+a_x,y+a_y) = \exp(ik_xa_x)\exp(ik_ya_y)\exp(ipK_yy)u_k(x,y)
\end{equation}
Despite these not actually being Bloch conditions, we can build Bloch-like states satisfying these conditions which bring the Hamiltonian to a form that is diagonal in momentum. Inspired by the usual Fourier series construction of Bloch functions, we use the following basis, which is labeled by a Landau level index (natural number), sub-band index ($1$ to $p$) and a momentum $(k_\parallel)$:
\begin{equation}
    \ket{\ell,n,k_\parallel} = \frac{1}{\sqrt{N_{qa}}}\sum_{s=-\infty}^\infty \exp(-ik_x\ell_B^2 (sp+\ell)K_y) \ket{n,k_y + (sp+\ell)K_y}
\end{equation}
This is for a "classical" (quadratic) dispersion. For the graphene case (linear dispersion), we may easily generalize this by taking analogous linear combinations of graphene Landau levels (in one valley):
\begin{equation}
    \ket{\ell,n,k_\parallel,\lambda} = \frac{1}{\sqrt{N_{qa}}}\sum_{s=-\infty}^\infty \exp(-ik_x\ell_B^2(sp+\ell)K_y) \begin{bmatrix}\ket{n,k_y + (sp+\ell)K_y}\\
   \lambda\ket{n-1,k_y + (sp+\ell)K_y} \end{bmatrix}
\end{equation}
Note that when expanding in this basis, we are using eigenstates of the magnetic Hamiltonian and hence implicitly taking the magnetic part of the Hamiltonian to be the unperturbed part. It is not surprising, then, that for strong external potentials, the basis is no longer useful. One would have to keep an impractical number of Landau levels to achieve convergence, even for low energies. The condition for a correct solution using this basis is roughly captured by 
\begin{equation}
   \frac{V}{\hbar v_F/\ell_B} \lesssim 1
\end{equation}
where $V$ is the strength of the first harmonic of the periodic potential, and $v_F$ is the Fermi velocity.

\section{Explicit Representation of the Basis}
Write $\Gamma$ for the collective sublattice, spin, LL and valley indices, and let $\psi_n$ be the Landau level in the Landau gauge $A = Bx\hat{y}$. Then our wave functions are
\begin{equation}
        \ket{\Psi_\Gamma(k)} = \frac{1}{\sqrt{N_{qa}}}\sum_{s\in \Z} \exp(2\pi i k_1s)\hat{t}(sa_1)\ket{\psi_\Gamma(k_2 + \ell/q)} = \frac{1}{\sqrt{N_{qa}}}\sum_{s\in \Z} \exp(2\pi i k_1s)\ket{\psi_n(k_2-s\frac{p-\ell}{q})}
\end{equation}

\section{Computing the Structure Factor}
\label{app:structure_factor}
Following the approach of Ref.~\cite{wangvafek2024}, we find that our Fourier matrix element is
\begin{align}
    \Lambda^q_{\Gamma_1,\Gamma_2}(k,p) &= \hal\delta_{K_1,K_2}\delta_{p_1+q_1,k_1}\exp(2\pi i(k_1-q_1)s)
    \exp(iq_x\tilde{k_2}\ell_B^2)\delta_{k2,p_2+s\frac{\phi}{\phi_0}+q_2} \\
    &\sqrt{2}^{\delta_{n_1,0}+\delta_{n_2,0}}(S_{n_2}^{n_1}(q)
    + \theta(n_1-1)\theta(n_2-1)\lambda_1\lambda_2 S_{n_1-1}^{n_2-1}(q))\evalat_{s = \frac{\ell_2-\ell_1-qq_2}{p}}
\end{align}
which is only nonzero if $\ell_2-\ell_1-qq_2$ is an integer multiple of $p$, as $s$ must be an integer.

\section{Computing the Coloumb matrix element}
We compute the Coloumb matrix element directly. We do this for each element of the spinor, and we then extend our calculation to the graphene case using the formula presented in the "Going from a classical dispersion to a relativistic dispersion" section. We compute
\begin{align}
    &V^{n_1,\ell_1,K_1,S_1;n_3,\ell_3,K_3,S_3}_{n_2,,\ell_2,K_2,S_2;n_4,\ell_4,K_4,S_4}(k^1_\parallel,k^2_\parallel,k^3_\parallel,k^4_\parallel) = \delta_{K_1,K_3}\delta_{K_2,K_4}\delta_{S_1,S_3}\delta_{S_2,S_4}\bra{\Psi_{n_1,\ell_1}(k^1_\parallel),\Psi_{n_2,\ell_2}(k^2_\parallel)}V\ket{\Psi_{n_3,\ell_3}(k^3_\parallel),\Psi_{n_4,\ell_4}(k^4_\parallel)}\\
    &= \frac{1}{N_{qa}^2}\sum_{s_1,s_3}\exp(2\pi i(\frac{k_x^3}{K_x}s_3 - \frac{k_x^1}{K_x}s_1))\sum_{s_2,s_4}\exp(2\pi i(\frac{k_x^4}{K_x}s_4 - \frac{k_x^2}{K_x}s_2))\\
    & \bra{\psi_{n_1}(k_y^1+\frac{\ell_1-ps_1}{q}K_y),\psi_{n_2}(k_y^2+\frac{\ell_2-ps_2}{q}K_y)}V\ket{\psi_{n_3}(k_y^3+\frac{\ell_3-ps_3}{q}K_y),\psi_{n_4}(k_y^4+\frac{\ell_4-ps_4}{q}K_y)}\\
    &= \frac{1}{N_{qa}^2}\sum_{s_1,s_3}\exp(2\pi i(\frac{k_x^3}{K_x}s_3 - \frac{k_x^1}{K_x}s_1))\sum_{s_2,s_4}\exp(2\pi i(\frac{k_x^4}{K_x}s_4 - \frac{k_x^2}{K_x}s_2))\\
    & \sum_{q,G}V(Q)S^{n_1}_{n_3}(Q)S^{n_2}_{n_4}(-Q)\exp(iQ_x\ell_B^2(k_y^4-k_y^3+\frac{\ell_4-\ell_3-ps_4+ps_3}{q}K_y))\delta_{k_y^1+\frac{l_1-ps_1}{q}K_y}^{k_y^3+\frac{l_3-ps_3}{q}K_y}(Q_y)\delta_{k_y^2+\frac{l_2-ps_2}{q}K_y}^{k_y^4+\frac{l_4-ps_4}{q}K_y}(-Q_y)
\end{align}
Write $Q=  q +G$, where $q$ is in the MBZ and $G$ is a magnetic reciprocal lattice vector. Using the delta function constraints, we now solve for $s_2$ in terms of $s_4$ in order to eliminate one of the sums. Compute
\begin{align}
    k_y^3 - k_y^2-Q_y + \frac{\ell_4-\ell_2-ps_4}{q}K_y &= \frac{-ps_2}{q}K_y\\
    \implies -q(\frac{k_y^4-k_y^2-q_y}{K_y}) + \frac{qG_y}{K_y} +  ps_4+\ell_2-\ell_4 &= ps_2
\end{align}
Note that the RHS is an integer, and so is $\frac{qG_y}{K_y} +ps_4+\ell_2-\ell_4$. Then $-q(\frac{k_y^4-k_y^2-q_y}{K_y})$ must also be an integer, and so $\frac{k_y^4-k_y^2-q_y}{K_y}$ must be a multiple of $1/q$. However, since our $k$-vectors have $y$ components in the MBZ, i.e from $0$ to $1/q$, this quantity is forced to be zero mod the MBZ. So we have that 
\begin{equation}
    ps_4 + \ell_2-\ell_4 + \frac{qG_y}{K_y} = ps_2
\end{equation}
Further note that the RHS is an integer multiple of $p$, as is $ps_4$. Thus $\ell_2-\ell_4-\frac{qG_y}{K_y}$ must also be a multiple of $p$ if the contribution to the matrix element is to not vanish. A very similar calculation yields
\begin{equation}
    ps_3 + \ell_1-\ell_3 - \frac{qG_y}{K_y}= ps_1
\end{equation}
Again we have that $\ell_1-\ell_3 - \frac{qG_y}{K_y}$ must be an integer multiple of $p$ for this summand to not vanish. We may now continue computing the Coloumb matrix elements. The sums over $s_2$ and $s_1$ are eliminated, and so we find
\begin{align}
    &= \frac{1}{N_{qa}^2}\sum_{q,G}\sum_{s_3}\exp(\frac{2\pi  i}{K_x}(k_x^2s_3 - k_x^1(s_3+\frac{\ell_1-\ell_3}{p} - \frac{qG_y}{pK_y})))\sum_{s_4}\exp(\frac{2\pi i}{K_x}(k_x^4s_4-k_x^2(s_4+\frac{\ell_2-\ell_4}{p} + \frac{qG_y}{pK_y})))\\
    & V(q+G)S^{n_1}_{n_3}(q+G)S^{n_2}_{n_4}(-q-G)\exp(i(q_x+G_x)\ell_B^2(k_y^4-k_y^3+\frac{\ell_4-\ell_3-ps_4+ps_3}{q}K_y))\\
    &= \frac{1}{N_{qa}^2} \sum_{q,G}V(q+G)S^{n_1}_{n_3}(q+G)S^{n_2}_{n_4}(-q-G)\exp(i(q_x+G_x)\ell_B^2(k_y^4-k_y^3+\frac{\ell_4-\ell_3}{q}K_y))\\
    & \exp(\frac{2\pi i}{pK_x}k_x^1(\ell_3-\ell_1+\frac{qG_y}{K_y}))\exp(\frac{2\pi i }{pK_x}k_x^2(\ell_4-\ell_2-\frac{qG_y}{K_y}))\\
    & \sum_{s_3}\exp(is_3(\frac{-2\pi ik_x^1}{K_x} - \frac{(q_x+G_x)\ell_B^2p}{q}K_y + \frac{2\pi k_x^3}{K_x})) \sum_{s_4}\exp(is_4(\frac{-2\pi ik_x^2}{K_x} +\frac{(q_x+G_x)\ell_B^2p}{q}K_y + \frac{2\pi k_x^4}{K_x}))\\
    &=  \sum_{q,G}V(q+G)S^{n_1}_{n_3}(q+G)S^{n_2}_{n_4}(-q-G)\exp(i(q_x+G_x)\ell_B^2(k_y^4-k_y^3+\frac{\ell_4-\ell_3}{q}K_y))\\
    & \exp(\frac{2\pi i}{pK_x}k_x^1(\ell_3-\ell_1 +\frac{qG_y}{K_y}))\exp(\frac{2\pi i }{pK_x}k_x^2(\ell_4-\ell_2-\frac{qG_y}{K_y}))\delta_{\frac{2\pi k_x^3}{K_x},\frac{2\pi k_x^1}{K_x}}(\frac{(q_x+G_x)\ell_B^2 p}{q}K_y)\delta_{\frac{2\pi k_x^4}{K_x},\frac{2\pi k_x^2}{K_x}}(-\frac{(q_x+G_x)\ell_B^2 p}{q}K_y)
\end{align}
We now use the delta function constraint to solve for $k_x^1$ in terms of $k_x^3$. Compute
\begin{equation}
    \frac{2\pi}{K_x}(k_x^3-k_x^1) - \frac{(q_x+G_x)\ell_B^2p}{q}K_y = 0 \implies k_x^1 = (k_x^3-q_x - G_x) 
\end{equation}
A very similar calculation yields
\begin{equation}
    k_x^2 = (k_x^4+q_x+G_x)
\end{equation}
Plugging back in, we find
\begin{align}
    &\sum_{q,G}V(q+G)S^{n_1}_{n_3}(q+G)S^{n_2}_{n_4}(-q-G)\exp(i(q_x+G_x)\ell_B^2(k_y^4-k_y^3+\frac{\ell_4-\ell_3}{q}K_y))\\
    &\exp(\frac{2\pi i}{pK_x}(k_x^3-q_x - G_x)(\ell_3-\ell_1 + \frac{qG_y}{K_y}))
    \exp(\frac{2\pi i }{pK_x}(k_x^4+q_x + G_x)(\ell_4-\ell_2 - \frac{qG_y}{K_y}))\delta_{k_x^1}^{k_x^3}(q_x + G_x)\delta_{k_x^2}^{k_x^4}(-q_x - G_x)
\end{align}
Recall that earlier we also deduced that $k_y^1 = k_y^3+q_y + G_y,k_y^2=k_y^4-q_y - G_y$, so we may add extra delta functions. We may further reduce the phases modulo reciprocal lattice vectors, as these phases have the periodicity of the reciprocal lattice.
\begin{align}
    &\sum_{q,G}V(q+G)S^{n_1}_{n_3}(q+G)S^{n_2}_{n_4}(-q-G)\exp(i(q_x+G_x)\ell_B^2(k_y^4-k_y^3+\frac{\ell_4-\ell_3}{q}K_y))\\
    &\exp(\frac{2\pi i}{pK_x}(k_x^3-q_x)(\ell_3-\ell_1 + \frac{qG_y}{K_y}))
    \exp(\frac{2\pi i }{pK_x}(k_x^4+q_x)(\ell_4-\ell_2 - \frac{qG_y}{K_y}))\delta_{k_x^1}^{k_x^3}(q_x+G_x)\\
    &\delta_{k_x^2}^{k_x^4}(-(q_x+G_x))\delta_{k_y^3,k_y^1}(q_y + G_y)\delta_{k_y^4,k_y^2}(-q_y - G_y)
\end{align}
We emphasize again that the summands are only nonzero for $\ell_2-\ell_4-\frac{qG_y}{K_y},\ell_3-\ell_1+\frac{qG_y}{K_y}\in p\Z$. 


\section{Computing the direct and exchange terms}

We again compute this term only for the classical dispersion, the relativistic dispersion is again an easy modification alla the "Going from a classical dispersion to a relativistic dispersion" section. We shall furthermore condense the spin, valley, subband and LL index degrees of freedom into a single index $\Gamma$. Recall that our formula for the exchange term in terms of the Coloumb matrix elements is 
\begin{equation}
    V_{ikj\ell}\Delta_{\ell k}
\end{equation}
Note that in our convention $\ell$ is the third index and $k$ is the second index. Thus we may use the expression derived in the previous section to write 
\begin{align}
    &E_{n_1,\ell_1,S_1,K_1}^{n_4,\ell_4,S_4,K_4}(k_\parallel^4) = \sum_{\Gamma_2,\Gamma_3}\sum_{k_\parallel^2,k_\parallel^3} \sum_{q,G}\Delta_{\Gamma_3,\Gamma_2}(k_\parallel^3,k_\parallel^2) V(q+G)S^{n_1}_{n_3}(q+G)S^{n_2}_{n_4}(-q-G)\\ &\exp(i(q_x+G_x)\ell_B^2(k_y^4-k_y^3+\frac{\ell_4-\ell_3 }{q}K_y))\exp(\frac{2\pi i}{pK_x}(k_x^3-q_x)(\ell_3-\ell_1 + \frac{qG_y}{K_y}))\exp(\frac{2\pi i }{pK_x}(k_x^4+q_x)(\ell_4-\ell_2 - \frac{qG_y}{K_y}))\\
    & \delta_{k_x^1}^{k_x^3}(q_x+G_x)\delta_{k_x^2}^{k_x^4}(-q_x - G_x)\delta_{k_y^3,k_y^1}(q_y + G_y)\delta_{k_y^4,k_y^2}(-q_y - G_y)\delta_{S_1,S_3}\delta_{S_2,S_4}\delta_{K_1,K_3}\delta_{K_2,K_4}\\
    &= \sum_{q,G}\sum_{\Gamma_2,\Gamma_3}\sum_{k_\parallel^2,k_\parallel^3}\Delta_{\Gamma_3,\Gamma_2}(k_\parallel^3,k_\parallel^2) \delta_{k_x^1}^{k_x^3}(q_x + G_x)\delta_{k_x^2}^{k_x^4}(-q_x - G_x)\delta_{k_y^3,k_y^1}(q_y)\delta_{k_y^4,k_y^2}(-q_y - G_y)\delta_{S_1,S_3}\delta_{S_2,S_4}\delta_{K_1,K_3}\delta_{K_2,K_4}\\
     &V(q+G)S^{n_1}_{n_3}(q+G)S^{n_2}_{n_4}(-q-G)\exp(i(q_x+G_x)\ell_B^2(k_y^4-k_y^3+\frac{\ell_4-\ell_3}{q}K_y))\\
     &\exp(\frac{2\pi i}{pK_x}(k_x^3-q_x)(\ell_3-\ell_1 + \frac{qG_y}{K_y}))\exp(\frac{2\pi i }{pK_x}(k_x^4+q_x)(\ell_4-\ell_2 - \frac{qG_y}{K_y}))
\end{align}

We also make the usual assumption that the density matrix is diagonal in momentum, i.e $\Delta(k_\parallel^2,k_\parallel^3)\propto \delta_{k_\parallel^2,k_\parallel^3}$. Thus we find
\begin{align}
      &= \sum_{q,G}\sum_{\Gamma_2,\Gamma_3}\sum_{k_\parallel^2,k_\parallel^3} \delta_{k_x^1}^{k_x^3}(q_x + G_x)\delta_{k_x^2}^{k_x^4}(-q_x - G_x)\delta_{k_y^3,k_y^1}(q_y + G_y)\delta_{k_y^4,k_y^2}(-q_y - G_y)\delta_{S_1,S_3}\delta_{S_2,S_4}\delta_{K_1,K_3}\delta_{K_2,K_4}\Delta_{\Gamma_3,\Gamma_2}(k_\parallel^3)\\
      &\delta_{k_\parallel^3,k_\parallel^2} \delta_{S_1,S_3}\delta_{S_2,S_4}\delta_{K_1,K_3}\delta_{K_2,K_4}  V(q+G)S^{n_1}_{n_3}(q+G)S^{n_2}_{n_4}(-q-G)\\
     &\exp(i(q_x+G_x)\ell_B^2(k_y^4-k_y^3+\frac{\ell_4-\ell_3}{q}K_y))\exp(\frac{2\pi i}{pK_x}(k_x^3-q_x)(\ell_3-\ell_1 + \frac{qG_y}{K_y}))\exp(\frac{2\pi i }{pK_x}(k_x^4+q_x)(\ell_4-\ell_2 - \frac{qG_y}{K_y}))\\
     &= \sum_{q,G}\sum_{\Gamma_2,\Gamma_3}\Delta_{\Gamma_3,\Gamma_2}(k_x^4+q_x + G_x,k_y^4+q_y + G_y) \delta_{S_1,S_3}\delta_{S_2,S_4}\delta_{K_1,K_3}\delta_{K_2,K_4}\\
      &V(q+G)S^{n_1}_{n_3}(q+G)S^{n_2}_{n_4}(-q-G)\exp(i(q_x+G_x)\ell_B^2(k_y^4-k_y^3+\frac{\ell_4-\ell_3}{q}K_y))\\
      &\exp(\frac{2\pi i}{pK_x}(k_x^3-q_x)(\ell_3-\ell_1 +\frac{qG_y}{K_y}))\exp(\frac{2\pi i }{pK_x}(k_x^4+q_x)(\ell_4-\ell_2 - \frac{qG_y}{K_y}))\\
        &= \sum_{q,G}\sum_{\Gamma_2,\Gamma_3}\Delta_{\Gamma_3,\Gamma_2}(k_x^4+q_x,k_y^4 + q_y) \delta_{S_1,S_3}\delta_{S_2,S_4}\delta_{K_1,K_3}\delta_{K_2,K_4} V(q+G)S^{n_1}_{n_3}(q+G)S^{n_2}_{n_4}(-q-G)\\
        &\exp(i(q_x+G_x)\ell_B^2(-q_y+\frac{\ell_4-\ell_3}{q}K_y))\exp(\frac{2\pi i}{pK_x}k_x^4(\ell_3-\ell_1 + \frac{qG_y}{K_y}))\exp(\frac{2\pi i }{pK_x}(k_x^4+q_x)(\ell_4-\ell_2 - \frac{qG_y}{K_y}))
\end{align}
Note that we have reduced the input to the density matrix mod reciprocal lattice vectors due to the periodicity of the system. Moreover, there is ample opportunity for simplifcation in this expression, in particular we may neglect inter-Landau level scattering, which would make the majority of terms in the sum vanish. Furthermore, though the $q$-sum/integral is apriori over all of $k$-space, we may truncate the domain of integration to be a disk with radius a small integer multiple of $1/\ell_B$, the reason being that the integrand is exponentially suppressed in $q\ell_B$ as well as oscillating rapidly for such large $q$. Furthermore, the summands are \textbf{extremely} sparse in the subband indices due to the aforementioned restriction. We may now derive an expression for the direct term with a very similar calculation. We also remind the reader that we have neglected the sublattice indices, but these can be introduced via an easy modification to the structure factor $S$. Recall that the direct term is given by   
\begin{equation}
    V_{ik\ell j}\Delta_{\ell k}
\end{equation}
Thus we compute 
\begin{align}
    &D_{n_1,\ell_1,S_1,K_1}^{n_3,\ell_3,S_3,K_3}(k_\parallel^3) = \sum_{\Gamma_2,\Gamma_4}\sum_{k_\parallel^2,k_\parallel^4} \sum_{q,G}\Delta_{\Gamma_4,\Gamma_2}(k_\parallel^4,k_\parallel^2) V(q+G)S^{n_1}_{n_3}(q+G)S^{n_2}_{n_4}(-q-G)\\
    &\exp(i(q_x+G_x)\ell_B^2(k_y^4-k_y^3+\frac{\ell_4-\ell_3}{q}K_y))\exp(\frac{2\pi i}{pK_x}(k_x^3-q_x)(\ell_3-\ell_1 + \frac{qG_y}{K_y}))\exp(\frac{2\pi i }{pK_x}(k_x^4+q_x)(\ell_4-\ell_2 - \frac{qG_y}{K_y}))\\
    & \delta_{k_x^1}^{k_x^3}(q_x+G_x)\delta_{k_x^2}^{k_x^4}(-q_x-G_x) \delta_{k_y^3,k_y^1}(q_y+G_y)\delta_{k_y^4,k_y^2}(-q_y-G_y) \delta_{S_1,S_3}\delta_{S_2,S_4}\delta_{K_1,K_3}\delta_{K_2,K_4}\\
    &= \sum_{q,G}\sum_{\Gamma_2,\Gamma_4}\sum_{k_\parallel^2,k_\parallel^4} \delta_{k_x^1}^{k_x^3}(q_x+G_x)\delta_{k_x^2}^{k_x^4}(-q_x-G_x)\delta_{k_y^3,k_y^1}(q_y)\delta_{k_y^4,k_y^2}(-q_y-G_y)\delta_{S_1,S_3}\delta_{S_2,S_4}\delta_{K_1,K_3}\delta_{K_2,K_4} V(q+G)S^{n_1}_{n_3}(q+G)\\
    &S^{n_2}_{n_4}(-q-G)\exp(i(q_x+G_x)\ell_B^2(k_y^4-k_y^3+\frac{\ell_4-\ell_3}{q}K_y))\exp(\frac{2\pi i}{pK_x}(k_x^3-q_x)(\ell_3-\ell_1 + \frac{qG_y}{K_y}))\\
    &\exp(\frac{2\pi i }{pK_x}(k_x^4+q_x)(\ell_4-\ell_2 - \frac{qG_y}{K_y}))
\end{align}
We again make the usual assumption that the density matrix is diagonal in momentum, i.e $\Delta(k_\parallel^2,k_\parallel^4)\propto \delta_{k_\parallel^2,k_\parallel^4}$. However, since $k_\parallel^4 = k_\parallel^2 - q$ as crystal momenta, we find that $q$ is set to zero. Hence
\begin{align}
      &= \sum_{q,G}\sum_{\Gamma_2,\Gamma_4}\sum_{k_\parallel^2,k_\parallel^4} \delta_{k_x^1}^{k_x^3}(q_x+G_x)\delta_{k_x^2}^{k_x^4}(-q_x-G_x)\delta_{k_y^3,k_y^1}(q_y+G_y)\delta_{k_y^4,k_y^2}(-q_y-G_y)\delta_{S_1,S_3}\delta_{S_2,S_4}\delta_{K_1,K_3}\delta_{K_2,K_4}\\
      &\Delta_{\Gamma_4,\Gamma_2}(k_\parallel^4)\delta_{k_\parallel^4,k_\parallel^2} \delta_{S_1,S_3}\delta_{S_2,S_4}\delta_{K_1,K_3}\delta_{K_2,K_4}V(q+G)S^{n_1}_{n_3}(q+G)S^{n_2}_{n_4}(-q-G)\\
      &\exp(iq_x\ell_B^2(k_y^3-k_y^3+\frac{\ell_4-\ell_3}{q}K_y))\exp(\frac{2\pi i}{pK_x}(k_x^3-q_x)(\ell_3-\ell_1 + \frac{qG_y}{K_y}))\exp(\frac{2\pi i }{pK_x}(k_x^4+q_x)(\ell_4-\ell_2 - \frac{qG_y}{K_y}))\\
     &= \sum_{k_\parallel^4}\sum_{\Gamma_2,\Gamma_4}\Delta_{\Gamma_3,\Gamma_2}(k_x^4,k_y^4) \delta_{S_1,S_3}\delta_{S_2,S_4}\delta_{K_1,K_3}\delta_{K_2,K_4}V(q+G)S^{n_1}_{n_3}(q+G)S^{n_2}_{n_4}(-q-G)\\
     &\exp(iG_x\ell_B^2(k_y^4-k_y^3+\frac{\ell_4-\ell_3}{q}K_y))\exp(\frac{2\pi i}{pK_x}k_x^3(\ell_3-\ell_1 + \frac{qG_y}{K_y}))\exp(\frac{2\pi i }{pK_x}k_x^4(\ell_4-\ell_2 - \frac{qG_y}{K_y}))\\
        &= \sum_{k_\parallel^4,G}\sum_{\Gamma_2,\Gamma_4}\Delta_{\Gamma_4,\Gamma_2}(k_x^4,k_y^4) \delta_{S_1,S_3}\delta_{S_2,S_4}\delta_{K_1,K_3}\delta_{K_2,K_4}V(G)S^{n_1}_{n_3}(G)S^{n_2}_{n_4}(-G)\\
        &\exp(\frac{2\pi i}{pK_x}k_x^3(\ell_3-\ell_1 +\frac{qG_y}{K_y})) \exp(\frac{2\pi i }{pK_x}k_x^4(\ell_4-\ell_2 - \frac{qG_y}{K_y}))\exp(iG_x\ell_B^2(k_y^4-k_y^3+\frac{\ell_4-\ell_3}{q}K_y))
\end{align}

To go from the case of the classical dispersion to the graphene dispersion, we simply make the following linear combination: 
\begin{align}
    V^{n_1,\ell_1,k_\parallel^1,\alpha^1;n_2,\ell_2,k_\parallel^2,\alpha^2}_{n_3,\ell_3,k_\parallel^3,\alpha^3;n_4,\ell_4,k_\parallel^4,\alpha^4} = \frac{\sqrt{2}^{\sum_{i=1}^4\delta_{i,0}}}{4} &(V^{n_1,\ell_1,k_\parallel^1;n_2,\ell_2,k_\parallel^2}_{n_3,\ell_3,k_\parallel^3;n_4,\ell_4,k_\parallel^4} + \alpha_1\alpha_3V^{n_1-1,\ell_1,k_\parallel^1;n_2,\ell_2,k_\parallel^2}_{n_3-1,\ell_3,k_\parallel^3;n_4,\ell_4,k_\parallel^4}\Theta(n_1-1)\Theta(n_3-1)\\
    &+ \alpha_2\alpha_4V^{n_1,\ell_1,k_\parallel^1;n_2-1,\ell_2,k_\parallel^2}_{n_3,\ell_3,k_\parallel^3;n_4-1,\ell_4,k_\parallel^4}\Theta(n_2-1)\Theta(n_4-1)\\
    &+ \alpha_1\alpha_2\alpha_3\alpha_4 V^{n_1-1,\ell_1,k_\parallel^1;n_2-1,\ell_2,k_\parallel^2}_{n_3-1,\ell_3,k_\parallel^3;n_4-1,\ell_4,k_\parallel^4}\Theta(n_1-1)\Theta(n_2-1)\Theta(n_3-1)\Theta(n_4-1) )
\end{align}

Due to the SU(4) symmetric nature of the long-range Coloumb interaction, the interaction matrix elements are completely independent of the valley index. Furthermore, the non-interacting Hamiltonian is block-diagonal in the valley index as well as the spin index owing to the decoupling of the valleys and the absence of Zeeman terms.

\section{Accounting for the superlattice ionic background}
\label{app:background}
We wish to account for the effect of the superlattice ions on the Hartree-Fock Hamiltonian. We thus replace the density operator $\rho_q$ with $\delta\rho_q = \rho_q - \bar{\rho}_q$, where
\begin{equation}
    \bar{\rho}_q = \sum_{G,k}\tr(\Lambda^G(k))\delta_{q,G}
\end{equation}
Then the Coloumb part of the Hamiltonian becomes
\begin{equation}
    \sum_{q}V(q)\delta\rho_q\delta\rho_{-q} 
\end{equation}
We consider expanding the product. The term proportional to $\rho_q\rho_{-q}$ will be the usual Coloumb matrix element, computed in a previous section. The term $\bar{\rho}_{q}\bar{\rho}_{-q}$ will not be proportional to any field operators, so will not affect the dynamics of the Hamiltonian in any way, and thus may be neglected. Therefore, we need only account for the cross terms $\rho_q\bar{\rho}_{-q}$ and $\bar{\rho_{q}}\rho_{-q}$. Note that these are quadratic in the field operators, and so will enter the Hamiltonian as one-body terms. We define 
\begin{equation}
  I^1(k)  \equiv \sum_q V(q)\rho_q(k)\bar{\rho}_{-q}, I^2(k) \equiv \sum_q V(q)\rho_{-q}(k)\bar{\rho}_{q}(k)
\end{equation}
We now compute the matrix elements of these one body terms. First off,
\begin{equation}
    I^1_{\Gamma_1,\Gamma_3}(k) = \frac{1}{2}\sum_GV(G)\Lambda_{\Gamma_1,\Gamma_3}^G(k)\sum_{p}\tr(\Lambda_{\Gamma_2,\Gamma,4}^{-G}(p))
\end{equation}
Moreover,
\begin{equation}
    I^2_{\Gamma_2,\Gamma_4}(k) = \frac{1}{2}\sum_GV(G)\Lambda_{\Gamma_2,\Gamma_4}^{-G}(k)\sum_{p}\tr(\Lambda_{\Gamma_1,\Gamma_3}^{G}(p))
\end{equation}
where the trace is understood to run over orbital ($\Gamma$) indices.

\section{Probing states that break translational invariance}
\label{app:translation_breaking}
For translation-invariance breaking states, we expect a part of the density matrix to have the form $\Delta(k,k') \propto \delta_{k}^{k'}(Q)$ for some non zero vector $Q$ in the MBZ. Then we may modify our calculation for the direct and exchange terms to allow for this possibility. For the direct term, we see that instead of $q$ being set to zero, it is instead set to $Q$. Thus the second half of the calculation will instead proceed as follows:
\begin{align}
      &= \sum_{q,G}\sum_{\Gamma_2,\Gamma_4}\sum_{k_\parallel^2,k_\parallel^4} \delta_{k_x^1}^{k_x^3}(q_x+G_x)\delta_{k_x^2}^{k_x^4}(-q_x-G_x)\delta_{k_y^3,k_y^1}(q_y+G_y)\delta_{k_y^4,k_y^2}(-q_y-G_y)\delta_{S_1,S_3}\delta_{S_2,S_4}\delta_{K_1,K_3}\delta_{K_2,K_4}\\
      &\Delta_{\Gamma_4,\Gamma_2}(k_\parallel^4)\delta_{k_\parallel^4,k_\parallel^2}(Q) \delta_{S_1,S_3}\delta_{S_2,S_4}\delta_{K_1,K_3}\delta_{K_2,K_4}V(q+G)S^{n_1}_{n_3}(q+G)S^{n_2}_{n_4}(-q-G)\\
      &\exp(i(q_x+G_x)\ell_B^2(k_y^4-k_y^3+\frac{\ell_4-\ell_3}{q}K_y))\exp(\frac{2\pi i}{pK_x}(k_x^3-q_x)(\ell_3-\ell_1 + \frac{qG_y}{K_y}))\exp(\frac{2\pi i }{pK_x}(k_x^4+q_x)(\ell_4-\ell_2 - \frac{qG_y}{K_y}))\\
     &= \sum_{k_\parallel^4}\sum_{\Gamma_2,\Gamma_4}\Delta_{\Gamma_3,\Gamma_2}(k_x^4,k_y^4) \delta_{S_1,S_3}\delta_{S_2,S_4}\delta_{K_1,K_3}\delta_{K_2,K_4}V(Q+G)S^{n_1}_{n_3}(Q+G)S^{n_2}_{n_4}(-Q-G)\\
     &\exp(i(G_x+Q_x)\ell_B^2(k_y^4-k_y^3+\frac{\ell_4-\ell_3}{q}K_y))\exp(\frac{2\pi i}{pK_x}(k_x^3-Q_x)(\ell_3-\ell_1 + \frac{qG_y}{K_y}))\exp(\frac{2\pi i }{pK_x}(k_x^4+Q_x)(\ell_4-\ell_2 - \frac{qG_y}{K_y}))
\end{align}
Similarly, for the exchange term, we find that the second half of the calculation becomes
\begin{align}
      &= \sum_{q,G}\sum_{\Gamma_2,\Gamma_3}\sum_{k_\parallel^2,k_\parallel^3} \delta_{k_x^1}^{k_x^3}(q_x + G_x)\delta_{k_x^2}^{k_x^4}(-q_x - G_x)\delta_{k_y^3,k_y^1}(q_y + G_y)\delta_{k_y^4,k_y^2}(-q_y - G_y)\delta_{S_1,S_3}\delta_{S_2,S_4}\delta_{K_1,K_3}\delta_{K_2,K_4}\Delta_{\Gamma_3,\Gamma_2}(k_\parallel^3)\\
      &\delta_{k_\parallel^3,k_\parallel^2} (Q)\delta_{S_1,S_3}\delta_{S_2,S_4}\delta_{K_1,K_3}\delta_{K_2,K_4}  V(q+G)S^{n_1}_{n_3}(q+G)S^{n_2}_{n_4}(-q-G)\\
     &\exp(i(q_x+G_x)\ell_B^2(k_y^4-k_y^3+\frac{\ell_4-\ell_3}{q}K_y))\exp(\frac{2\pi i}{pK_x}(k_x^3-q_x)(\ell_3-\ell_1 + \frac{qG_y}{K_y}))\exp(\frac{2\pi i }{pK_x}(k_x^4+q_x)(\ell_4-\ell_2 - \frac{qG_y}{K_y}))\\
     &= \sum_{q,G}\sum_{\Gamma_2,\Gamma_3}\Delta_{\Gamma_3,\Gamma_2}(k_x^4+q_x - Q_x,k_y^4+q_y - Q_y) \delta_{S_1,S_3}\delta_{S_2,S_4}\delta_{K_1,K_3}\delta_{K_2,K_4}\\
      &V(q+G)S^{n_1}_{n_3}(q+G)S^{n_2}_{n_4}(-q-G)\exp(i(q_x+G_x)\ell_B^2(Q_y-q_y+\frac{\ell_4-\ell_3}{q}K_y))\\
      &\exp(\frac{2\pi i}{pK_x}(k_x^4-Q_x)(\ell_3-\ell_1 +\frac{qG_y}{K_y}))\exp(\frac{2\pi i }{pK_x}(k_x^4+q_x)(\ell_4-\ell_2 - \frac{qG_y}{K_y}))
\end{align}

\section{Matrix elements of the sublattice operators}
Our basis in the sublattice space has the form of the graphene Landau levels, i.e for the K valley it is
\begin{equation}
   \frac{1}{\sqrt{2}} \begin{bmatrix}
        \lambda\ket{n}\\
        \ket{n-1}
    \end{bmatrix},\begin{bmatrix}
        \ket{0}\\
        0
    \end{bmatrix}
\end{equation}
and for the K' valley it is
\begin{equation}
    \frac{1}{\sqrt{2}} \begin{bmatrix}
    \ket{n-1}\\
        \lambda\ket{n}
    \end{bmatrix},\begin{bmatrix}
    0\\
        \ket{0}
    \end{bmatrix}
\end{equation}
We begin by computing the matrix elements for the K valley $z$ sublattice operator:
\begin{align}
    1/2 \begin{bmatrix}
        \lambda\ket{n}\\
        \ket{n-1}
    \end{bmatrix}^\dagger
    \begin{bmatrix}
        1 & 0\\
        0 & -1
    \end{bmatrix}
    \begin{bmatrix}
        \lambda'\ket{m}\\
        \ket{m-1}
    \end{bmatrix} &= \frac{1}{2}(1-\lambda\lambda')\delta_{n,m}\\
   \frac{1}{\sqrt{2}}
    \begin{bmatrix}
        \lambda\ket{n}\\
        \ket{n-1}
    \end{bmatrix}^\dagger
    \begin{bmatrix}
        1 & 0\\
        0 & -1
    \end{bmatrix}
    \begin{bmatrix}
        \ket{0}\\
        0
    \end{bmatrix} &= \frac{1}{\sqrt{2}} \lambda\delta_{n,0} \\
    \begin{bmatrix}
        \ket{0}\\
        0
    \end{bmatrix}^\dagger
    \begin{bmatrix}
        1 & 0\\
        0 & -1
    \end{bmatrix}
    \begin{bmatrix}
        \ket{0}\\
        0
    \end{bmatrix} &= 1 
\end{align}
We now repeat the calculation for the $K'$ valley. The $z$ sublattice operator matrix elements in this valley are as follows:
\begin{align}
    1/2 \begin{bmatrix}
    \ket{n-1}\\
        \lambda\ket{n}
    \end{bmatrix}^\dagger
    \begin{bmatrix}
        1 & 0\\
        0 & -1
    \end{bmatrix}
    \begin{bmatrix}
    \ket{m-1}\\
        \lambda\ket{m}
    \end{bmatrix} &= \frac{1}{2}(\lambda\lambda'-1)\delta_{n,m} \\
   \frac{1}{\sqrt{2}}
    \begin{bmatrix}
    \ket{n-1}\\
        \lambda\ket{n}
    \end{bmatrix}^\dagger
    \begin{bmatrix}
        1 & 0\\
        0 & -1
    \end{bmatrix}
    \begin{bmatrix}
    0\\
        \ket{0}
    \end{bmatrix} &= \frac{-1}{\sqrt{2}} \lambda\delta_{n,0} \\
    \begin{bmatrix}
        0\\
        \ket{0}
    \end{bmatrix}^\dagger
    \begin{bmatrix}
        1 & 0\\
        0 & -1
    \end{bmatrix}
    \begin{bmatrix}
        0\\
        \ket{0}
    \end{bmatrix} &= -1 
\end{align}

These are matrix elements of the physical sublattice operator $\sigma_z$ acting on the original spinors $\psi$. In the valley-symmetric basis of Eq.~\eqref{eq:psi_tilde}, used for $H_{an}$ in Eq.~\eqref{eq:H_an}, the $K$ results are unchanged, while the $K'$ results for $\sigma_x$ and $\sigma_z$ change sign, since $\sigma_y\sigma_{x,z}\sigma_y = -\sigma_{x,z}$. With this sign, both valleys give $+1$ for the zeroth Landau level, consistent with valley-sublattice locking in that basis.

\section{Renormalization Group Flow in Finite Potential: Basic Setup}
\label{app:RG_setup}
First define  $w_0=\hbar v_F, w = \hbar v, G = \frac{2\pi}{L}$.
\begin{equation}
 E_G=wG,\qquad u=\frac{V}{wG},\qquad
 u_0=\frac{V}{w_0G}.\label{eq:udef}
\end{equation}
$u_0$ corresponds to the bare Fermi velocity. For $w_0=0.6582\,\mathrm{eV\,nm}$, $V=35\,\mathrm{meV}$, and $L=35\,\mathrm{nm}$, we find $E_{G,0}=118.16\,\mathrm{meV}$ and $u_0\simeq0.2962$.

For the RG calculation we use a continuum linear dispersion, in which the kinetic operator is $w\bm\sigma\cdot\mathbf k$ in both valleys. The sublattice labels in this basis must be transformed consistently when comparing to the physical-sublattice Landau-level convention used earlier. Define Pauli matrices $\tau_a$ in valley, $\sigma_b$ in sublattice, and $s_i$ in spin.
We define the dimensionless Coulomb strength by $\mathcal V_C(q)=2\pi\kappa w/q$, with $\kappa=e^2/(4\pi\epsilon_0\epsilon_r w)$. The anisotropy interaction takes the following form in position space: 
\begin{equation}
 H_{an}=\frac12\int\dd^2r\sum_{a=x,y,z}\sum_{b=x,y,z}
 g_{ab}:(\tilde\psi^\dagger(r)\tau_a\sigma_b\tilde\psi(r))^2: .\label{eq:contacts}
\end{equation}
The notation $\perp$ is implicitly taken to mean that $x$ and $y$ have the same coefficients, but both are included in the Hamiltonian. In particular $g_{\perp\perp}$ multiplies four bilinear squares, whereas $g_{zz}$ multiplies one. For each fixed valley channel use
\begin{equation}
 \bg_a=\begin{pmatrix}g_{az}\\g_{a\perp}\end{pmatrix},\qquad a=z,\perp .\label{eq:pair}
\end{equation}
The $\perp$ entries are coefficients per operator, not sums of coefficients. This convention is responsible for several factors of two below. The bare couplings (from Ref.~\cite{xuhuang2025}) are given by
\begin{equation}
 (g_{zz},g_{z\perp},g_{\perp z},g_{\perp\perp})^{\ee}=(184,25,43,269),
 \qquad g_{z\perp}^{\elph}=-52,\quad g_{\perp z}^{\elph}=-69.\label{eq:inputs}
\end{equation}

\section{Zero Potential Logarithmic Fermi Velocity Renormalization}
\label{app:RG_velocity}

The logarithmic renormalization of the Fermi velocity by the long-range Coulomb interaction is well established \cite{gonzalez1994,kotov2012,elias2011}; here we rederive the RPA rate used in the flow. We use the positive screening convention $W=\mathcal V_C/(1+\mathcal V_C\Pi_0)$.
There are $N_f=4$ physical spin--valley species, or $N=N_f/2=2$ in the
notation of the velocity flow. Temporarily setting $w=1$, the density bubble is
\begin{equation}
 \Pi_0(q,i\omega)=-N_f\int\frac{\dd p_0\,\dd^2p}{(2\pi)^3}
 \operatorname{tr}_\sigma[G_0(\mathbf p,ip_0)G_0(\mathbf p+\mathbf q,i(p_0+\omega))]
 \bigg|_{\rm subtracted}.
\end{equation}
The trace is over two sublattice components; spin and valley are counted by
$N_f$. The subtraction removes the
momentum-independent ultraviolet term, corresponding to the ionic correction. With $G_0(p)=-[ip_0+\bm\sigma\cdot\mathbf p]/(p_0^2+|\mathbf p|^2)$,
we get
\begin{equation}
 \Pi_0(q,i\omega)=\frac{N_fq^2}{2\pi\sqrt{q^2+\omega^2}}
 \int_0^1\sqrt{x(1-x)}\dd x
 =\frac{N_fq^2}{16\sqrt{q^2+\omega^2}}.
\end{equation}
Here the parameter integral equals $\pi/8$. Restoring $w$ gives
$\Pi_0=N_fq^2/[16\sqrt{w^2q^2+\omega^2}]$.
The bubble series for the screened line is
$W=\mathcal V_C-\mathcal V_C\Pi\mathcal V_C+\mathcal V_C\Pi\mathcal V_C\Pi\mathcal V_C-\cdots$.
This can easily be summed as a geometric series. With $x=\omega/(wq)$ and the homogeneous bubble,
\begin{equation}
 W_0(q,i\omega)=\frac{2\pi\kappa w/q}{1+\lambda/\sqrt{1+x^2}},
 \qquad\lambda=\frac{N\pi\kappa}{4}.\label{eq:RPAline}
\end{equation}
Now to understand the renormalization of the Fermi velocity, we need to expand the exchange self-energy. Let $\omega$ and $\mathbf q$ be the internal variables. Then
\begin{equation}
 \Sigma(\mathbf k,i\varepsilon)
 =-\int_{\rm shell}\frac{\dd^2q}{(2\pi)^2}\frac{\dd\omega}{2\pi}
 W_0(q,i\omega)\,G_0(\mathbf q+\mathbf k,i\omega+i\varepsilon).
 \label{eq:selfenergy_routing}
\end{equation}
The constant $\Sigma(\mathbf0,0)$ vanishes with a symmetric neutral reference. It is helpful to write the Green function as a function of its complex energy argument $z$:
\begin{equation}
 G_0(\mathbf q,z)=[z-w\bm\sigma\cdot\mathbf q]^{-1},\qquad
 G_0(\mathbf q,i\omega)=-\frac{i\omega+w\bm\sigma\cdot\mathbf q}{D},
 \quad D=\omega^2+w^2q^2.
\end{equation}
Its expansion about zero external variables is
\begin{equation}
 G_0(\mathbf q+\mathbf k,i\omega+i\varepsilon)
 ={}G_0(\mathbf q,i\omega)
 +\left.i\varepsilon\,\partial_zG_0(\mathbf q,z)\right|_{z=i\omega} +\nabla_k G_0(\mathbf q,i\omega)\cdot \mathbf{k} +\text{h.o.t}
\end{equation}
For the spatial derivative, we find
\begin{equation}
 \left.\partial_{k_i}G_0(\mathbf q+\mathbf k,i\omega)\right|_{\mathbf k=0}
 =-\frac{w\sigma_i}{D}
 +\frac{2w^2q_i(i\omega+w\bm\sigma\cdot\mathbf q)}{D^2}.
 \label{eq:selfenergy_spatial_derivative}
\end{equation}
Differentiating $(z-w\bm\sigma\cdot\mathbf q)G_0=\mathbf 1$ gives $G_0+(z-w\bm\sigma\cdot\mathbf q)\partial_zG_0=0$, hence $\partial_zG_0=-G_0^2$. Using this identity and $(\bm\sigma\cdot\mathbf q)^2=q^2$ we find
\begin{equation}
 \left.\partial_zG_0\right|_{z=i\omega}
 =-\frac{(i\omega+w\bm\sigma\cdot\mathbf q)^2}{D^2}
 =\frac{(\omega^2-w^2q^2)-2i\omega w\bm\sigma\cdot\mathbf q}{D^2}.
 \label{eq:selfenergy_frequency_derivative}
\end{equation}
The homogeneous screened interaction depends on $q=|\mathbf q|$, not its direction. We can therefore average each derivative over the polar angle $\phi$ before performing the radial and frequency integrals. With $\langle\cdot\rangle_\phi=(2\pi)^{-1}\int_0^{2\pi}\dd\phi\,(\cdot)$,
\begin{equation}
 \langle q_i\rangle_\phi=0,\qquad
 \langle q_iq_j\rangle_\phi=\frac{q^2}{2}\delta_{ij},\qquad
 \langle q_i\bm\sigma\cdot\mathbf q\rangle_\phi=\frac{q^2}{2}\sigma_i.
\end{equation}
The odd terms disappear, while the two spatial terms combine as
\begin{align}
 \left\langle\left.\partial_{k_i}G_0\right|_{\mathbf k=0}\right\rangle_\phi
 &=-\frac{w\sigma_i}{D}+\frac{w^3q^2\sigma_i}{D^2}
 =-\frac{w\sigma_i\omega^2}{D^2},\\
 \left\langle\left.\partial_zG_0\right|_{z=i\omega}\right\rangle_\phi
 &=\frac{\omega^2-w^2q^2}{D^2}.
\end{align}

Keeping the minus sign of the exchange self-energy in Eq.~\eqref{eq:selfenergy_routing}, we write
\begin{align}
 \Sigma(\mathbf k,i\varepsilon)-\Sigma(\mathbf0,0)
 &=a_\omega\,i\varepsilon+a_k\,w\bm\sigma\cdot\mathbf k
 +\text{h.o.t}\\
 a_\omega&=\int_{\rm shell}\frac{q\,\dd q}{2\pi}\frac{\dd\omega}{2\pi}
 W_0(q,i\omega)\frac{w^2q^2-\omega^2}{D^2},\\
 a_k&=\int_{\rm shell}\frac{q\,\dd q}{2\pi}\frac{\dd\omega}{2\pi}
 W_0(q,i\omega)\frac{\omega^2}{D^2}.
\end{align}
The physical velocity depends on both coefficients. The inverse propagator is written as
\begin{equation}
 G^{-1}=(1-a_\omega)i\varepsilon-(1+a_k)w\bm\sigma\cdot\mathbf k+\cdots.
\end{equation}
Normalizing the coefficient of $i\varepsilon$ to one gives
\begin{equation}
 \begin{aligned}
 w_{\rm new}&=w\frac{1+a_k}{1-a_\omega},\\
 \delta\ln w&=a_k+a_\omega+O(a_k^2,a_k a_\omega,a_\omega^2).
 \end{aligned}
\end{equation}
In their sum the frequency numerators combine and give
\begin{equation}
 \delta\ln w=\int_{\rm shell}\frac{q\,\dd q}{2\pi}\frac{\dd\omega}{2\pi}
 W_0(q,i\omega)\frac{w^2q^2}{(\omega^2+w^2q^2)^2}
 +\text{h.o.t}
\end{equation}
Setting $x=\omega/(wq)$ leaves a radial logarithm $\int_{\rm shell}\dd q/q=\dd t$, with $t$ increasing toward longer lengths. Dividing by $\dd t$ gives
\begin{equation}
 A=\frac{\kappa}{2\pi}\int_{-\infty}^{\infty}
 \frac{\dd x}{(1+x^2)^2[1+\lambda/\sqrt{1+x^2}]}
 =\frac{\kappa}{\pi}\int_0^{\pi/2}\frac{\sin^2\theta\,\dd\theta}{1+\lambda\sin\theta}.
 \label{eq:Aintegral}
\end{equation}
We can evaluate this integral with some work. For $\lambda>0$,
\begin{equation}
 \frac{\lambda\sin^2\theta}{1+\lambda\sin\theta}
 =\sin\theta-\frac1\lambda+\frac1{\lambda(1+\lambda\sin\theta)}.
\end{equation}
Hence $A=4[1-\pi/(2\lambda)+I(\lambda)/\lambda]/(\pi^2N)$, where
\begin{equation}
 I(\lambda)=\int_0^{\pi/2}\frac{\dd\theta}{1+\lambda\sin\theta}
 =2\int_0^1\frac{\dd z}{z^2+2\lambda z+1}
 =\begin{cases}
 \arccos\lambda/\sqrt{1-\lambda^2},&0\leq\lambda<1,\\
 1,&\lambda=1,\\
 \operatorname{arccosh}\lambda/\sqrt{\lambda^2-1},&\lambda>1.
 \end{cases}
\end{equation}
Using the definition of $w$ in terms of the renormalized Fermi velocity, we deduce that 
\begin{align}
 \lambda&=\frac{N\pi\kappa}{4},\qquad N=2,\label{eq:lambda}\\
 \frac{\dd\ln v}{\dd t}&=A(\lambda),\qquad
 \frac{\dd\lambda}{\dd t}=-A(\lambda)\lambda,\label{eq:velocity}\\
 A(\lambda)&=\frac{4}{\pi^2N}\left[1-\frac{\pi}{2\lambda}
 +\frac{\arccos\lambda}{\lambda\sqrt{1-\lambda^2}}\right].\label{eq:A}
\end{align}

For $\lambda>1$, the last expression is understood with
$\arccos\lambda/\sqrt{1-\lambda^2}$ replaced by
$\operatorname{arccosh}\lambda/\sqrt{\lambda^2-1}$, as in the preceding
piecewise formula; the ratio equals one at $\lambda=1$.

\section{Zero Potential Anisotropy Loop Integrals}

In this section we calculate the contact matrix, the linear generator
sending the instantaneous electronic anisotropy couplings to their mixed
one-loop Coulomb--contact correction at zero potential. We retain only terms
linear in the contact vertex. Scale
invariance makes the dimensionless bare-Coulomb matrix independent of the
cutoff. First define
\begin{equation}
 M_{\alpha\beta}=\tau_\alpha\otimes\mathbf 1_s\otimes\sigma_\beta,
 \qquad(T_M)_{ab;cd}=M_{ac}M_{bd}-M_{ad}M_{bc}.\label{eq:tensor}
\end{equation}
Here $a,b,c,d=1,\ldots,8$ are composite valley, spin, and sublattice indices: $a,b$ label the two outgoing fermions and $c,d$ the two incoming fermions. We allow $\alpha,\beta=0,x,y,z$ in the sixteen-contact projection, with $0$ denoting the identity. Normal ordering and fermionic anticommutation give
\begin{align}
 :\!(\psi^\dagger M\psi)^2\!:
 &=\sum_{abcd}M_{ac}M_{bd}\,
   \psi_a^\dagger\psi_b^\dagger\psi_d\psi_c,\\
 \sum_{abcd}(T_M)_{ab;cd}\,
   \psi_a^\dagger\psi_b^\dagger\psi_d\psi_c
 &=2:\!(\psi^\dagger M\psi)^2\!:.
\end{align}
For the second identity, interchange $c$ and $d$ in the exchange term and use $\psi_c\psi_d=-\psi_d\psi_c$. Thus the contact Hamiltonian can equivalently be written as
\begin{equation}
 H_g=\frac14\int\dd^2r\sum_{abcd}\Gamma_{ab;cd}\,
 \psi_a^\dagger\psi_b^\dagger\psi_d\psi_c,
 \qquad \Gamma=\sum_I g_I T_I.
 \label{eq:vertex_contact_dictionary}
\end{equation}
Let $\mathcal L$ denote the local mixed Coulomb--contact loop map, with its overall factor $\kappa$ removed. It includes the momentum and frequency integrals and all retained contractions, evaluated at the external projection. Because precisely one input contact vertex occurs in each retained diagram, this map is linear in that vertex. For an input tensor $T_J$, write its output as
\begin{equation}
 X_J\equiv\mathcal L[T_J]
 =\sum_I\widehat{\mathcal K}_{IJ}T_I+X_J^{\rm residual}
 \label{eq:matrix_column_dictionary}
\end{equation}
where $X_J^{\rm residual}$ is the projection of the output onto the subspace orthogonal to the one spanned by the $16$ couplings. Our goal is then to calculate the components of $\widehat{\mathcal K}_{IJ}$, from which we will extract the $2\times2$ contact matrix. For an arbitrary input $\Gamma=\sum_Jg_JT_J$, an infinitesimal fixed-background step then gives
\begin{align}
 \delta\Gamma
 &=\kappa\,\dd\tau\sum_Jg_JX_J,\\
 \delta g_I
 &=\kappa\,\dd\tau\sum_J\widehat{\mathcal K}_{IJ}g_J,
 \qquad
 \frac{\dd g_I}{\dd\tau}
 =\kappa\sum_J\widehat{\mathcal K}_{IJ}g_J.
 \label{eq:vertex_to_beta}
\end{align}
or in matrix-vector form
\begin{equation}
    \frac{d\mathbf g}{d\tau} = \kappa \widehat{\mathcal K} \mathbf g
\end{equation}
Define $\langle X,Y\rangle=\sum_{abcd}X_{ab;cd}^*Y_{ab;cd}$.
The coefficient extraction is the normal equation for orthogonal projection:
\begin{equation}
 \begin{aligned}
 X&=\sum_I c_I T_I+X_{\rm residual},\qquad
 \mathcal G_{IJ}=\langle T_I,T_J\rangle,\\
 c_I&=\sum_J(\mathcal G^{-1})_{IJ}\langle T_J,X\rangle.
 \end{aligned}\label{eq:gram}
\end{equation}
The tensors $T_I$ are not necessarily orthogonal. For the Hermitian Pauli products above,
\begin{equation}
 \mathcal G_{IJ}=2\{[\operatorname{Tr}(M_IM_J)]^2-
 \operatorname{Tr}(M_IM_JM_IM_J)\}.
\end{equation}
Using this definiton of $X_J$ gives 
\begin{equation}
     \widehat{\mathcal K}_{IJ}
 =\sum_R(\mathcal G^{-1})_{IR}\langle T_R,\mathcal L[T_J]\rangle = \sum_R(\mathcal G^{-1})_{IR}\langle T_R,X_J\rangle 
\end{equation}
To evaluate these coefficients, we must calculate the loop normalization coefficients. Note that the particularities of this calculation are specified to the pure Dirac cone case. However, the same setup is used in the finite-potential case, differing in that the band sums and momentum integrals must generally be evaluated numerically. We first use an instantaneous bare Coulomb line, with its factor $\kappa$ removed, and set $w=1$. Start with the Dirac propagator:
\begin{equation}
 G_0(\mathbf k,i\omega)=-\frac{i\omega+\sigma_xk_x+\sigma_yk_y}{\omega^2+k^2}
\end{equation}
At fixed background velocity define $\tau=-\ln(\Omega/\Omega_{\rm ref})$,  and put
\begin{equation}
    \chi_\Omega(\omega)=\frac{\omega^2}{\omega^2+\Omega^2},\quad
 R_\Omega^{\rm loop}\equiv\chi_\Omega\partial_\tau\chi_\Omega
 =\frac{2\Omega^2\omega^4}{(\omega^2+\Omega^2)^3}.
\end{equation}
The particle-hole part multiplies components at the same $(\mathbf k,i\omega)$, whereas the particle-particle part uses the partner $(-\mathbf k,-i\omega)$. Both nonzero components reverse sign under this simultaneous reversal. Consequently the diagonal products are
\begin{equation}
 \begin{array}{c|cc}
 &\mathrm{ph}&\mathrm{pp}\\ \hline
 \mu=0&-\omega^2/(\omega^2+k^2)^2&+\omega^2/(\omega^2+k^2)^2\\[2pt]
 \mu=i\in\{x,y\}&+k_i^2/(\omega^2+k^2)^2&-k_i^2/(\omega^2+k^2)^2
 \end{array}
\end{equation}
This reduces the calculation to two positive integrals. Define
\begin{align}
 I_0&=\int\frac{\dd\omega}{2\pi}R_\Omega^{\rm loop}(\omega)
 \int\frac{\dd^2k}{(2\pi)^2}\frac{2\pi}{k}
 \frac{\omega^2}{(\omega^2+k^2)^2},\\
 I_i&=\int\frac{\dd\omega}{2\pi}R_\Omega^{\rm loop}(\omega)
 \int\frac{\dd^2k}{(2\pi)^2}\frac{2\pi}{k}
 \frac{k_i^2}{(\omega^2+k^2)^2},\qquad i=x,y.
\end{align}
Once these magnitudes are known, the sign table gives
$L^{\pp}_{00}=I_0$, $L^{\ph}_{00}=-I_0$,
$L^{\pp}_{ii}=-I_i$, and $L^{\ph}_{ii}=I_i$.
All components involving $z$ vanish because $g_z=0$. Note
\begin{align}
 \int_0^\infty\frac{\omega^2\dd k}{(\omega^2+k^2)^2}
 &=\frac1{|\omega|}\int_0^{\pi/2}\cos^2\theta\,\dd\theta
 =\frac{\pi}{4|\omega|},\\
 \int_0^\infty\frac{k^2\dd k}{(\omega^2+k^2)^2}
 &=\frac1{|\omega|}\int_0^{\pi/2}\sin^2\theta\,\dd\theta
 =\frac{\pi}{4|\omega|}.
\end{align}
Using the evenness of the integrand gives
\begin{equation}
 I_0=\int_{-\infty}^{\infty}\frac{\dd\omega}{2\pi}
 \frac{2\Omega^2\omega^4}{(\omega^2+\Omega^2)^3}\frac{\pi}{4|\omega|}
 =\frac{\Omega^2}{2}\int_0^\infty
 \frac{\omega^3\dd\omega}{(\omega^2+\Omega^2)^3} = \frac{1}{8}
\end{equation}
We then deduce
\begin{equation}
 L^{\pp}_{00}=\frac18,\qquad L^{\ph}_{00}=-\frac18,
 \qquad L^{\pp}_{xx}=L^{\pp}_{yy}=-\frac1{16},\qquad
 L^{\ph}_{xx}=L^{\ph}_{yy}=\frac1{16},
\end{equation}
The independence of $\Omega$ is physically significant: in zero potential there is no intrinsic scale in this calculation. With these coefficients in hand, we turn back to the calculation of the contact matrix elements. 

\section{Calculating the contact matrix}
\label{app:contact_matrix}

The previously calculated $L_{\mu\nu}$  coefficients specify the propagation of the two internal fermions.
However, it remains to calculate the actual operators, and the contact matrix.
To get at the contact matrix, we must contract the internal indices of one Coulomb
vertex and one contact vertex. We keep both possible placements of the Coulomb
vertex. This is why the coefficients computed above are not, by themselves,
entries of the contact matrix.

We make this contraction as explicit as humanly possible. Write
$F_\mu=\mathbf 1_\tau\otimes\mathbf 1_s\otimes\sigma_\mu$.
For a fixed pair $\mu,\nu$, define the following maps on two four-leg tensors
$X,Y$:
\begin{align}
 [\mathcal C_0(X,Y)]_{ab;cd}
 &=-\sum_{ijkl}(F_\mu)_{ij}(F_\nu)_{kl}X_{ab;ik}Y_{jl;cd},\label{eq:contC0}\\
 [\mathcal C_1(X,Y)]_{ab;cd}
 &=\sum_{ijkl}(F_\mu)_{ij}(F_\nu)_{kl}X_{al;ci}Y_{jb;kd},\\
 [\mathcal C_2(X,Y)]_{ab;cd}
 &=\sum_{ijkl}(F_\mu)_{ij}(F_\nu)_{kl}X_{aj;ck}Y_{lb;id},\\
 [\mathcal C_3(X,Y)]_{ab;cd}
 &=-\sum_{ijkl}(F_\mu)_{ij}(F_\nu)_{kl}X_{bl;ci}Y_{ja;kd},\\
 [\mathcal C_4(X,Y)]_{ab;cd}
 &=-\sum_{ijkl}(F_\mu)_{ij}(F_\nu)_{kl}X_{bj;ck}Y_{la;id}.
 \label{eq:contC4}
\end{align}
The first contraction is particle--particle. The other four are the direct
and exchange particle--hole contractions. For example, $\mathcal C_0$ connects
the incoming pair $i,k$ of $X$ to the outgoing pair $j,l$ of $Y$; the indices
$ab;cd$ are the remaining free indices. The minus signs in $\mathcal C_3,\mathcal C_4$ follow
by exchanging the outgoing fermionic legs of $\mathcal C_1,\mathcal C_2$. 
In terms of their proportionality to the Coulomb weight, the relevant Coulomb tensors are given as
\begin{equation}
 C^{\pp}_{ab;cd}=\delta_{ac}\delta_{bd}-\delta_{ad}\delta_{bc},
 \qquad C^{\ph}_{ab;cd}=-\delta_{ad}\delta_{bc}.
\end{equation}
In the particle-hole channel the direct Coulomb transfer equals the zero external
transfer is removed by the ionic background correction. Note that the left and right legs are equal by re-indexing the sum. Then
the complete mixed correction to an input tensor $T$, divided by $\kappa$, is
\begin{align}
 \mathcal L[T]=\sum_{\mu\nu}\Big\{&
 2L^{\pp}_{\mu\nu}\mathcal C_0(C^{\pp},T)\\
 &+L^{\ph}_{\mu\nu}\big[
 \mathcal C_1(C^{\ph},T)+\mathcal C_3(C^{\ph},T)
 +\mathcal C_2(T,C^{\ph})+\mathcal C_4(T,C^{\ph})\big]\nonumber\\
 &+L^{\ph}_{\mu\nu}\big[
 \mathcal C_1(T,C^{\ph})+\mathcal C_3(T,C^{\ph})
 +\mathcal C_2(C^{\ph},T)+\mathcal C_4(C^{\ph},T)\big]\Big\}.
 \label{eq:contfull}
\end{align}
For zero potential only $\mu=\nu$ contributes, but it is useful to do this
algebra with arbitrary diagonal loop coefficients, since the same calculation
will also supply the finite-potential matrix. Define
\begin{equation}
 P_\mu=L^{\pp,L}_{\mu\mu}+L^{\pp,R}_{\mu\mu},\qquad
 H_\mu=L^{\ph,L}_{\mu\mu}+L^{\ph,R}_{\mu\mu},\qquad
 P_\perp=P_x+P_y,\quad H_\perp=H_x+H_y.
 \label{eq:contPH}
\end{equation}
Thus $P,H$ include both Coulomb placements. Note that the $L,R$ placements are written for clarity; they are in fact equal, which can be seen by re-indexing the sums that define them. In the homogeneous calculation
each placement has the value calculated above, so
\begin{equation}
 P_0=\frac14,\quad P_x=P_y=-\frac18,\quad
 H_0=-\frac14,\quad H_x=H_y=\frac18,\quad P_z=H_z=0.
 \label{eq:contPHzero}
\end{equation}
The sum in $P_\perp$ is a sum of loop weights. It is distinct from the
per-operator coupling $g_{a\perp}$.

Fix an input $M=M_{\alpha\beta}$ and put $F=F_\mu$.
Since $F^2=\mathbf 1$, Pauli conjugation gives
\begin{equation}
 FMF=\eta_{\mu\beta}M,\qquad
 \eta_{\mu\beta}=\begin{cases}
 +1,&\mu=0\text{ or }\mu=\beta,\\
 -1,&\mu\ne0,\ \mu\ne\beta,
 \end{cases}\qquad \beta=x,y,z.
\end{equation}
Here $T_{FM}$ means the tensor construction specified above and applied to the matrix product $FM$. Inserting $C^{\pp}$ in the first contraction gives
\begin{align}
 [\mathcal C_0(C^{\pp},T_M)]_{ab;cd}
 &=-\sum_{jl}(F_{aj}F_{bl}-F_{bj}F_{al})(T_M)_{jl;cd}\\
 &=-2\big[(FM)_{ac}(FM)_{bd}-(FM)_{ad}(FM)_{bc}\big]
 =-2(T_{FM})_{ab;cd}.
\end{align}
The factor two follows from antisymmetry of $T_M$ in its outgoing pair.
The other placement gives $-2T_{MF}$, which equals $-2T_{FM}$ because
$MF=\eta_{\mu\beta}FM$. Hence the total particle-particle contribution at this $\mu$ is
$-2P_\mu T_{FM}$. For the particle-hole contribution, a representative contraction is
\begin{equation}
 [\mathcal C_1(C^{\ph},T_M)]_{ab;cd}
 =-(FMF)_{ac}M_{bd}+(FM)_{ad}(MF)_{bc}.
\end{equation}
Adding its outgoing-leg exchange and the two terms from $\mathcal C_2,
\mathcal C_4$ in the same placement gives
\begin{align}
 Q_\mu(M)_{ab;cd}={}&-(FMF)_{ac}M_{bd}+(FMF)_{bc}M_{ad}
 -M_{ac}(FMF)_{bd}+M_{bc}(FMF)_{ad}\nonumber\\
 &+(FM)_{ad}(MF)_{bc}-(FM)_{bd}(MF)_{ac}
 +(MF)_{ad}(FM)_{bc}-(MF)_{bd}(FM)_{ac}\nonumber\\
 ={}&-2\eta_{\mu\beta}(T_M+T_{FM})_{ab;cd}.
\end{align}
The other placement reduces to the same expression. We have therefore
obtained the useful rule
\begin{equation}
 \mathcal L_{\rm diag}[T_M]
 =\sum_{\mu=0,x,y,z}
 \left[-2P_\mu T_{F_\mu M}
 -2\eta_{\mu\beta}H_\mu(T_M+T_{F_\mu M})\right].
 \label{eq:contcompact}
\end{equation}
We can now pratically read off the entries of the contact matrix. Fix a valley index $a=x,y,z$. To avoid confusing it with a four-leg index,
in the rest of this subsection $T_{ab}$ denotes the contact tensor labeled
by valley $a$ and sublattice $b$. First take the input $T_{az}$. The relevant
products are
\begin{equation}
 F_0M_{az}=M_{az},\quad F_xM_{az}=-iM_{ay},\quad
 F_yM_{az}=iM_{ax},\quad F_zM_{az}=M_{a0}.
\end{equation}
For example, $T_{F_xM_{az}}=-T_{ay}$, which explains the sign of the mixing
from a $z$ input into a perpendicular output. Substitution yields
\begin{align}
 \mathcal L_{\rm diag}[T_{az}]={}&
 (-2P_0-4H_0+2H_x+2H_y-2H_z)T_{az}\nonumber\\
 &+2(P_y-H_y)T_{ax}+2(P_x-H_x)T_{ay}
 -2(P_z+H_z)T_{a0}.
 \label{eq:contcolumnz}
\end{align}
For square symmetry the $ax$ and $ay$ coefficients are equal. Their common
value is their average, $P_\perp-H_\perp$. Consequently the  first column is
\begin{equation}
 \begin{pmatrix}
 -2P_0-4H_0+2H_\perp-2H_z\\ P_\perp-H_\perp
 \end{pmatrix}.
\end{equation}

For the second column, the unit input is $T_{ax}+T_{ay}$, since
$g_{a\perp}$ multiplies both operators. Applying the same rule separately gives
\begin{align}
 \mathcal L_{\rm diag}[T_{ax}]={}&
 (-2P_0-4H_0-2H_x+2H_y+2H_z)T_{ax}\nonumber\\
 &+2(P_z-H_z)T_{ay}+2(P_y-H_y)T_{az}-2(P_x+H_x)T_{a0},\\
 \mathcal L_{\rm diag}[T_{ay}]={}&
 (-2P_0-4H_0+2H_x-2H_y+2H_z)T_{ay}\nonumber\\
 &+2(P_z-H_z)T_{ax}+2(P_x-H_x)T_{az}-2(P_y+H_y)T_{a0}.
\end{align}
Adding them and using $H_x=H_y$ gives
\begin{align}
 \mathcal L_{\rm diag}[T_{ax}+T_{ay}]={}&
 2(P_\perp-H_\perp)T_{az}
 +(-2P_0+2P_z-4H_0)(T_{ax}+T_{ay})\nonumber\\
 &-2(P_\perp+H_\perp)T_{a0}.
 \label{eq:contcolumnperp}
\end{align}
Reading the coefficients of $T_{az}$ and $T_{ax}+T_{ay}$ now gives
\begin{equation}
 \widehat{\mathsf K}=
 \begin{pmatrix}
 -2P_0-4H_0+2H_\perp-2H_z&2(P_\perp-H_\perp)\\
 P_\perp-H_\perp&-2P_0+2P_z-4H_0
 \end{pmatrix}.
 \label{eq:contK}
\end{equation}
At zero potential, the preivously calculated integrals give
\begin{equation}
 \widehat{\mathsf K}(0)=
 \begin{pmatrix}
 -\frac12+1+\frac12&2(-\frac14-\frac14)\\
 -\frac14-\frac14&-\frac12+1
 \end{pmatrix}
 =\begin{pmatrix}1&-1\\-\frac12&\frac12\end{pmatrix}
 \equiv\mathsf K_0.
 \label{eq:contKzero}
\end{equation}
The $T_{a0}$ outputs cancel because $P_z+H_z=P_\perp+H_\perp=0$.
The anisotropy flow (neglected the renormalized velocity) is therefore given as
\begin{equation}
 \frac{\dd\bg_a^{\ee}}{\dd\tau}=\kappa\mathsf K_0\bg_a^{\ee},
 \qquad
 \frac{\dd g_{az}^{\ee}}{\dd\tau}=\kappa(g_{az}^{\ee}-g_{a\perp}^{\ee}),\quad
 \frac{\dd g_{a\perp}^{\ee}}{\dd\tau}=-\frac\kappa2(g_{az}^{\ee}-g_{a\perp}^{\ee}).
\end{equation}
This reproduces the contact matrix used in Xu-Huang \cite{xuhuang2025}.

\section{Incorporating the renormalized velocity in the flow}

To account for the zero-potential velocity flow, the prescription of Ref.~\cite{xuhuang2025} is to modify the flow equations in the following way:
\begin{equation}
 \frac{\dd\bg_a^{\ee}}{\dd t}=4A[\lambda(t)]\mathsf K_0\bg_a^{\ee},
 \qquad t=\ln(\ell/a_0).
 \label{eq:contRPA}
\end{equation}
Here $a_0$ is the microscopic length at which the bare couplings are specified. Since  $A=\kappa/4+O(\kappa^2)$,
we can at least understand this generalization as matching at zero flow.

\section{Finite Potential Flow Equations}
\label{app:RG_finiteV}

We now retain a smooth square scalar potential
\begin{equation}
 U(\mathbf r)=-2V[\cos(Gx)+\cos(Gy)].
 \label{eq:contpotential}
\end{equation}
Thus $V$ is the magnitude of one Fourier coefficient, not the peak-to-peak
variation of $U$, which is $8|V|$. The potential is proportional to the
identity in spin, valley, and sublattice. We continue to evaluate zero-field
loops; this, of course, another approximation.
At finite potential the contact vertex will generally acquire momentum
dependence. Here we define its local part by setting all external original
plane-wave momenta and frequencies to zero. This is yet another approximation. For $\mathbf k$ in the mini-Brillouin zone and reciprocal vectors
$\mathbf Q=G(m,n)$, the one-body matrix diagonalizing the zero-field Hamiltonian gives eigenvalues $E_{n\mathbf k}$,
and normalized eigenvectors $u_{n\mathbf k}(\mathbf Q)$. Its propagator is
\begin{equation}
 G_{U,\mathbf Q\mathbf Q'}(\mathbf k,i\omega)=
 \sum_n\frac{u_{n\mathbf k}(\mathbf Q)u_{n\mathbf k}(\mathbf Q')^\dagger}
 {i\omega-E_{n\mathbf k}}.
\end{equation}
We measure energies in $wG$ and momenta in $G$ in the following loop formulas,
so that $w=G=1$, $u=V/(wG)$, and $s=\Omega/(wG)$.
The Coulomb weight with $\kappa$ removed is $D(\mathbf p)=2\pi/|\mathbf p|$.
Writing $G_U=\sum_\mu g_\mu\sigma_\mu$ gives
\begin{equation}
 g_\mu(\mathbf Q,\mathbf Q';\mathbf k,i\omega)=
 \sum_n\frac{m_{n\mathbf k}^\mu(\mathbf Q,\mathbf Q')}{i\omega-E_{n\mathbf k}},
 \quad m_{n\mathbf k}^\mu(\mathbf Q,\mathbf Q')=
 \frac12u_{n\mathbf k}(\mathbf Q')^\dagger\sigma_\mu u_{n\mathbf k}(\mathbf Q).
 \label{eq:contspectral}
\end{equation}
This decomposition retains both excitation energies and wavefunction overlaps.
Notably, replacing $v$ by a miniband slope alone would not only produce different numerators, but would give different signs of the final RG flow lines. At fixed $w$, use the same frequency regulator as before. Suppressing only
the already explicit arguments of $g_\mu$, the left-placement arrays are
\begin{align}
 L^{\pp}_{\mu\nu}(u,s)
 &=\int_{\rm MBZ}\frac{\dd^2k}{(2\pi)^2}\int\frac{\dd\omega}{2\pi}
 R_s^{\rm loop}(\omega)\sum_{\mathbf Q\mathbf Q'}D(\mathbf k+\mathbf Q)
 g_\mu(\mathbf Q,\mathbf Q';\mathbf k,i\omega)
 g_\nu(-\mathbf Q,-\mathbf Q';-\mathbf k,-i\omega),\label{eq:contLpp}\\
 L^{\ph}_{\mu\nu}(u,s)
 &=\int_{\rm MBZ}\frac{\dd^2k}{(2\pi)^2}\int\frac{\dd\omega}{2\pi}
 R_s^{\rm loop}(\omega)\sum_{\mathbf Q\mathbf Q'}D(\mathbf k+\mathbf Q)
 g_\mu(\mathbf Q,\mathbf Q';\mathbf k,i\omega)
 g_\nu(\mathbf Q',\mathbf Q;\mathbf k,i\omega).\label{eq:contLph}
\end{align}
This generalization is easy to understand. At finite potential, the eigenstates are linear combinations of momentum states, i.e expressible as Fourier transforms. Then the resultant weights are roughly speaking Fourier transforms of the zero-potential ones, weighted by the appropriate structure factors.\\\\
The momentum integrals and band sums require numerical evaluation, but the
frequency integrals remain analytically doable.
For any two band energies $E,F$, define
\begin{equation}
 W^{\ph}_s(E,F)=\int\frac{\dd\omega}{2\pi}
 \frac{R_s^{\rm loop}(\omega)}{(i\omega-E)(i\omega-F)},\qquad
 W^{\pp}_s(E,F)=\int\frac{\dd\omega}{2\pi}
 \frac{R_s^{\rm loop}(\omega)}{(i\omega-E)(-i\omega-F)}.
\end{equation}
To evaluate them, first consider
\begin{equation}
 J_s(E)=\int\frac{\dd\omega}{2\pi}\frac{R_s^{\rm loop}(\omega)}{i\omega-E}
 =-\frac{2Es^2}{\pi}\int_0^\infty
 \frac{\omega^4\dd\omega}{(\omega^2+s^2)^3(\omega^2+E^2)}.
\end{equation}
The odd part vanishes. Partial fractions, or differentiating integrals of
$(\omega^2+s^2)^{-1}$ with respect to $s^2$, gives
\begin{equation}
 \int_0^\infty\frac{\omega^4\dd\omega}
 {(\omega^2+s^2)^3(\omega^2+a^2)}
 =\frac{\pi(3a+s)}{16s(a+s)^3},\qquad a\geq0.
\end{equation}
Consequently $J_s(E)=-\varphi_s(E)$, where
\begin{equation}
 \varphi_s(E)=\frac{Es(3|E|+s)}{8(|E|+s)^3}.
\end{equation}
Using partial fractions in the two propagator denominators now yields
\begin{equation}
 W^{\ph}_s(E,F)=-\frac{\varphi_s(E)-\varphi_s(F)}{E-F},\qquad
 W^{\pp}_s(E,F)=\frac{\varphi_s(E)+\varphi_s(F)}{E+F}.
 \label{eq:contweights}
\end{equation}

The contact matrix is then calculated in the same way as in the zero-potential case. We again combine it with the velocity flow to get
\begin{equation}
 \frac{\dd\bg_a^{\ee}}{\dd t}
 =4A[\lambda(t)]\widehat{\mathsf K}\bg_a^{\ee},\qquad
 \frac{\dd\ln v}{\dd t}=A[\lambda(t)],\quad
 \lambda(t)=\frac{\pi\kappa_0}{2}\frac{v_F}{v(t)}.
 \label{eq:contmaster}
\end{equation}
These are our finite-potential RG flow equations.

\clearpage
\twocolumngrid

\end{document}